\documentclass[%
 aip,
 amsmath,amssymb,
 reprint,%
]{revtex4-1}

\usepackage{makecell}
\usepackage{graphicx}
\usepackage{dcolumn}
\usepackage{bm}

\usepackage[utf8]{inputenc}
\usepackage[T1]{fontenc}
\usepackage{mathptmx}
\usepackage{etoolbox}
\usepackage{mathtools}

\makeatletter
\def\@email#1#2{%
 \endgroup
 \patchcmd{\titleblock@produce}
  {\frontmatter@RRAPformat}
  {\frontmatter@RRAPformat{\produce@RRAP{*#1\href{mailto:#2}{#2}}}\frontmatter@RRAPformat}
  {}{}
}%
\makeatother
\begin{document}

\preprint{AIP/123-QED}

\title
{Quantifying non-equilibrium pressure-gradient turbulent boundary layers through a symmetry-based framework}
\author{Wei-Tao Bi}%
\email{weitaobi@pku.edu.cn}
\affiliation{
School of Mechanics and Engineering Science, Peking University, Beijing 100871, China
}
\affiliation{
State Key Laboratory for Turbulence and Complex Systems, Peking University, Beijing 100871, China
}
\author{Ke-Xin Zheng}
\affiliation{
School of Mechanics and Engineering Science, Peking University, Beijing 100871, China
}
\affiliation{
State Key Laboratory for Turbulence and Complex Systems, Peking University, Beijing 100871, China
}
\author{Jun Chen}%
\affiliation{
State Key Laboratory for Turbulence and Complex Systems, Peking University, Beijing 100871, China
}
\affiliation{
College of Engineering, Peking University, Beijing 100871, China
}

\author{Zhen-Su She}%
\affiliation{
School of Mechanics and Engineering Science, Peking University, Beijing 100871, China
}%
\affiliation{
State Key Laboratory for Turbulence and Complex Systems, Peking University, Beijing 100871, China
}
\date{\today}

\begin{abstract}
This study establishes a symmetry-based framework to quantify non-equilibrium processes in complex pressure-gradient (PG) turbulent boundary layers (TBLs), using a Lie-group-informed dilation-symmetry-breaking formalism. We derive a universal multilayer defect scaling law for the evolution of total shear stress (TSS). The law shows that gradually varying adverse pressure gradients (APGs) break the dilation symmetry in the two-layer defect scaling of equilibrium TSS, leading to three-layer TSS structures. For abrupt PG transitions, we identify boundary-layer decoupling into: 1) an equilibrium internal boundary layer, and 2) a history-dependent outer flow, arising from disparate adaptation timescales. The framework introduces a unified velocity scale $u_{**}$ mapping non-equilibrium TSS to canonical zero-PG scaling. Validation spans different aerodynamic systems, including developing APG on airfoils, APG-to-favorable-PG transition on a Gaussian bump, and favorable PG to rapidly amplifying APG in a converging-diverging channel. The work enables improved prediction of non-equilibrium PG TBL behavior through unified characterization of stress evolution dynamics, providing new physics-based parameterizations that could promote machine learning of complex wall-bounded turbulent flows.
\end{abstract}

\maketitle

\section{Introduction}\label{sec:Introduction}
Turbulent boundary layers (TBLs) driven by pressure gradients (PGs) under non-equilibrium conditions are ubiquitous in aerodynamic surfaces, marine vehicles, and industrial systems where fluid-structure interactions dictate performance. Predicting such flows is critical for mitigating flow separation in aircraft wings, optimizing energy efficiency, and improving equipment durability under unsteady loads. However, despite decades of research, the physics of PG-dominated TBL under non-equilibrium conditions remains a fundamental challenge. This challenge exposes persistent gaps between multiscale turbulence dynamics and engineering requirements.

Here, non-equilibrium refers to flows that deviate from those driven by PG under equilibrium conditions. According to Devenport and Lowe,\cite{devenport2022equilibrium} equilibrium flows are self-preserving, such that their statistical profiles admit invariant representations when normalized by locally-defined length and velocity scales. Equilibrium PG TBLs occur when the freestream velocity varies as $u_{\infty}=A|x - x_0|^{m}$, where $A$, $m$, and $x_0$ are constants.\cite{townsend1961equilibrium} Conventional analyses \cite{clauser1954turbulent,townsend1961equilibrium,Mellor1966,bradshaw1967turbulence,castillo2001similarity, maciel2006self} predict that this condition leads to constant Clauser PG parameter, $\beta\equiv (\delta_1/\tau_w)(\partial p/\partial x)_{\delta}={\rm const.}$, where $\delta_1$ is the displacement thickness, $\tau_w$ is the wall shear stress, and $(\partial p/\partial x)_{\delta}$ is the streamwise PG at the boundary-layer edge $\delta$. However, Vigdorovich \cite{Vigdorovich2013,Vigdorovich2014} demonstrated that $\beta$ alone is insufficient to characterize all equilibrium states. He identified four distinct similarity regimes characterized by $\beta$ and three other parameters, respectively. In real scenarios, TBLs are seldom in equilibrium states due to surface curvature or non-canonical variations in freestream velocity.

Numerous two-dimensional (2D) flow configurations have been investigated for studying non-equilibrium PG TBLs, including planar TBLs under varying PG conditions,\cite{Marusic1995,Bobke2017les,Coleman2018,abe2019direct,Volino2020,Romero2022jfm,Rkein2023dns,Parthasarathy2023,Gungor2024} TBLs past airfoils,\cite{Vinuesa2017wing,Vinuesa2018wing,Tanarro2020JFM-Wing} surface-mounted 2D bumps,\cite{Cavar2011,Matai2019,balin2021direct,Uzun2022AIAA,balin2024direct} periodic hills,\cite{Kahler2016JFM-exp} and flows through converging-diverging channels.\cite{Laval2012JoT} Additionally, studies have examined more complex flows involving three-dimensional effects,\cite{Abe2020JFM,QiWG2022PoF,Zuo2023JFM-SWBLI} compressibility,\cite{Tonicello2022JFM-Ramp} and surface roughness. \cite{Vishwanathan2023,Volino2023} Several key questions arise: Are there standard non-equilibrium processes shared among these diverse flows? Can we establish a general framework to quantify such processes? What reliable assumptions can inform this framework, \cite{Perry1994taumodel} and which flow parameters are most appropriate for characterization? 

These studies have documented the primary effects of non-equilibrium PG on mean-flow characteristics and turbulence statistics.\cite{devenport2022equilibrium} When TBLs are exposed to sustained developing adverse PG (APG), they exhibit transitional features toward equilibrium APG TBLs.\cite{Marusic1995,Bobke2017les,abe2019direct} These features include thinning of the inner region, strengthening of the wake region in the mean velocity profiles, and amplification of the Reynolds shear and normal stresses (when normalized by the friction velocity $u_\tau$). Notably, this amplification occurs particularly through the formation of an outer peak in the streamwise velocity variance.\cite{Bobke2017les} The peak Reynolds stresses in such flows remain significantly lower than those in equilibrium APG TBLs at equivalent $\beta$.\cite{Marusic1995,Bobke2017les} This discrepancy is attributed to a pronounced hysteresis effect in the outer region, where the slow response to increasing APG delays complete adaptation to equilibrium conditions.\cite{devenport2022equilibrium} Open questions regarding these flows include: What are the parameters suitable to characterize the strength of departure of a TBL from the equilibrium state? \cite{JoTreview2024} How can these non-equilibrium behaviors be quantified?

When TBLs experience sudden APG removal or rapid PG sign alternation, an internal boundary layer (IBL) forms near the wall. Above this IBL lies outer turbulence inherited from upstream conditions. This phenomenon is more commonly found in TBLs with abrupt changes of wall conditions,\cite{Smits1985} such as surface curvature, roughness, and wall heat flux, and it represents a radical departure from equilibrium conditions.\cite{devenport2022equilibrium} For transitions from favorable PG (FPG) to APG, the IBL manifests on the leeward side of bumps \cite{Baskaran1987,Cavar2011,balin2021direct,Uzun2022AIAA} and within converging-diverging channels.\cite{Tsuji1976,Laval2012JoT,Parthasarathy2023} Conversely, for APG-FPG transitions, it occurs on the windward side of bumps. In such flows, Reynolds stress profiles exhibit a distinct ``knee'' shape,\cite{Tsuji1976,Baskaran1987,Cavar2011} signaling IBL formation. This characteristic reflects rapid near-wall adaptation to local PG changes, contrasting with the slow response of outer-region turbulence, as proposed by Smits and Wood.\cite{Smits1985} Open questions regarding such dual-boundary-layer flows include: How can the coupling and decoupling between IBL and outer flow be parameterized such that the evolution dynamics of the TBLs are unveiled and quantified? What are the implications of such radical non-equilibrium conditions for Reynolds-averaged Navier-Stokes (RANS) modelling, as well as for the overall performance of aero/naval vehicles? \cite{JoTreview2024}

Although investigating these critical issues is the primary aim of this paper, some of these issues are already widely studied in the existing literature. As demonstrated by Bobke et al., \cite{Bobke2017les} $\beta$ and Reynolds number ($Re$) insufficiently characterize non-equilibrium TBLs. Their analysis revealed that the less pronounced wake region observed in wing-section TBLs, when compared with equilibrium TBLs at matched $\beta$ and $Re$, could be attributed to diminished cumulative effects of the $\beta$ history. Several authors identified deviation of the Clauser shape factor, $G=(u_\infty/u_\tau)(H-1)/H$ (where $H$ is the shape factor), from Mellor and Gibson's equilibrium curve \cite{Mellor1966} as a marker of TBL departure from equilibrium.\cite{Knopp2021,Volino2023} Utilizing the stress balance and a hybrid inner velocity scale,\cite{Romero2022fluid} Romero et al. \cite{Romero2024} explored a scenario where an APG TBL at nominally fixed $\beta$ undergoes an abrupt APG removal. They linked the failure of the hybrid velocity scale in this flow with the alteration of the internal leading-order stress balance, hypothesizing this change as an indicator of crossover from equilibrium to non-equilibrium states. Gungor et al. \cite{Gungor2024} observed significant cumulative effects of continuous PG variations, demonstrating that parameters based solely on local variables ($\beta$ and its gradient) cannot fully describe the physics of non-equilibrium TBLs. This finding urges the identification of appropriate parameters and characteristic response length scales specific to non-equilibrium PG TBLs.

Kianfar and Johnson \cite{Kianfar2025} applied an angular momentum integral (AMI) equation to analyze TBLs over airfoils, Gaussian bumps, and flat plates. Their analysis quantifies how turbulence and PG competitively alter skin friction via torque-like terms. A new Clauser-like parameter, $\beta_\ell$, based on AMI, better captures PG history effects compared to $\beta$, revealing distinct mechanisms in APG/FPG and relaminarization dynamics. The framework offers interpretable insights for complex boundary layer physics. Wei et al. \cite{WeiT2024} proposed a new momentum integral equation for TBLs under arbitrary PGs by incorporating wall-normal momentum advection. Validated via direct numerical simulation (DNS) data, their study overcomes classical K\'{a}rm\'{a}n integral's limitations in strong APGs, introduces a new PG parameter $\beta_k$ that remains stable compared to the traditional $\beta$ in separated flows, and offers an approximate method for experimental wall shear stress determination.

Many researchers have tried to identify appropriate velocity scales for quantifying mean velocity and Reynolds stresses in both equilibrium and non-equilibrium PG TBLs. Several authors \cite{skote2002direct,Romero2022fluid,balin2024direct,McKeonJFM2025} employed a hybrid inner velocity scale, which incorporates both $Re$ and PG effects, to quantify Reynolds shear stress in the overlap region. Through a scaling patch analysis of the mean momentum equations, Wei and Knopp \cite{wei_knopp_2023} proposed an outer scaling in which the characteristic velocity and length scales are determined from the location of peak Reynolds shear stress. This outer scaling was validated to effectively collapse datasets on APG TBLs over a wide range of $Re$ and PG. Ma et al. \cite{MaYan2024} proposed a hybrid outer velocity scale which captures the variations of peak Reynolds shear stress with $\beta$ and $Re$ in both equilibrium APG TBLs and non-equilibrium TBLs approaching separation. Prakash et al. \cite{balin2024direct} obtained a similar scaling with an outer velocity scale derived from the integrated mean momentum equation. Maciel et al. \cite{Maciel2018} identified the Zagarola-Smits (ZS) velocity \cite{zagarola1998mean} $u_{zs}$ as the optimal outer velocity scale for APG TBLs, deriving consistent dimensionless parameters ($\beta_{ZS}$, $\alpha_{ZS}$, $Re_{ZS}$) that precisely track force balances in the outer region across all velocity-defect conditions. Han et al. \cite{YanJFM2024} extended the Zagarola-Smits scaling with history-effect correction, consistently collapsing velocity defects and Reynolds stresses in APG TBLs across diverse databases.
  
Modeling non-equilibrium PG TBLs remains a formidable challenge. Perry, Marusic, and Jones  \cite{perry2002streamwise} integrated the log-law, the Coles' wake law, and the boundary-layer equations into a four-parameter framework to predict mean velocity, Reynolds shear stress, and total shear stress (TSS) profiles under arbitrary PG conditions. By modeling the parameters using empirical closure and the attached eddy hypothesis, the framework was successfully applied to describe diverse flows from APG TBLs in relaxing and developing states to FPG TBLs approaching equilibrium sink flow. The researchers emphasized the critical need for improved closure models and experimental validation under extreme non-equilibrium conditions. 

Yang et al. \cite{YangX2024} developed a predictive near-wall model for TBLs with arbitrary PGs. By inverting a Navier-Stokes-based velocity transformation,\cite{ChenShiYang2023} the model introduces a transport equation to track the Lagrangian-integrated TSS, eliminating empirical calibration. Validated against experimental and numerical data, it outperforms classical RANS models and Kays' empirical correction under strong PGs, while matching equilibrium models in mild/moderate conditions. Results reveal that history effects primarily matter in extreme PGs, and momentum equations inherently capture non-equilibrium physics via PG and material-derivative terms. The framework offers robustness without added parameters, advancing wall-modeled simulations for complex flows. 

Through asymptotic expansions, Ma et al. \cite{MaPoF2024-2} established multilayered descriptions for mean velocity profiles in APG TBLs up to separation. They developed a unified log-pressure law, which demonstrates progressive breakdown of the log-law into PG-dominated scaling. A critical Reynolds number ($Re_c=30$) defines overlapping regions, while a new half-power law emerges in outer layers under strong APGs. The multilayer framework, validated by experiments and DNS data, clarifies APG flow physics and provides foundational insights for turbulence modeling. 

While TBLs under non-equilibrium PG conditions have constituted a persistent research focus for decades, this discussion examines selective seminal advances from recent investigations. For a comprehensive synthesis of historical developments and contemporary debates, readers are directed to the authoritative review by Devenport and Lowe.\cite{devenport2022equilibrium} Broadly speaking, despite sustained efforts, current findings remain fragmented: preliminary insights into localized flow phenomena coexist with persistent ambiguities in governing mechanisms. At the same time, the absence of unifying theories underscores the field's conceptual stagnation. Novel conceptual frameworks combined with advanced methodologies thus become imperative to reconcile multiscale interactions and advance predictive capabilities.

This study builds upon a fundamental paradox in turbulence research:\cite{Josserand2020} While mean-flow properties form the cornerstone of fluid dynamic analysis, they remain primarily irreducible to first principles, with few exceptions rooted in classical theories. This paradox stems from turbulence's inherent nonlinearity and multiscale coupling, preventing direct statistical derivation from the Navier-Stokes equations. To address this, we apply statistical symmetry principles,\cite{Frisch} particularly scale invariance via power-law scaling, to decode turbulent mean-flow laws. In TBLs, however, the coexistence of multiple characteristic scales (e.g., viscous-sublayer, buffer-layer, logarithmic-layer, and boundary-layer thicknesses) inherently disrupts scale invariance. This multiscale hierarchy demands explicit modeling of crossover scaling behavior across distinct flow regimes.

To address this challenge, we adopt the structural ensemble dynamics (SED) framework developed by She et al.,\cite{she2010new,she2017quantifying} which employs Lie-group analysis to construct a dilation-breaking ansatz for multiscale crossover modeling. In a recent work,\cite{ZhengBi2025} we have demonstrated that this compact yet physically consistent scaling formulation accurately captures the multilayer characteristics of TSS in equilibrium PG TBLs. In this study, we extend the formulation to describe the TSS evolutions in 2D non-equilibrium PG TBLs.

TSS serves as a pivotal quantity bridging mean-flow dynamics and turbulent transport. The scaling of TSS in the overlap region is critical for deriving both the generalized overlap law \cite{townsend1961equilibrium,skote2002direct} and the half-power law \cite{stratford1959prediction} of mean velocity. Chen et al. \cite{ChenShiYang2023} derived a velocity transformation that maps the mean velocity of TBLs with arbitrary PGs to the canonical law of the
wall. In this transformation, history effects are incorporated through a Lagrangian integral of TSS originating from an initial equilibrium state. Their results suggest that TSS plays a key role in the hysteresis of the mean flow.\cite{ChenShiYang2023} In our recent studies,\cite{BiWT2024,ZhengBi2025} we further validated that TSS exhibits a multilayer defect scaling, thereby serving as an ``order function'' for TBLs within the SED framework - a concept analogous to order parameters in statistical physics.\cite{she2017quantifying}

In this study, we propose a general symmetry-based framework to describe complete TSS profiles in non-equilibrium PG TBLs. The formulation assumes that non-equilibrium effects induce dilation-breaking mechanisms in the two-layer structure of equilibrium TSS. When introducing a single dilation breaking, a three-layer model successfully captures both the TSS profiles and their streamwise evolution in TBLs subjected to gradually increasing/decreasing APG. While additional dilation-breaking mechanisms could be incorporated to describe more complex non-equilibrium flows, a dual-boundary-layer model demonstrates the capability to capture TSS profiles in TBLs with an IBL emerging near the wall. This model combines an equilibrium TSS profile for the IBL with a Reynolds-shear-stress profile featuring multilayer scaling for the outer history-dependent flow. The proposed models are validated against published high-fidelity numerical simulation datasets spanning diverse flow configurations, including wing sections, a Gaussian bump, and a converging-diverging channel.

The paper is organized as follows. Section \ref{sec:tau_model_noneq} presents the general symmetry-based framework and the dual-boundary-layer model for TSS of non-equilibrium PG TBLs. Section \ref{sec:ValidationDNS} validates the models with numerical simulation datasets of diverse flows. Section \ref{sec:conclusion} discusses and concludes the study.

\section {Theory for TSS in non-equilibrium PG TBLs}\label{sec:tau_model_noneq}
\subsection{TSS in 2D PG TBLs}
We focus on incompressible, attached, and statistically 2D TBLs. This restriction can be considered as ``Assumption 0'' of our theory. The streamwise mean momentum equation of a 2D incompressible planar TBL reads
 \begin{equation}
  u^+\frac{\partial u^+}{\partial x^+}+v^+\frac{\partial u^+}{\partial y^+}  = -P^++\frac{\partial^2 u^+}{\partial {x^+}^2}+\frac{\partial^2 u^+}{\partial {y^+}^2}-\frac{\partial \left<{u'u'}\right>^+}{\partial x^+}-\frac{\partial \left<{u'v'}\right>^+}{\partial y^+},
  \label{eq:x_momentum}
\end{equation}
where superscript plus denotes wall-unit normalization, angle bracket denotes Reynolds averaging, $x$ and $y$ represent the streamwise and wall-normal Cartesian coordinates, $u$ and $v$ represent the streamwise and wall-normal mean velocity components, respectively, with $u'$ and $v'$ representing the corresponding velocity fluctuations, $P^+=\nu/(\rho u_\tau^3){\partial p}/{\partial x}$, $\rho$ is density, $p$ is mean static pressure, $\nu$ is kinematic molecular viscosity, and $u_\tau$ is friction velocity. The wall-normal integration of (\ref{eq:x_momentum}) gives
\begin{align}
  &\tau^+ \equiv\frac{\partial u^+}{\partial y^+}-\left<{u'v'}\right>^+ \nonumber\\
&=1 +\int_{0}^{y^+}{\left(u^+\frac{\partial u^+}{\partial x^+}+v^+\frac{\partial u^+}{\partial y^+}+P^+ -\frac{\partial^2 u^+}{\partial {x^+}^2}+\frac{\partial \left<{u'u'}\right>^+}{\partial x^+}\right){\rm d}y^+},
  \label{eq:tau}
\end{align}
where $\tau^+$ is the TSS. Within the integral on the right-hand side of (\ref{eq:tau}), the advection terms (i.e., the first and second terms) and the streamwise PG provide the dominant contributions to the $\tau^+$ profile. During incipient separation and in detached TBLs, the evolution of the streamwise turbulent kinetic energy plays a finite role.\cite{devenport2022equilibrium}

In the viscous sublayer, the advection is negligible, and $\tau^+$ has the approximation of
\begin{equation}
  \tau^+=1 +P_w^+ y^+,\quad y^+\ll\delta^+,
  \label{eq:tau_nearwall}
\end{equation}
where $P_w^+=\nu/(\rho u_\tau^3)({\partial p}/{\partial x})_w$ (subscript $w$ denotes the variable on the wall) is the PG parameter,\cite{Mellor1966} and $\delta$ represents the boundary layer thickness. Above the viscous sublayer and in the inner region of quasi-equilibrium TBLs, $\tau^+$ approximately keeps linear, and limitedly deviates from (\ref{eq:tau_nearwall}) due to the inertial contribution. Thus, Skote and Henningson \cite{skote2002direct} proposed a hybrid inner velocity scale, ${u_*}^2={u_\tau}^2+y^+u_p^3/u_\tau$ where $u_p^3=(\nu/\rho)(\partial p/\partial y)_w$,\cite{Mellor1966} such that $\tau^*=\tau/(\rho u_*^2)\approx1$ in the inner region as that of zero PG (ZPG) TBLs. Above the viscous sublayer, the TSS profile is a modeling target because of turbulence effects, particularly for TBLs under non-equilibrium PG conditions. This study attacks this problem by exploring the multiscale defect scaling properties of TSS in non-equilibrium PG TBLs. 

\subsection {A general symmetry-based framework for modeling TSS in PG TBLs}\label{subsec:tau_model_framework}
The theoretical framework rests on several explicitly defined postulates:

\noindent \textbf{Assumption 1 (Order Function Privilege):} The TSS operates as an ``order function'' characterized by a multiscale/multilayer defect power law, where the dilation-breaking ansatz from the SED theory formally governs the scaling transitions/crossovers between different scales/layers.

The dilation-breaking ansatz for describing crossover scaling between two power laws reads:\cite{she2010new,she2017quantifying} $[1+(y/y_0)^{\zeta} ]^{\Delta/\zeta}$, where $y_0$ is the crossover scale, $\Delta$ is the scaling exponent increment after the crossover, and $\zeta$ is a positive integer characterizing the crossover steepness. The ansatz has been constructed through a Lie-group method in the SED theory,\cite{she2017quantifying} and Gluzman and Yukalov have developed its mathematical foundation \cite{Gluzman1998} via an algebraic self-similar renormalization method. The ansatz is a solution of the following differential equation regarding the scaling exponent $\alpha$:
\begin{equation}
\frac{{\rm d}\alpha}{{\rm d}{\rm ln}y}=\zeta\Delta\frac{\left(y/y_0\right)^{\zeta}}{{\left[1+\left(y/y_0\right)^{\zeta}\right]}^2},
  \label{eq:flow_eq}
\end{equation}
which resembles a renormalization-group flow equation. At $y=y_0$, (\ref{eq:flow_eq}) reduces to $\frac{{\rm d}\alpha}{{\rm d}{\rm ln}y}|_{y_0}=\zeta\Delta/4$. Notably, when $\zeta=4$, (\ref{eq:flow_eq}) further reduces to $\frac{{\rm d}\alpha}{{\rm d}\ln y}|_{y_0}=\Delta$, suggesting that $\zeta=4$ represents a universal crossover steepness, as observed in TBLs \cite{she2010new,she2017quantifying,chen2018quantifying} and diverse physical systems.\cite{Gluzman1998,Sethna2015,BiWT2023}

A multilayer scaling can be formulated with a successive multiplication of the ansatz.\cite{she2010new,she2017quantifying} An $m$-layer scaling law reads
\begin{equation}
  f(y)=cy^{p_0}\prod_{n=1}^{m-1} \left[1+\left(y/y_n\right)^{\zeta_n} \right]^{\Delta_n/\zeta_n},
  \label{eq:multilayermodel}
\end{equation}
where $c$ and $p_0$ are, respectively, the scaling coefficient and exponent of the zeroth layer. As an illustration of this framework, She et al. \cite{she2017quantifying} applied multilayer scaling to the mixing length profile in the inner regions of canonical wall-bounded turbulent flows:
\begin{equation}
  \ell_m^+=\frac{\kappa {y_s^+}^2}{y_b^+}{y^+}^{1.5}\left[1+\left(y^+/y_s^+\right)^4 \right]^{0.5/4}\left[1+\left(y^+/y_b^+\right)^4 \right]^{-1/4},
  \label{eq:ellm_sed}
\end{equation}
where $\kappa$ is the K\'{a}rm\'{a}n constant, $y_s^+$ and $y_b^+$ are constant crossover thicknesses representing the viscous-sublayer thickness and buffer-layer thickness, respectively. In (\ref{eq:ellm_sed}) the crossover steepnesses take the universal value of 4, and the scaling exponents are theoretically determined by Chen et al. \cite{chen2018quantifying} via a random dilation analysis on the Reynolds stress transport equations. (\ref{eq:ellm_sed}) captures a three-layer scaling for $\ell_m^+$ in the inner region of canonical flows. Particularly in the overlap region where $y_b^+\ll y^+\ll\delta^+$, $\ell_m^+=\kappa y^+$, adhering to Prandtl's mixing-length hypothesis.\cite{prandtl1925ausgebildete}

Now, we formulate multilayer defect scaling with the dilation-breaking ansatz, because TSS possesses a multilayer defect scaling law. There are at least two methods for this formulation. In the first case, the defect origin is a constant for all layers, and the multilayer defect scaling can be written as 
\begin{equation}
  g(y)=g(0)\left[1+f(y)\right],
  \label{eq:defectmultilayermodel_1}
\end{equation}
where g(0) is the defect origin, and $f(y)$ possesses a multilayer scaling described with (\ref{eq:multilayermodel}). 
Zheng et al. \cite{ZhengBi2025} have employed (\ref{eq:defectmultilayermodel_1}) to describe the multilayer defect scaling of TSS in equilibrium-FPG (E-FPG) TBLs: 
\begin{equation}
\tau_{\rm {E\mbox{-}FPG}}^+=1+P_w^+{y^+}\left[{1 + {\left(y^+/y_P^+\right)^4}} \right]^{(\gamma-1)/4}\left[{1 + {\left(y^+/\delta^+\right)^{4}}} \right]^{-\gamma/4},
  \label{eq:tau_equili_FPG}
\end{equation}
where $\gamma$ depends on $\beta$, and crossover-thickness $y_P^+$ is a PG-induced critical scale which correlates with $\gamma$, $\delta^+$, and $P_w^+$ via $y_P^+(\delta^+/y_P^+)^\gamma=-(1/{P_w^+})$. (\ref{eq:tau_equili_FPG}) captures a three-layer defect scaling: $\tau^+=1+P_w^+{y^+}$ when $y^+\ll y_P^+$ in the viscous sublayer; $\tau^+=1-{(y^+/\delta^+)}^\gamma$ when $y_P^+\ll y^+\ll \delta^+$; and $\tau^+=1-{y^+}^0=0$ when $y^+\gg \delta^+$.

In the second case, the defect origin varies across layers. For this scenario, an $m$-layer defect scaling law can be formulated as
\begin{equation}
  h(y)=h(0)\prod_{n=1}^{m} \left\{1+c_n\left(y/y_n\right)^{p_n}\left[1+\left(y/y_n\right)^{\zeta_n} \right]^{-p_n/\zeta_n}\right\}.
  \label{eq:defectmultilayermodel_2}
\end{equation}
Here, the scales satisfy $y_1^+\ll y_2^+ \ll \cdots \ll y_m^+$, and $p_1\leq p_2 \leq \cdots \leq p_m$, ensuring that the upper-layer defect scaling does not interfere with the lower-layer defect scaling. 
Zheng et al. \cite{ZhengBi2025} have applied (\ref{eq:defectmultilayermodel_2}) to describe the multilayer defect scaling of TSS in equilibrium-APG (E-APG) TBLs: 
\begin{align}
\tau_{\rm {E\mbox{-}APG}}^+&=\left\{1+P_0^+\left(y^+/y_P^+\right)\left[{1 + {\left(y^+/y_P^+\right)^4}} \right]^{-1/4} \right\} \nonumber\\
&\times\left\{{1-\left(y^+/\delta^+\right)}^{1.5}\left[{1 + {\left(y^+/\delta^+\right)^{20}}} \right]^{-1.5/20}\right\},
  \label{eq:tau_equili_APG}
\end{align}
where $P_0^+=P_w^+y_P^+$, representing an APG-induced shear stress parameter that depends on $\beta$. In (\ref{eq:tau_equili_APG}), the crossover steepness at $\delta^+$ is empirically set to 20, characterizing a quick decay to zero for TSS near the boundary layer edge. This steep crossover scaling can be removed if $y^+$ is restricted to be no more than $\delta^+$. Thus, (\ref{eq:tau_equili_APG}) captures a two-layer defect scaling: $\tau^+=1+P_w^+{y^+}$ when $y^+\ll y_P^+$; and $\tau^+=(1+P_0^+)[1-{(y^+/\delta^+)}^{1.5}]$ when $y_P^+\ll y^+ \leq \delta^+$. Note that the defect origin of TSS in the outer layer becomes $1+P_0^+$. Zheng et al. \cite{ZhengBi2025} established that $P_0^+=1.5\beta$ for equilibrium APG TBLs. They derived $y_P^+=0.491\left[3\beta/(2+3\beta)\right]^{2/3}\delta^+$. However, finite $Re$ effects cause deviations from these correlation relationships. The functional dependence between $y_P^+$ and $\beta$ reveals that $y_P^+$ rapidly asymptotes toward $0.491\delta^+$ when $\beta$ exceeds unity.

Now, we introduce the second assumption to formulate TSS in general non-equilibrium PG TBLs:

\noindent \textbf{Assumption 2 (Dilation-Breaking Mechanism):} Non-equilibrium PG effects activate dilation-breaking mechanisms, altering the multilayer structure of TSS in equilibrium TBLs through the emergence of new scales/layers. 

Assuming only one dilation-breaking process occurring in a non-equilibrium PG (NE-PG) TBL, the TSS profile can be described with 
 \begin{align}
\tau_{\rm {NE\mbox{-}PG}}^+&=\left\{1+c_m\left(y^+/y_m^+\right)^{p_m}\left[{1 + {\left(y^+/y_m^+\right)^{\zeta_m}}} \right]^{-p_m/\zeta_m} \right\}\tau_{\rm {E\mbox{-}APG}}^+ \nonumber\\
&=\left\{1+c_m\left[{1 + {\left(y_m^+/y^+\right)^{\zeta_m}}} \right]^{-p_m/\zeta_m} \right\}\tau_{\rm {E\mbox{-}APG}}^+,
  \label{eq:tau_NonEPG_threelayer} 
\end{align}
where a new crossover scaling (within the brace) is introduced at $y_m^+$ in the intermediate region between the inner and outer flows. This new layer revises the defect origin of the outer-layer scaling in $\tau_{\rm {E\mbox{-}APG}}^+$ with a coefficient $(1+c_m)$, thus capturing a hysteresis to PG variations above $y_m^+$. For gradually-increasing APG, $c_m<0$ indicates underdeveloped Reynolds shear stress in the outer flow; for gradually-decreasing APG, $c_m>0$ reflects a stress overshoot compared with the equilibrium state. Therefore, we call the dilation-breaking function in the brace of (\ref{eq:tau_NonEPG_threelayer}) the ``delay function'', which quantifies the response of TSS to PG variation across the boundary layer. 

Having fewer parameters is almost always desirable. Determining the scaling exponent $p_m$ ($1< p_m\leq1.5$) in the delay function is a theoretical challenge. In TBLs subjected to gradually strengthening APG, $y_m^+<y_P^+$. To preserve the near-wall linear law $\tau^+=1+P_w^+{y^+}$, we set $p_m$ to its upper limit 1.5. In this case, $\zeta_m$, assumed a constant, may be different from the universal crossover steepness 4 to ensure a smooth scaling transition behavior. As validated in section \ref{subsubsec:TBL_wing}, $\zeta_m=2$ well captures the TSS profiles in all investigated wing section flows, indicating a slower transition in the delay function than that when $\zeta_m$ takes the universal value of 4. This small $\zeta_m$ value reflects a specific property of PG TBLs: namely, a relatively slow transition between the near-wall instant-response region and the outer delayed-response zone. 

Regarding $\tau_{\rm {E\mbox{-}APG}}^+$ in (\ref{eq:tau_NonEPG_threelayer}), $P_w^+$ and $\delta^+$ are assumed to be known at a specific streamwise location. However, the definition of $\delta^+$ becomes problematic in non-equilibrium flows,\cite{vinuesa2016determining,Griffin2021,WeiT2023a} a point we will discuss later. $y_P^+$ has been formulated as a function of $\beta$ for equilibrium flows in Ref.~\onlinecite{ZhengBi2025}, but is unknown here because $\beta$ of the hypothetical equilibrium TBL is unknown. Since $y_P^+$ quickly approaches $0.491\delta^+$ when $\beta$ exceeds unity, we assume $y_P^+$ takes this limit value and remains independent of both $Re$ and $\beta$ when the APG is above moderate, which occurs in TBLs subjected to continually-intensifying APG. Otherwise, $y_P^+$ should be empirically determined, such as in TBLs undergoing APG-FPG transitions where $\beta$ is small. Consequently, the streamwise evolution of TSS in moderate non-equilibrium TBLs can be captured using only two, at most three, empirical parameters in (\ref{eq:tau_NonEPG_threelayer}): $y_m^+$, $c_m$, and perhaps $y_P^+$. 

TBLs characterized by (\ref{eq:tau_NonEPG_threelayer}) should be considered in moderate non-equilibrium states because of the single dilation breaking. While employing more dilation-breaking mechanisms could theoretically enable the description of more complex non-equilibrium flows, such an approach inevitably leads to a proliferation of model parameters. This parametric expansion exposes the current framework to substantial criticism regarding its susceptibility to overfitting. To circumvent this challenge while preserving modeling capability, we subsequently introduce a dual-boundary-layer formulation that achieves enhanced physical fidelity without excessive parameterization.

\subsection {A dual-boundary-layer model for TSS in radical non-equilibrium PG TBLs}\label{subsec:tau_model_noneq}
Many authors have observed the formation of an IBL within the near-wall region of TBLs subjected to abrupt PG transitions. This distinctive flow structure emerges when the boundary layer experiences either a rapid APG development following FPG conditions or a sudden APG removal in relaxation scenarios. The observed flow decoupling stems from the differential response timescale between inner and outer regions. The viscous-dominated near-wall flow adjusts quasi-instantaneously to PG variations, with a characteristic timescale of $\nu/u_\tau^2$. In contrast, the outer inertia-dominated flow retains memory effects through a much larger timescale: $\frac{\rho\theta u_\delta}{\delta_1 (\partial p/\partial x)_\delta}$ (where $\theta$ is the momentum thickness). This timescale corresponds to an adjustment length of tens of boundary layer thickness.\cite{devenport2022equilibrium} The multiscale response mechanism justifies the conceptual decomposition of non-equilibrium PG TBLs into two dynamically distinct regions: a local equilibrium IBL governed by instantaneous PG and a history-dependent outer layer carrying cumulative flow development information.

Therefore, we postulate the following force decomposition assumption for modeling TSS of non-equilibrium PG TBLs with the emergence of an IBL:

\noindent \textbf{Assumption 3 (Force Decomposition):} In such dual-boundary-layer flows, the TSS can be decomposed into two physically distinct components:
 \begin{equation}
  \tau_{\rm {NE\mbox{-}PG}}^+=\tau_{\rm{in}}^++W_{\rm {out}}^+,
  \label{eq:tau_noneq}
\end{equation}
where $\tau_{\rm{in}}^+$ describes the equilibrium TSS contribution from the IBL, and $W_{\rm {out}}^+$ quantifies the history-dependent Reynolds shear stress in the non-equilibrium outer region.

According to (\ref{eq:tau_equili_APG}), $\tau_{\rm{in}}^+$ of a non-equilibrium APG TBL reads
 \begin{align}
\tau_{\rm{in}\mbox{-}APG}^+&=\left\{1+P_w^+y^+\left[{1 + {\left(y^+/y_P^+\right)^4}} \right]^{-1/4}\right\} \nonumber\\
&\times\left\{{1-\left(y^+/\delta_i^+\right)}^{1.5}\left[{1 + {\left(y^+/\delta_i^+\right)^{4}}} \right]^{-1.5/4}\right\},
  \label{eq:tau_noneq_in_APG}
\end{align}
where $\delta_i^+$ is the thickness of the IBL. According to (\ref{eq:tau_equili_FPG}), $\tau_{\rm{in}}^+$ of a non-equilibrium FPG TBL reads
\begin{align}
  \tau_{\rm{in}\mbox{-}FPG}^+  =& 1+P_w^+y^+\left[{1 + {\left(y^+/y_P^+\right)^4}} \right]^{(\gamma-1)/4}\nonumber\\
&\times\left[{1 + {\left(y^+/\delta_i^+\right)^4}} \right]^{-\gamma/4},
  \label{eq:tau_noneq_in_FPG}
\end{align}
where $y_P^+$, $\gamma$, $\delta_i^+$, and $P_w^+$ are correlated with $y_P^+(\delta_i^+/y_P^+)^\gamma=-(1/{P_w^+})$.

$W_{\rm {out}}^+$ is formulated as below. Near the boundary layer edge $W_{\rm {out}}^+$ should obey the $3/2$ defect power law: $W_{\rm {out}}^+=W_{\rm {max}}^+[1-{({y^+/\delta^+})}^{1.5}]$, where $W_{\rm {max}}^+$ characterizes the maximum Reynolds shear stress of the outer flow. Close to the wall, $W_{\rm {out}}^+$ adheres to a power-law decay: $W_{\rm {out}}^+\propto{y^+}^{p_w}$. According to the asymptotic wall condition, $p_w=3$.\cite{Bradshaw1995} However, because we focus on correctly separating the $W_{\rm {out}}^+$ contribution from the gross TSS in the intermediate region, which is away from the wall, this asymptotic wall condition is not necessarily useful. Instead, we set $p_w=1.5$, as setting $p_m=1.5$ in Eq. (\ref{eq:tau_NonEPG_threelayer}), and consistent with the asymptotic behavior of the $3/2$ defect power law near the wall. Then, a dilation-breaking ansatz could be introduced to capture the crossover scaling from this near-wall power law to the near-edge $3/2$ defect law. Consequently, a uniform profile of $W_{\rm {out}}^+$ is constructed as below:
\begin{align}
  W_{\rm{out}}^+=&{W_{\rm{max}}^+}\times\left(y^+/\delta_w^+\right)^{1.5}\left[1+\left(y^+/\delta_w^+\right)^4\right]^{-1.5/4}\nonumber\\
&\times\left[1-\left(y^+/\delta^+\right)^{1.5}\right],
  \label{eq:tau_residue_stress} 
\end{align}
where $\delta_w^+$ denotes the crossover thickness that can be reckoned as the lower boundary of the outer history-dependent flow.

Equation (\ref{eq:tau_noneq}) thus establishes a dual-boundary-layer framework for modeling TSS in PG-driven TBLs experiencing radical non-equilibrium conditions. The model's predictive power originates from four physically interpretable parameters that resolve distinct flow regimes: the near-wall dynamics governed by $y_P^+$ (local-PG-induced critical thickness) and $\delta_i^+$ (IBL thickness) through $\tau_{\rm{in}}^+$, coupled with outer wake characterization via $W_{\rm{max}}^+$ (wake strength) and $\delta_w^+$ (wake lower boundary) in $W_{\rm{out}}^+$. While the four-parameter configuration might initially raise concerns about overparameterization (notably recalling von Neumann's caution that ``with four parameters I can fit an elephant''), the formulation demonstrates irreducible minimalism when confronted with the inherent complexity of non-equilibrium PG flows. This parametrization is essential to capture the dynamic interplay between inner and outer flow regions. It is particularly critical in relaxation scenarios with abrupt APG removal, where the IBL and outer layer transition from overlapping to separation. The model's physical fidelity comes from its explicit inclusion of multi-timescale processes: the inner layer adjusts rapidly through $\delta_i^+$, while the outer wake dissipates more slowly, governed by $W_{\rm{max}}^+$. These competing dynamics drive the system toward an equilibrium attractor state, where the boundary layer thickness becomes canonical ($\delta_i^+ \rightarrow \delta^+$) and the wake fully decays ($W_{\rm{max}}^+\rightarrow0$). The parameter set's parsimony thus reflects not mathematical convenience, but rather a necessary compromise to resolve the fundamental scale separation inherent in non-equilibrium PG TBL physics.

\section{Validation in typical non-equilibrium processes}\label{sec:ValidationDNS}
Non-equilibrium flows in engineering applications are diverse. Thus, we decompose general non-equilibrium PG TBLs into distinct archetypal processes: moderately varying APG, rapid FPG-APG or APG-FPG transitions, abrupt APG removal, and reattached TBLs. These processes are essential building blocks, collectively characterizing the intricate non-equilibrium PG TBLs in engineering systems. Analyzing these archetypal processes provides insights into systematic turbulence modulation mechanisms, enabling predictive model development for real-world scenarios.

In this section, we analyze three typical non-equilibrium PG processes with the current TSS models: (1) TBL with gradually varying APG on the suction side of wing section; (2) relaxing TBL with abrupt APG-FPG transition on the windward surface of Gaussian bump; (3) TBL subjected to rapid FPG-APG transition in a converging-diverging channel. The recovery process of a reattached TBL is critical and interesting. It includes a particular history effect owing to strong turbulence advected from the upstream separated shear layer. This flow is not discussed here, but an analysis regarding the multilayer scaling of its TSS and mixing length function is documented in Ref.~\onlinecite{BiWT2024}.

\subsection{TBLs subjected to gradually varying APG}\label{subsubsec:TBL_wing}
TBLs subjected to gradually varying APG are ubiquitous on surfaces of streamlined bodies such as airplanes, ships, and submarines. Here we investigate TBLs over suction surfaces of NACA4412 and NACA0012 airfoils based on the high-resolution large-eddy simulation (LES) datasets from Vinuesa et al.\cite{Vinuesa2017wing,Vinuesa2018wing} For the NACA4412 airfoil, three chord Reynolds numbers ($Re_c=2.0\times 10^5$, $4.0\times 10^5$, and $1.0\times10^6$) are considered at $5^\circ$ angle of attack (AoA). For the NACA0012 airfoil, $Re_c=4.0\times 10^5$ and AoA$=0^\circ$.

\begin{figure*}
    \centering
    \includegraphics[width=0.35\linewidth]{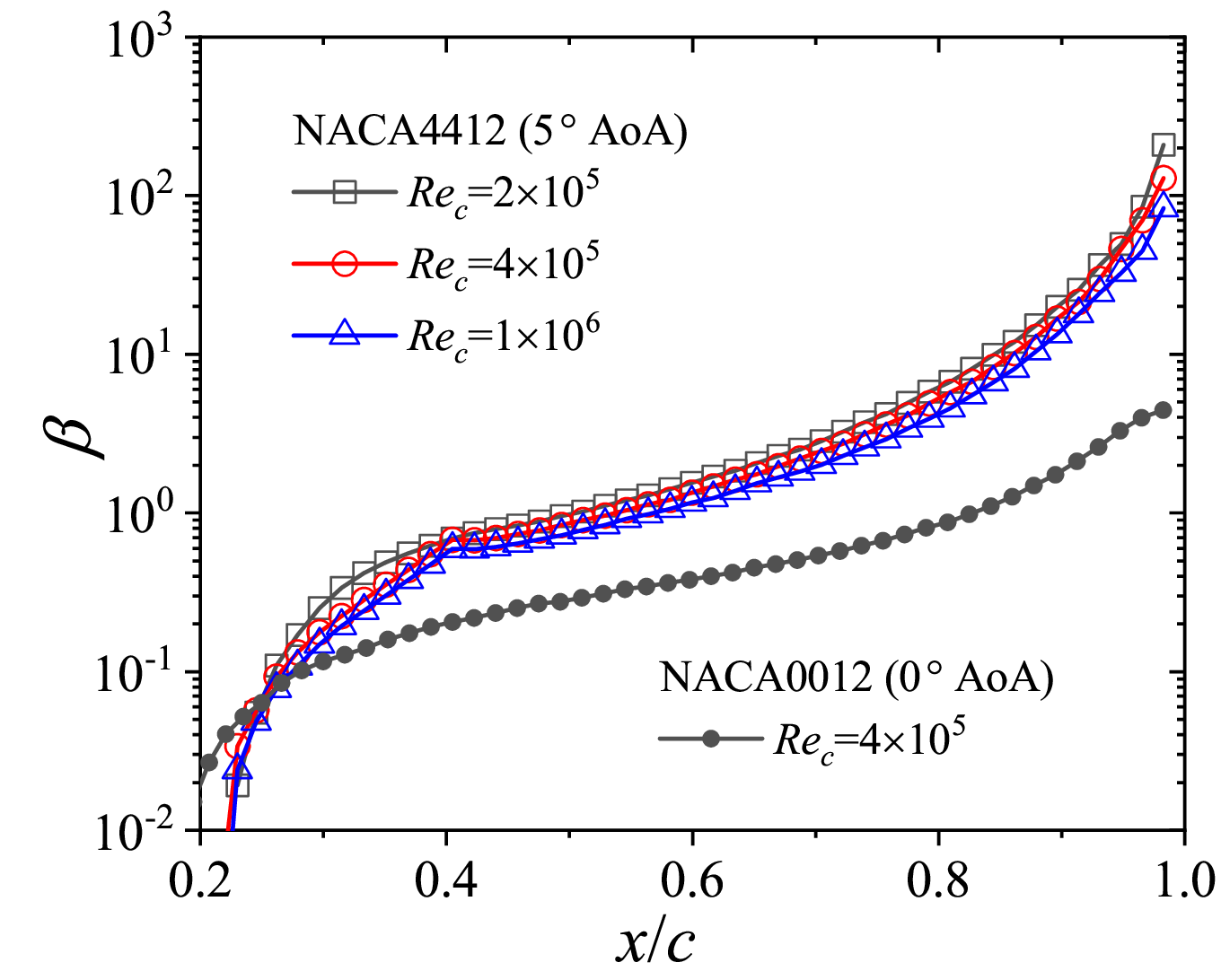}
    \includegraphics[width=0.35\linewidth]{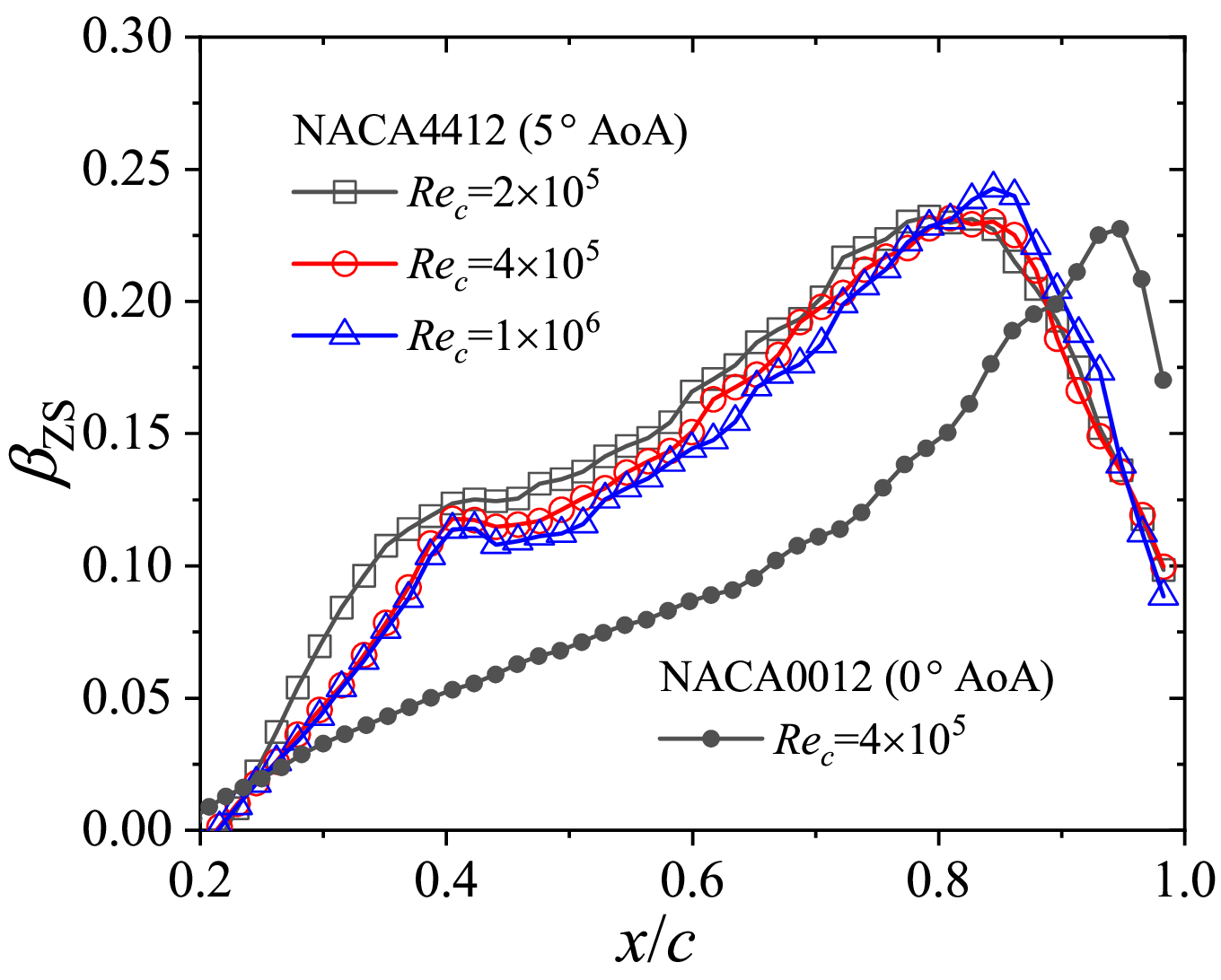}
    \\\quad\quad(a)\quad\quad\quad\quad\quad\quad\quad\quad\quad\quad\quad\quad\quad\quad\quad\quad\quad\quad\quad(b)\\
    \includegraphics[width=0.35\linewidth]{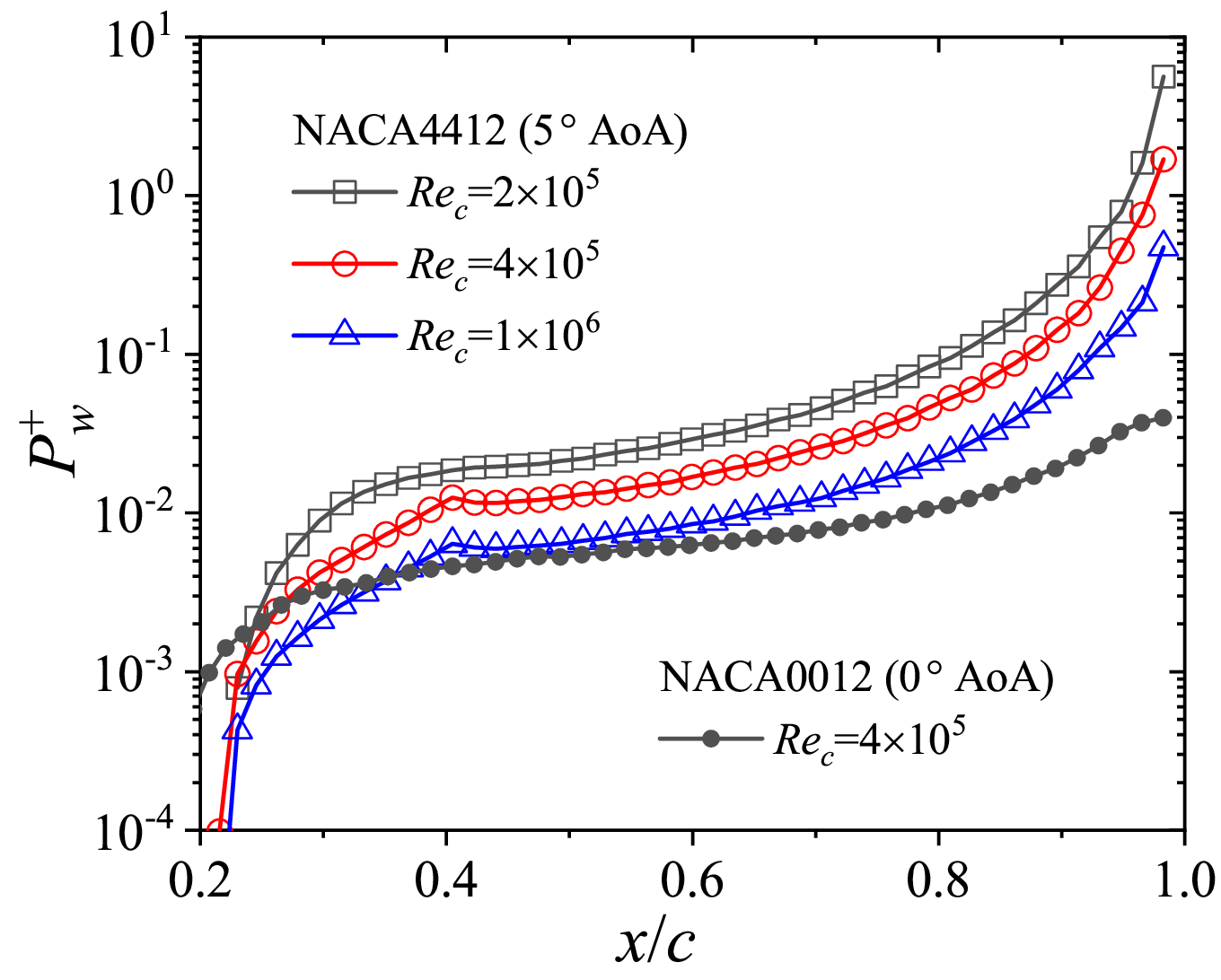}
    \includegraphics[width=0.35\linewidth]{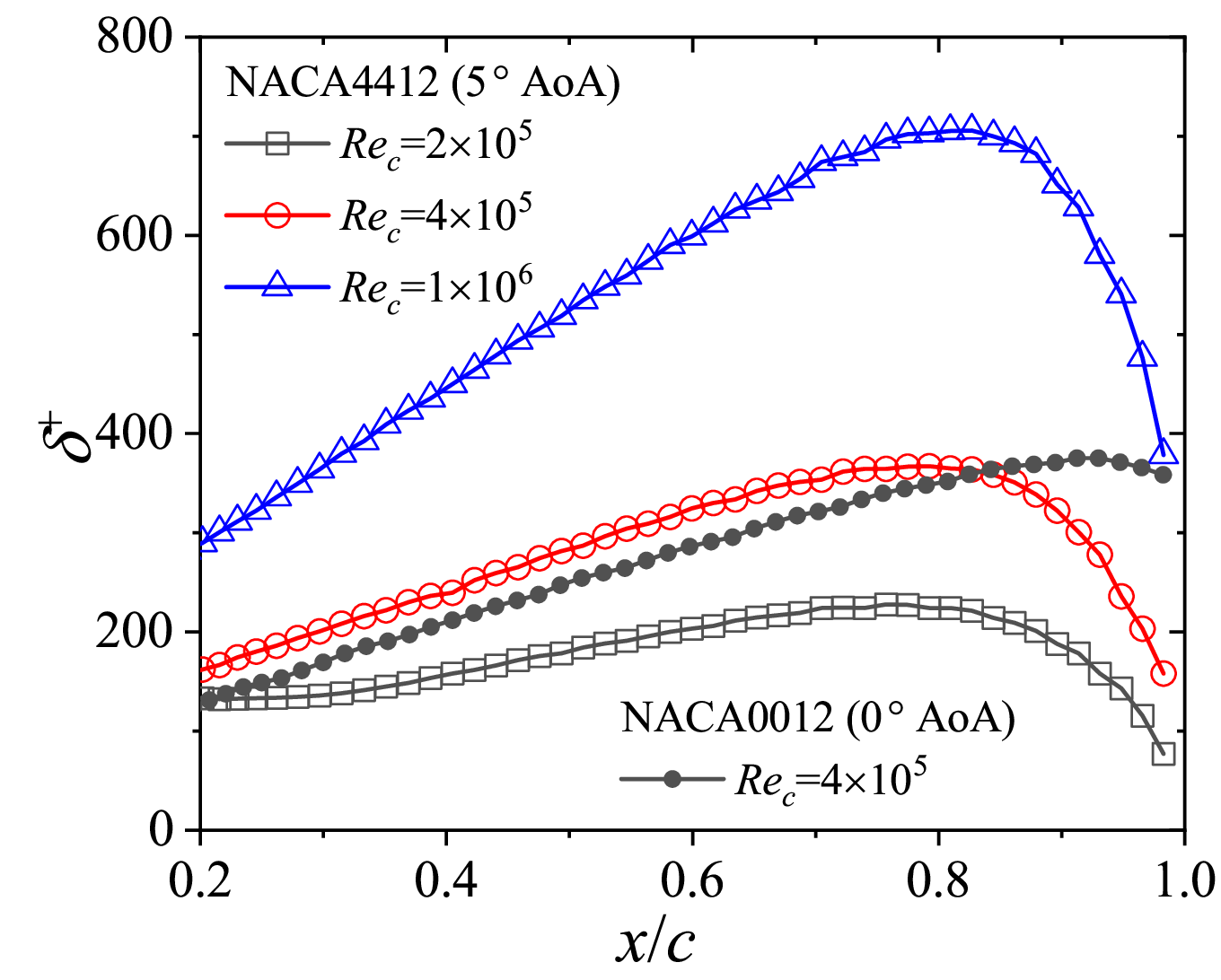}
    \\\quad\quad(c)\quad\quad\quad\quad\quad\quad\quad\quad\quad\quad\quad\quad\quad\quad\quad\quad\quad\quad\quad(d)\\
  \caption{Distributions of (a) the Clauser PG parameter $\beta$, (b) the Zagarola-Smits PG parameter $\beta_{ZS}$, (c) the PG parameter $P_w^+$, and (d) the dimensionless boundary layer thickness $\delta^+$ along the chord direction of the airfoils. Data for the NACA4412 suction surface correspond to $5^\circ$ AoA and three chord Reynolds numbers ($Re_c=2.0\times 10^5$, $4.0\times 10^5$, and $1.0\times10^6$), obtained from the LES by Vinuesa et al. in Ref.~\onlinecite{Vinuesa2018wing}. Compared data for the NACA0012 airfoil at $0^\circ$ AoA and $Re_c=4.0\times 10^5$ are from the LES study by Vinuesa et al. in Ref.~\onlinecite{Vinuesa2017wing}.}
  \label{fig:wingstat}
\end{figure*}
 
Fig. \ref{fig:wingstat} compares the streamwise evolutions of $\beta$, $\beta_{ZS}$, $P_w^+$, and $\delta^+$ of the TBLs over the suction surfaces of the airfoils. Here, $\beta_{ZS}\equiv\delta/(\rho u_{ZS}^2)(\partial p/\partial x)_\delta$, where the Zagarola-Smits velocity $u_{ZS}$ is defined as $u_{ZS}\equiv\frac{1}{\delta}\int_{0}^{\delta}{(u_\delta-u){\rm d}y}=u_\delta \delta_1/\delta$. As shown in Fig. \ref{fig:wingstat}(a), the NACA4412 airfoil exhibits a $\beta$ distribution essentially independent of $Re_c$ at this moderate AoA. Near the NACA4412 trailing edge, $\beta$ reaches significantly large values, indicating incipient separation. In contrast, the APG on the NACA0012 suction surface remains weak and increases gradually in the chord direction. 

Maciel et al. \cite{Maciel2018} argued that $\beta$ is not an inner or outer PG parameter, but rather a global one that encompasses the entire boundary
layer. They proposed that the Zagarola-Smits PG parameter $\beta_{ZS}$ reflects the PG effects on the outer flow. As illustrated in Fig. \ref{fig:wingstat}(a) and (b), near the trailing edge of the NACA4412 airfoil, while $\beta$ continues to increase due to the decrease in $u_\tau$, $\beta_{ZS}$ drops quickly after its peak at about $x/c=0.85$. Researchers have observed that the APG does not reach a local maximum and remains relatively small near flow separation,\cite{Alving1996} with a tendency toward equilibrium before separation.\cite{CastilloASME} The distribution of $\beta_{ZS}$ agrees with these observations. 

The evolutions of $P_w^+$ and $\delta^+$ are displayed in Fig. \ref{fig:wingstat}(c) and (d), respectively. They are treated as known quantities in the present modeling framework. Here, $\delta$ represents the 99\% boundary-layer thickness obtained via the diagnostic scaling method proposed by Vinuesa et al.\cite{vinuesa2016determining} For all NACA4412 boundary-layer cases, the maximum $\delta^+$ occurs at $x/c\approx0.8$, which differs from the NACA0012 airfoil, where it appears near the trailing edge (Fig. \ref{fig:wingstat}(d)). Readers are directed to Refs.~\onlinecite{Vinuesa2017wing,Vinuesa2018wing} for more flow details.

\begin{figure}
    \centering
    \includegraphics[width=0.75\linewidth]{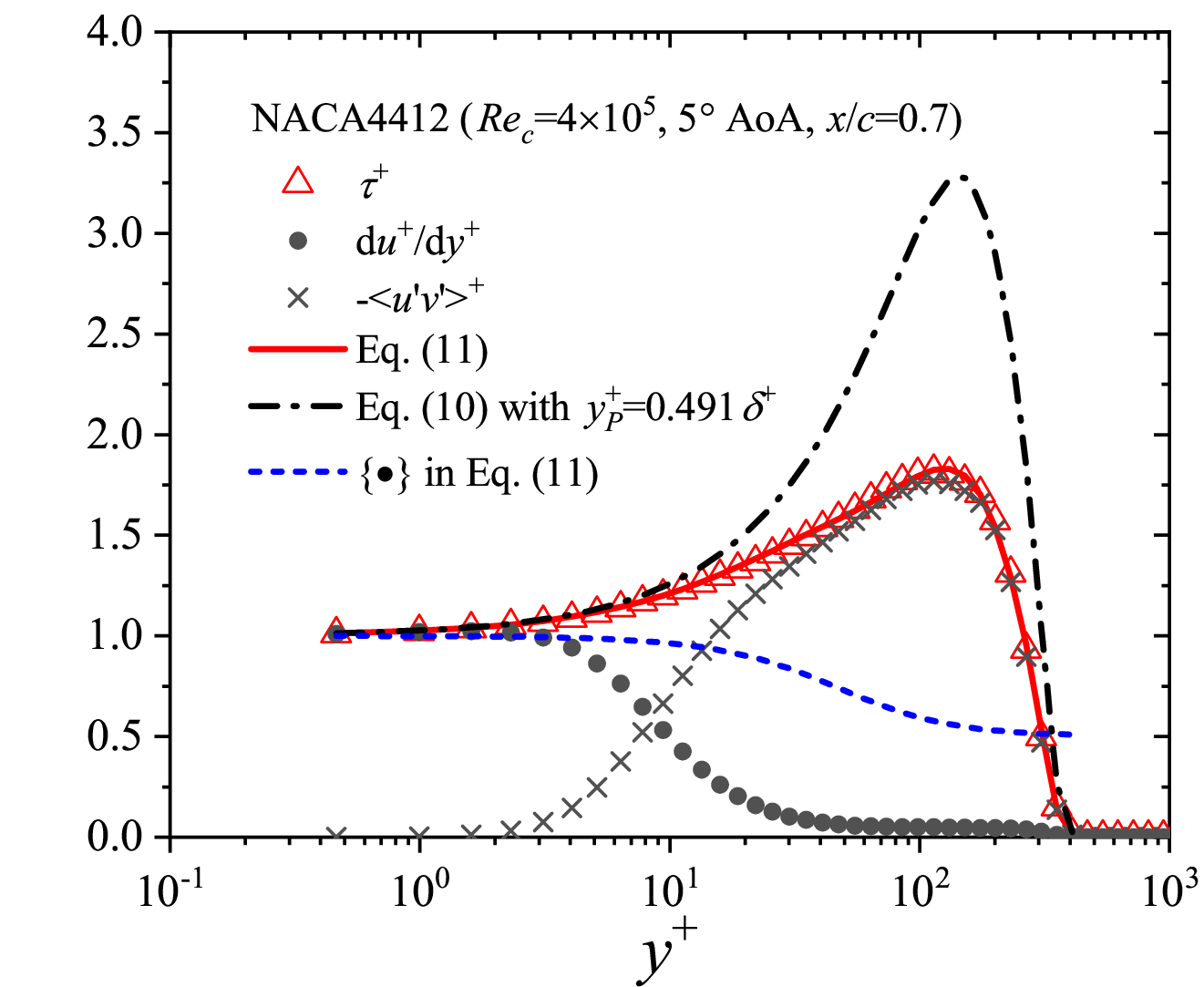}
  \caption{Profiles of the mean viscous shear stress (solid circles), the Reynolds shear stress (cross symbols), and the TSS (open triangles) at $x/c=0.7$ on the suction surface of the NACA4412 airfoil at $5^\circ$ AoA with $Re_c=4.0\times 10^5$. $y$ is the wall-normal coordinate. The solid line represents descriptions from the current TSS model (Eq. (\ref{eq:tau_NonEPG_threelayer})), while the dashed-dotted line indicates the reference equilibrium TSS solution (Eq. (\ref{eq:tau_equili_APG})) calculated using $y_P^+=0.491\delta^+$. The delay function contained in the brace term of Eq. (\ref{eq:tau_NonEPG_threelayer}) is separately shown by the short-dashed curve. To calculate this function, $p_m=1.5$, $\zeta_m=2$, and $c_m$ and $y_m^+$ are acquired from the LES data of TSS via a least-squares method.}
  \label{fig:wing_tau_model_decomp}
\end{figure}

\begin{figure*}
    \centering
    \includegraphics[width=0.35\linewidth]{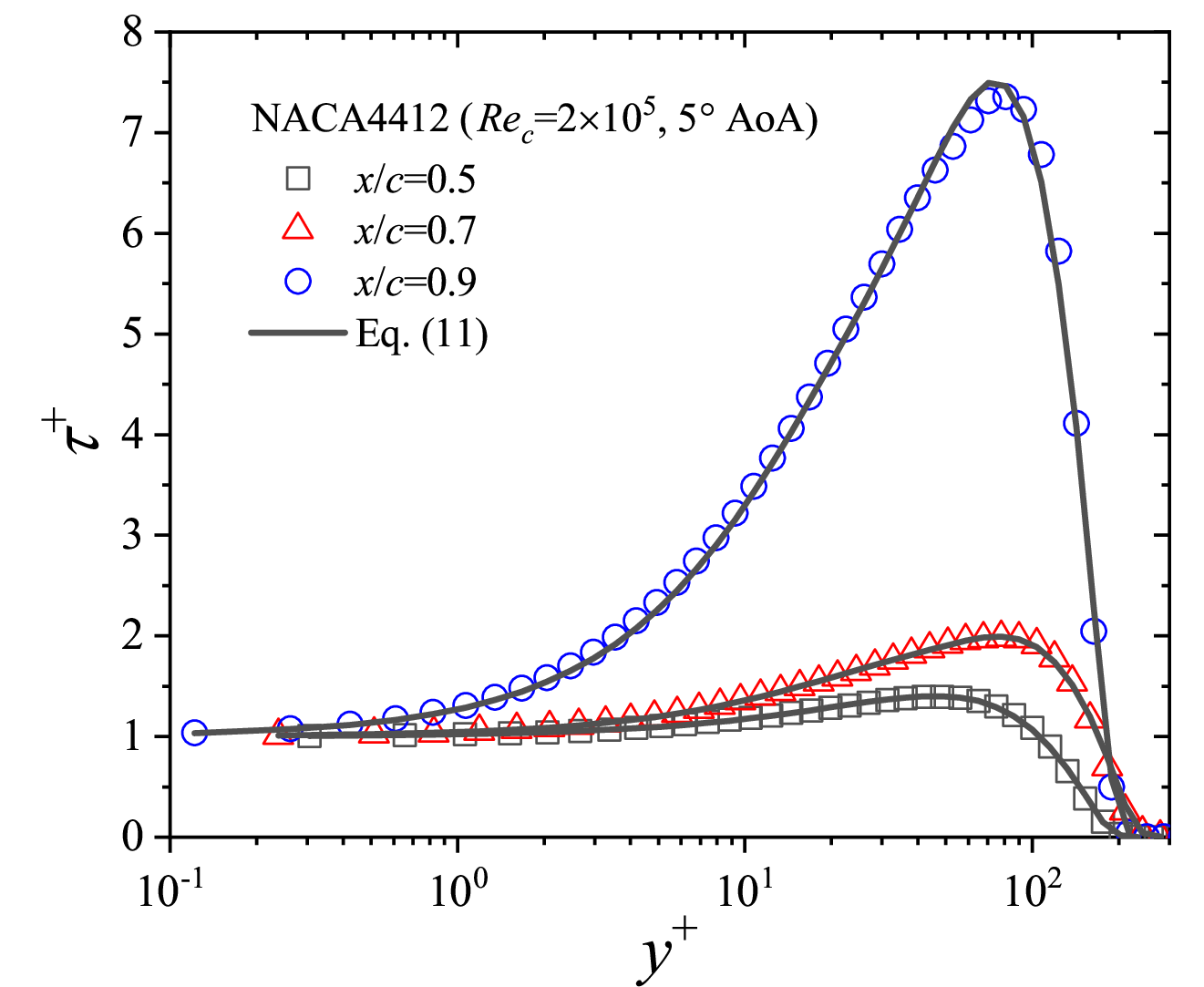}
    \includegraphics[width=0.35\linewidth]{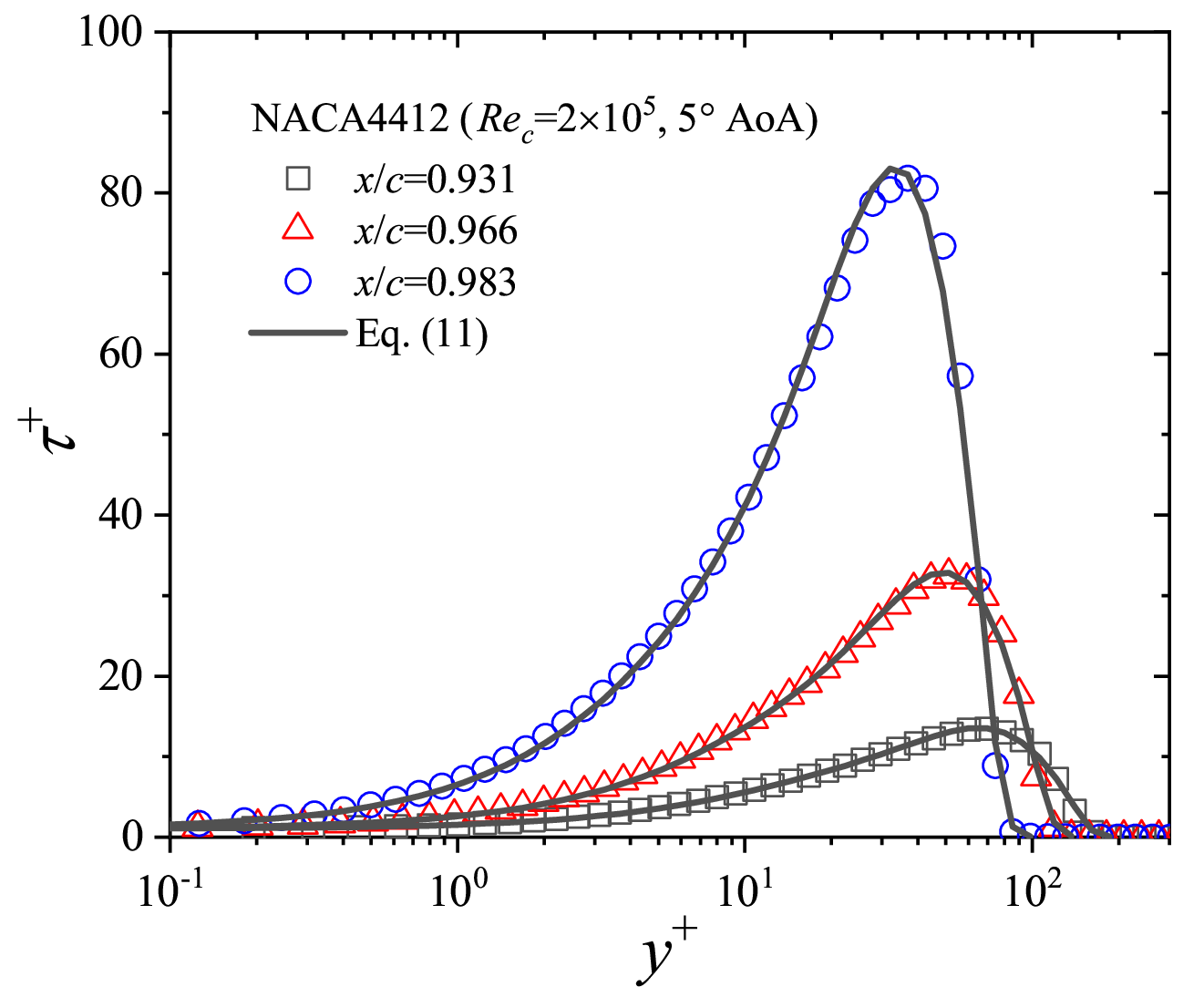}
    \\\quad\quad(a)\quad\quad\quad\quad\quad\quad\quad\quad\quad\quad\quad\quad\quad\quad\quad\quad\quad\quad\quad(b)\\
    \includegraphics[width=0.35\linewidth]{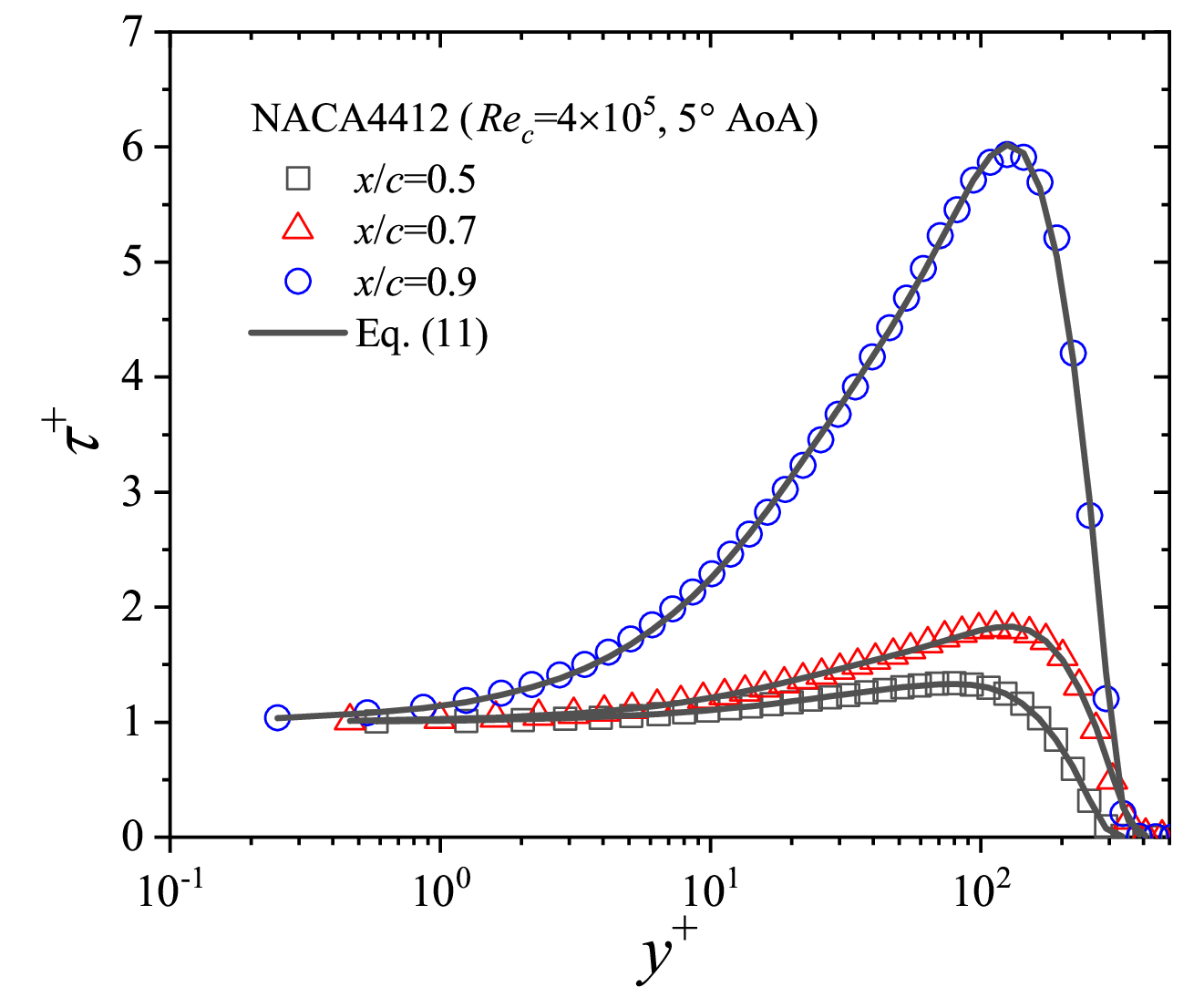}
    \includegraphics[width=0.35\linewidth]{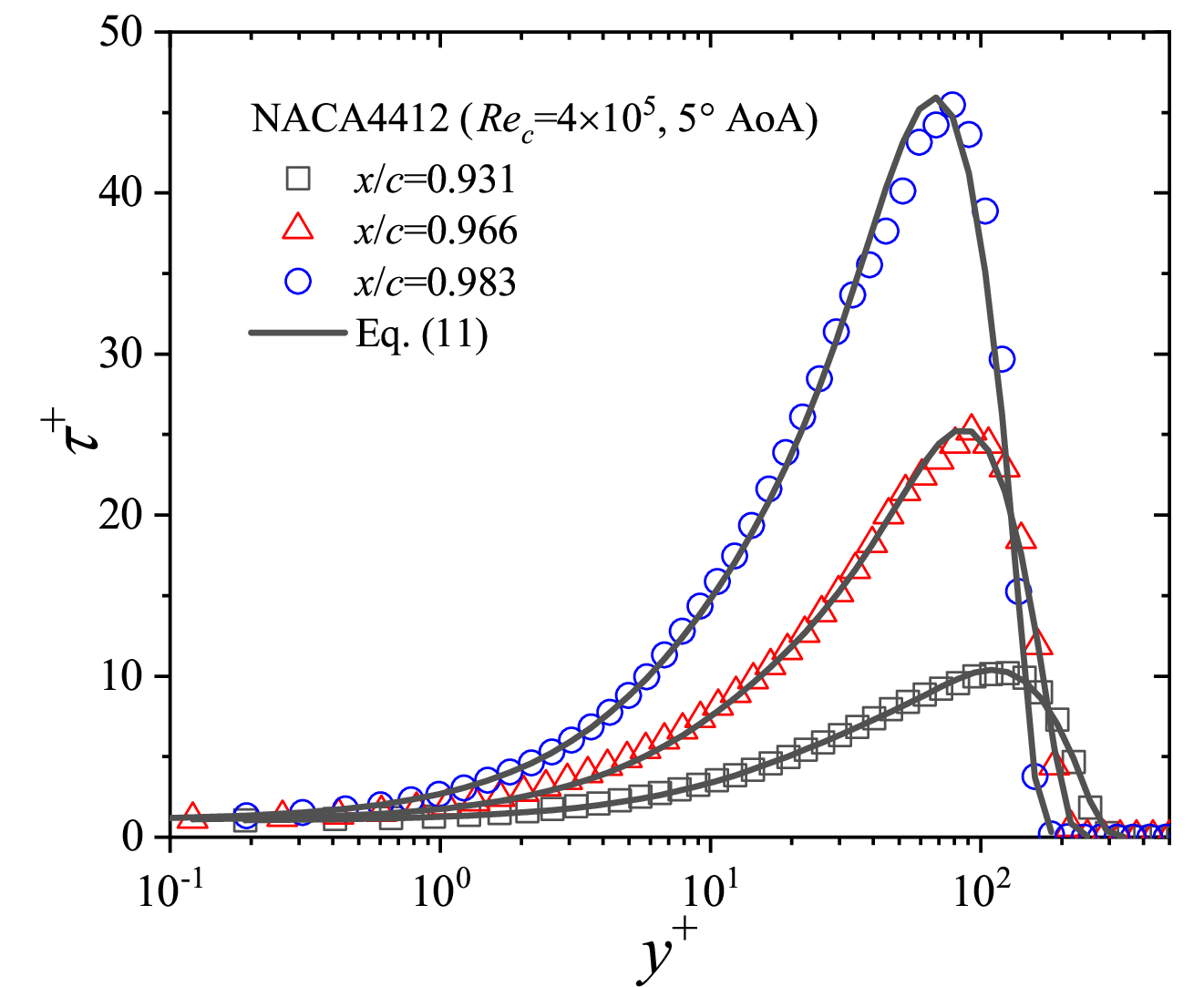}
    \\\quad\quad(c)\quad\quad\quad\quad\quad\quad\quad\quad\quad\quad\quad\quad\quad\quad\quad\quad\quad\quad\quad(d)\\
    \includegraphics[width=0.35\linewidth]{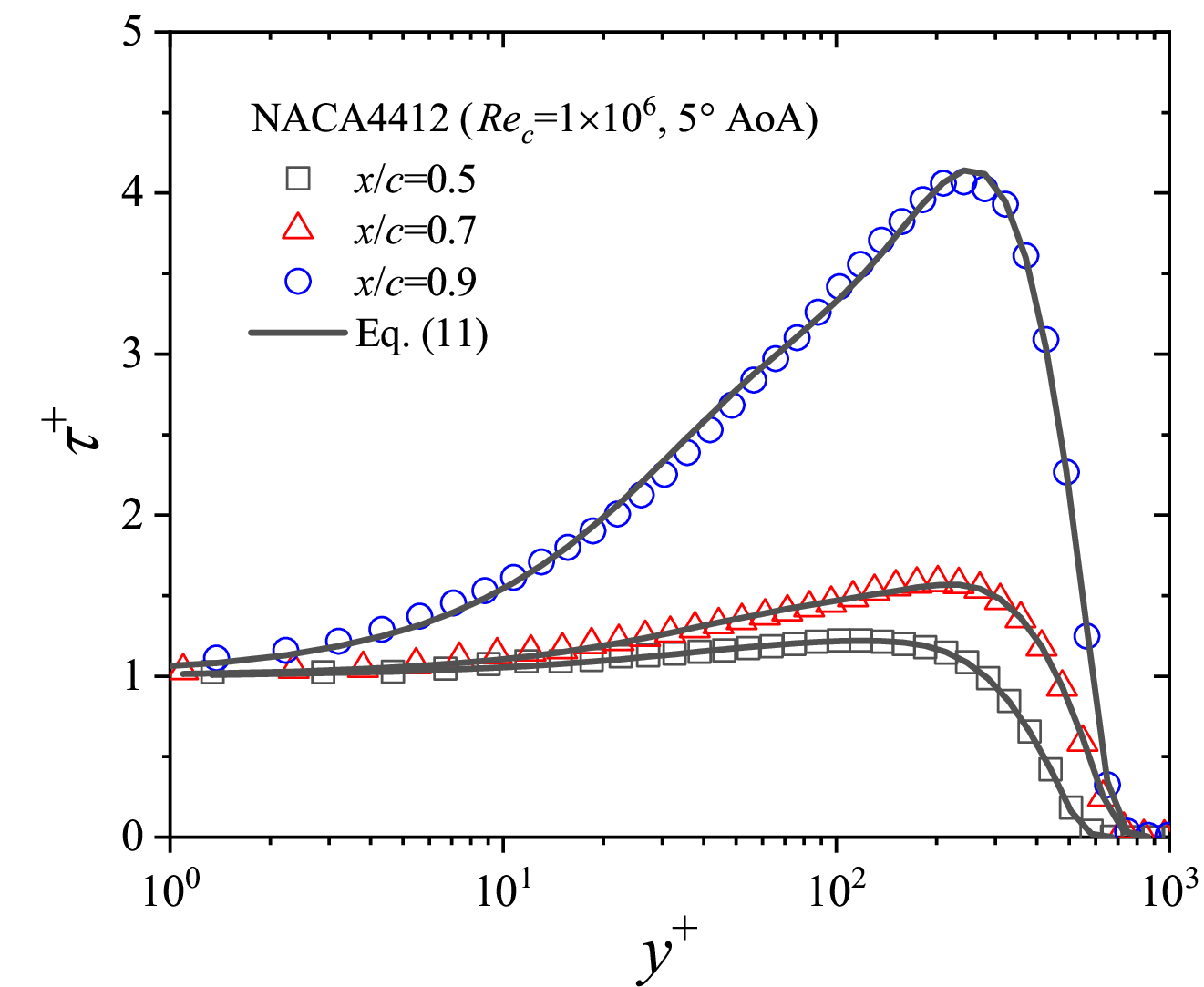}
    \includegraphics[width=0.35\linewidth]{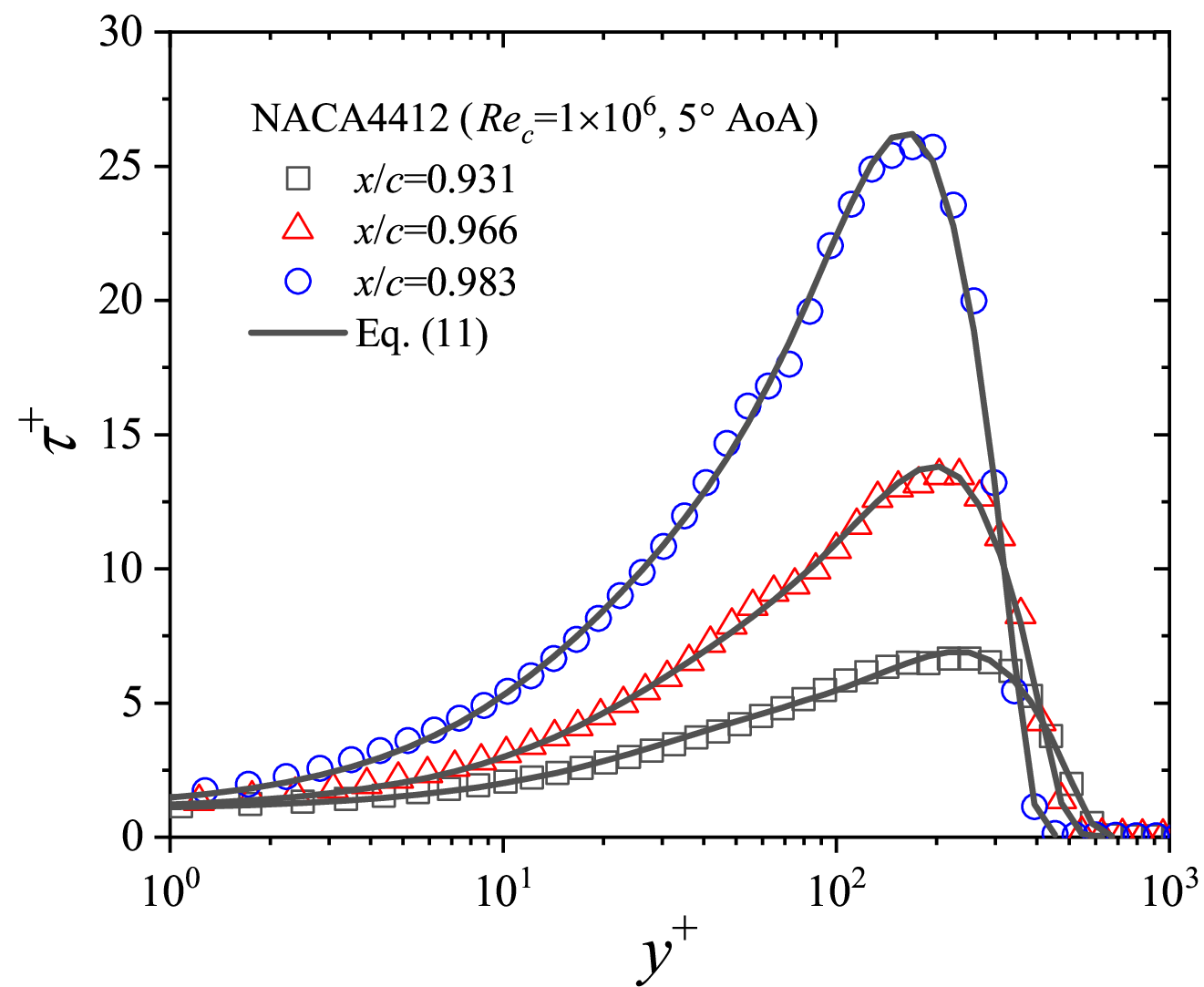}
    \\\quad\quad(e)\quad\quad\quad\quad\quad\quad\quad\quad\quad\quad\quad\quad\quad\quad\quad\quad\quad\quad\quad(f)\\
    \includegraphics[width=0.35\linewidth]{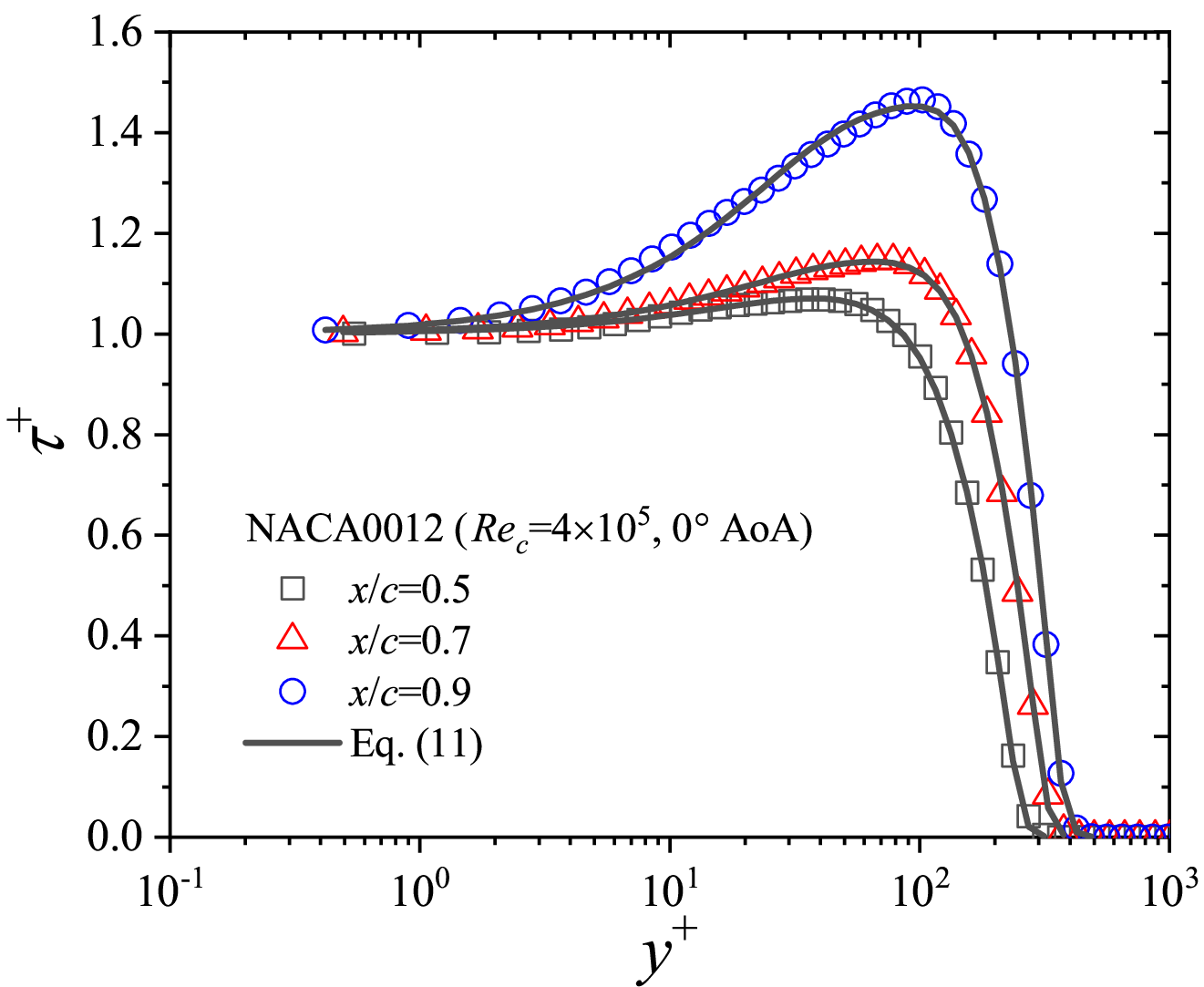}
    \includegraphics[width=0.35\linewidth]{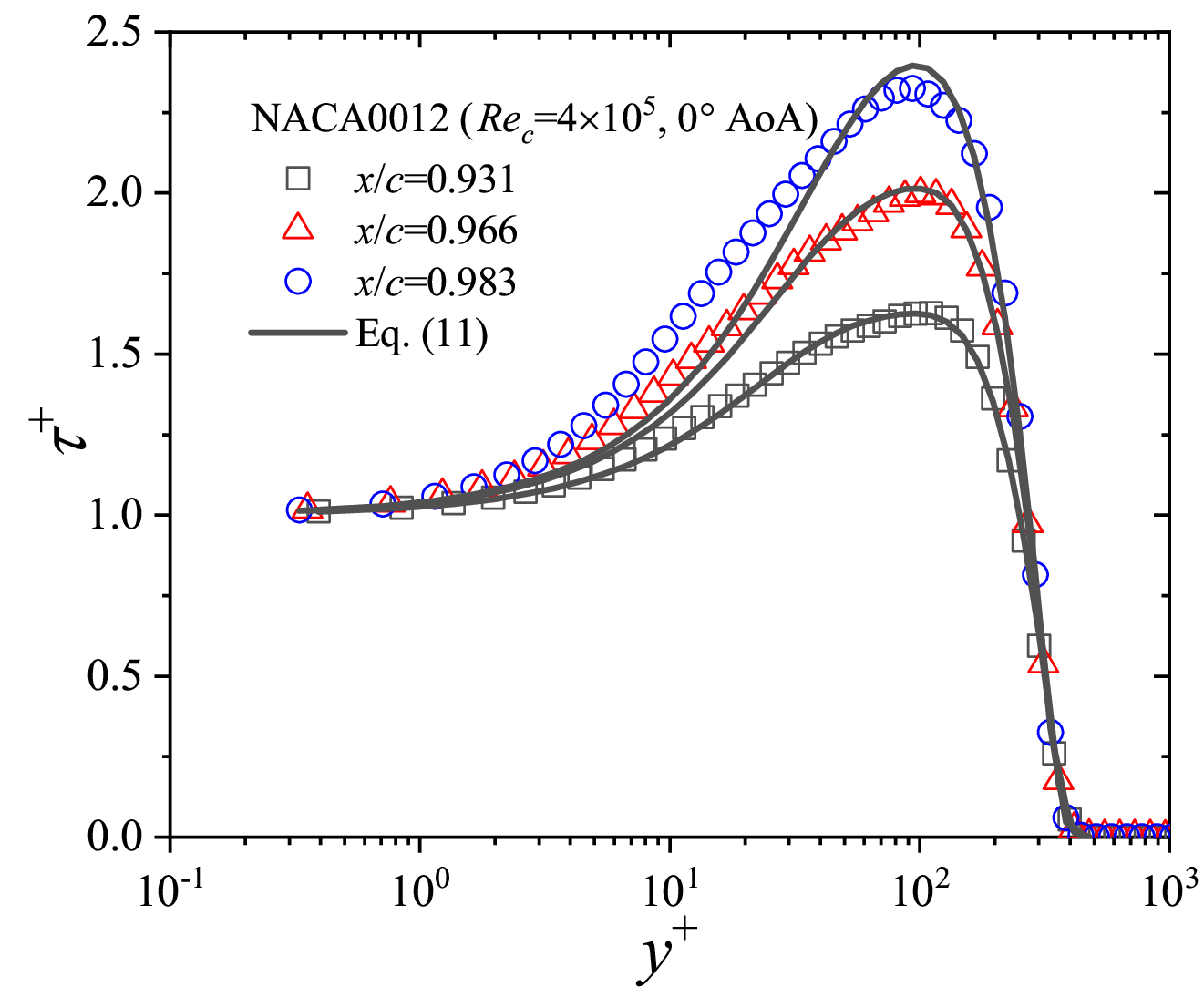}
    \\\quad\quad(g)\quad\quad\quad\quad\quad\quad\quad\quad\quad\quad\quad\quad\quad\quad\quad\quad\quad\quad\quad(h)
  \caption{TSS profiles at multiple chord locations on the suction surfaces of the NACA4412 and NACA0012 airfoils under different $Re_c$ and AoA conditions. (a) and (b): NACA4412 at $Re_c=2.0\times 10^5$. (c) and (d): NACA4412 at $Re_c=4.0\times 10^5$. (e) and (f): NACA4412 at $Re_c=1.0\times10^6$. (g) and (h): NACA0012 at $Re_c=4.0\times 10^5$. The LES data (symbols) are compared with predictions from the current TSS model (solid lines, Eq. (\ref{eq:tau_NonEPG_threelayer})). To calculate (\ref{eq:tau_NonEPG_threelayer}), $y_P^+=0.491\delta^+$, $p_m=1.5$, $\zeta_m=2$, and $c_m$ and $y_m^+$ are acquired from the LES data via a least-squares method.}
  \label{fig:wing_tau_model_valid}
\end{figure*}

\begin{figure*}
    \centering
    \includegraphics[width=0.33\linewidth]{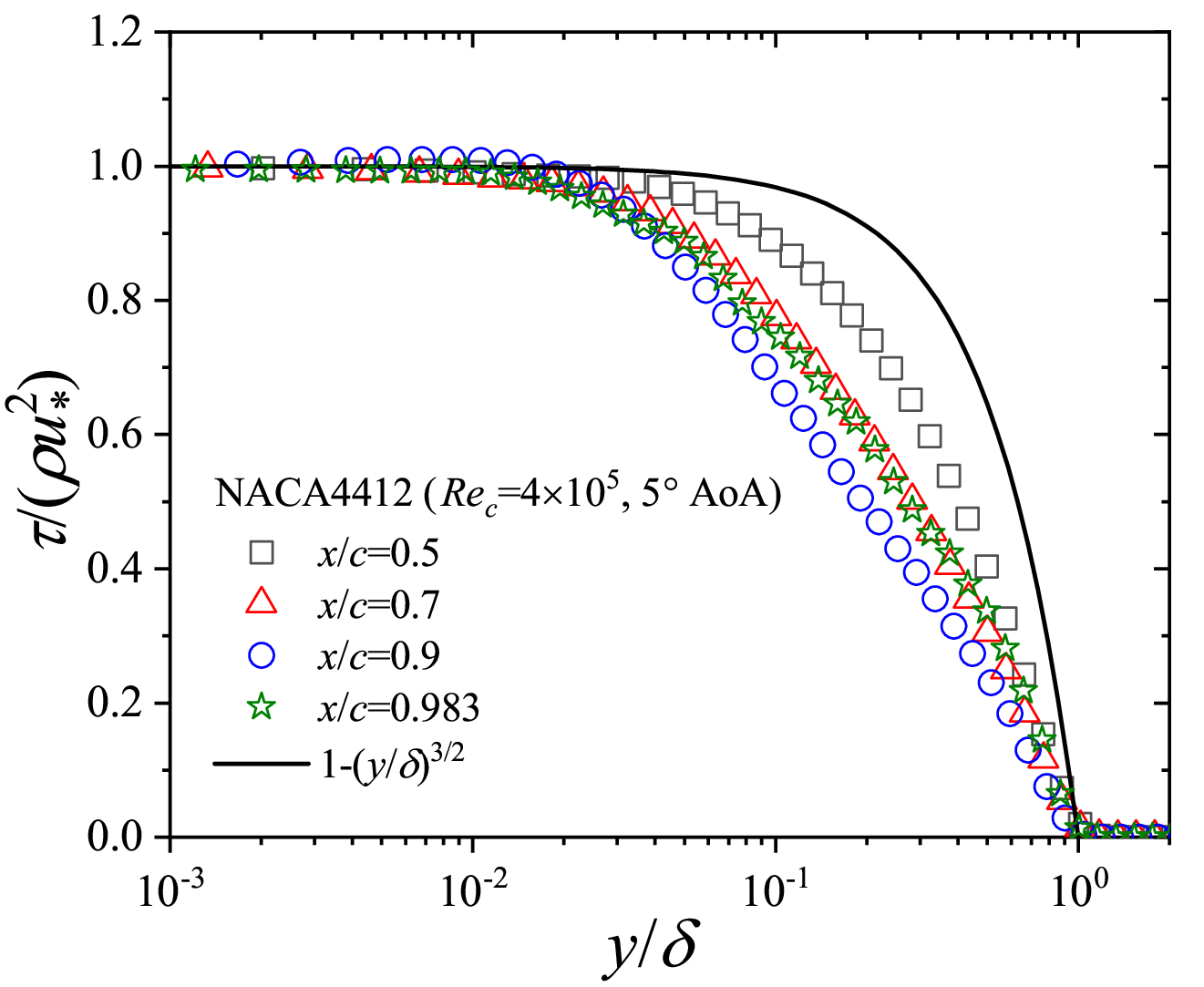}
    \includegraphics[width=0.33\linewidth]{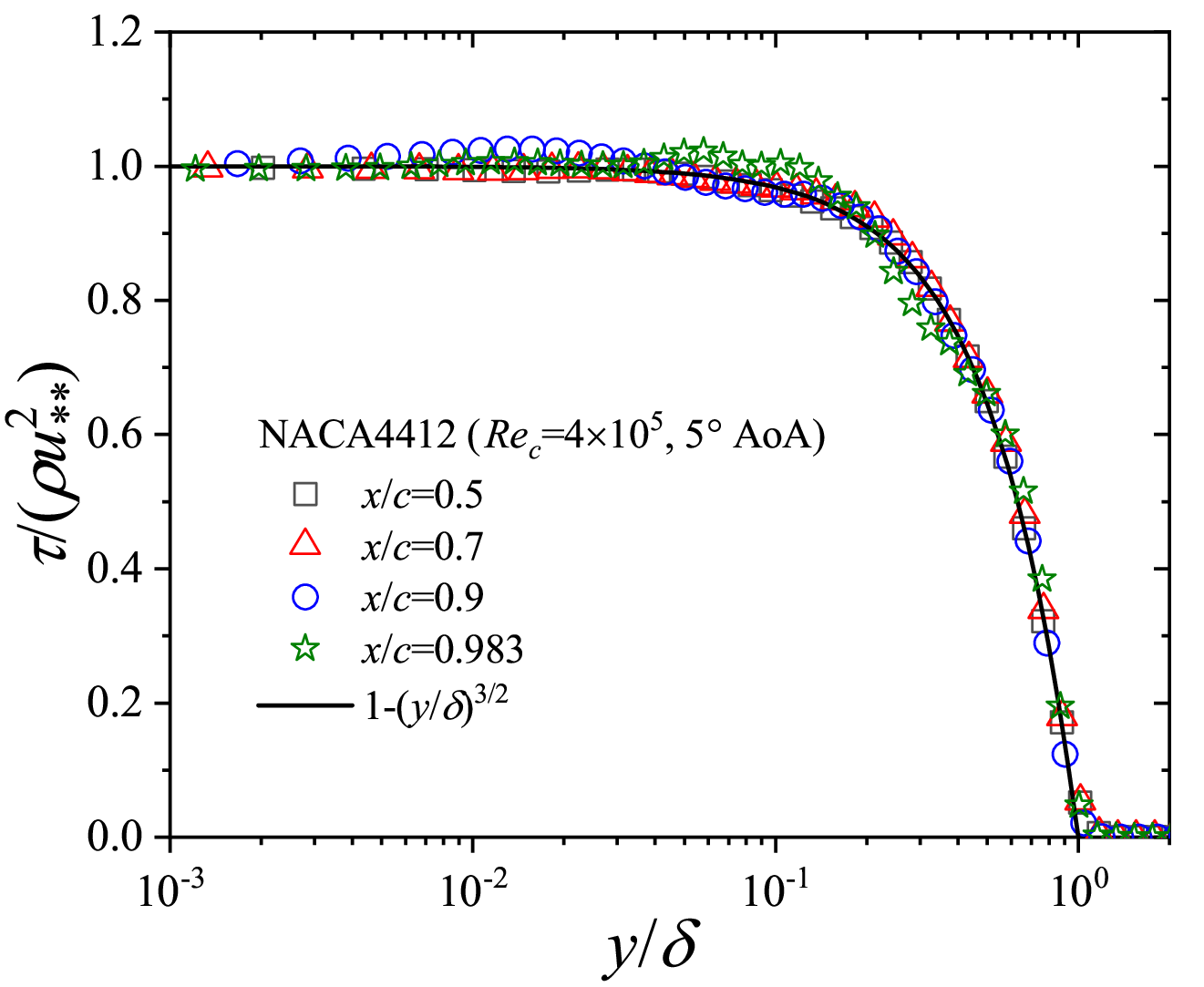}
    \includegraphics[width=0.33\linewidth]{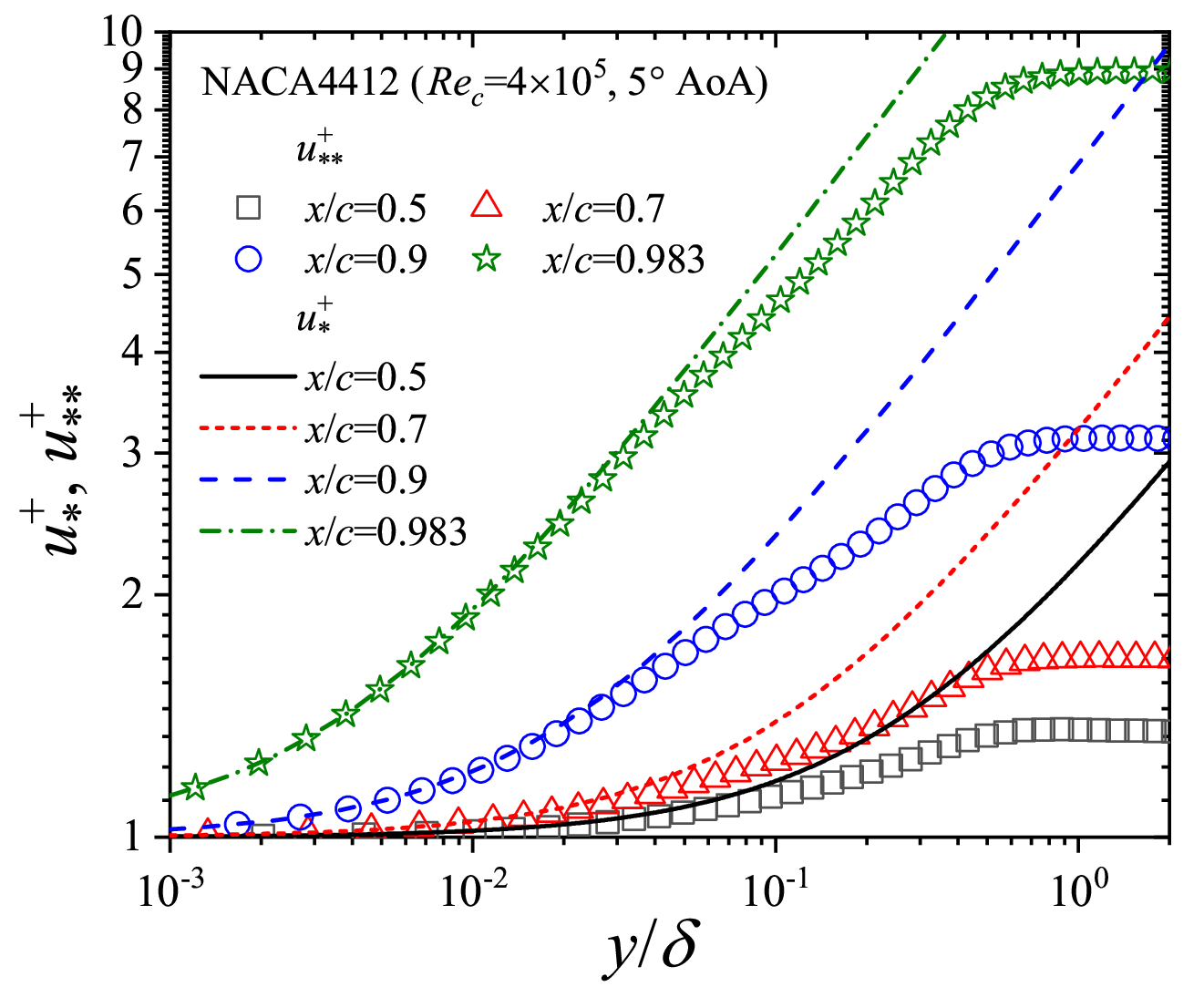}
    \\\quad(a)\quad\quad\quad\quad\quad\quad\quad\quad\quad\quad\quad\quad\quad\quad\quad\quad\quad\quad(b)\quad\quad\quad\quad\quad\quad\quad\quad\quad\quad\quad\quad\quad\quad\quad\quad\quad(c)
  \caption{(a) Comparison between the $u_*$-scaling of TSS in NACA4412 wing section TBLs (symbols) and the 3/2 defect power law of ZPG TBLs (solid line).
(b) The $u_{**}$-scaling of TSS in NACA4412 wing section TBLs (symbols) collapses with the 3/2 defect power law of ZPG TBLs (solid line). (c) Wall-normal profiles of $u_{*}^+$ (lines) and $u_{**}^+$ (symbols) used in panels (a) and (b).}
  \label{fig:Skote_scaling}
\end{figure*}

For TBLs subjected to gradually varying APG, the TSS profiles exhibit moderate non-equilibrium features. We therefore implement the three-layer formulation (Eq. (\ref{eq:tau_NonEPG_threelayer})) to model these flows. The two empirical parameters, $c_m$ and $y_m^+$, are calibrated via least-squares optimization against LES-derived TSS profiles, with $y_P^+$ fixed to $0.491\delta^+$ and $p_m=1.5$. A modest yet universal crossover steepness $\zeta_m=2$ effectively characterizes the scaling transition between inner and outer layers across all cases. Fig. \ref{fig:wing_tau_model_decomp} validates the model predictions (solid line) against LES data (open triangles) at $x/c=0.7$ on the NACA4412 suction surface ($5^\circ$ AoA and $Re_c=4.0\times 10^5$), demonstrating excellent agreement throughout the boundary layer. The peak TSS magnitude remains notably lower than that of the hypothesized equilibrium APG TBL (dashed-dotted line) due to flow hysteresis in the outer layer under progressive APG forcing. The delay function (short-dashed line) quantifies the flow hysteresis effects through a smooth transition (centered at $y_m^+$) from instant near-wall response at about $y^+<10$ to delayed adaptation (characterized by a negative $c_m$) at about $y^+>100$. As a multiplicative factor on the hypothesized equilibrium TSS profile, the delay function ultimately reproduces the LES-observed TSS distribution with high fidelity.

Fig. \ref{fig:wing_tau_model_valid} demonstrates the validity of the three-layer TSS model (Eq. (\ref{eq:tau_NonEPG_threelayer})) against LES data at multiple chord positions on the suction surface of both airfoils under varying aerodynamic conditions. The model achieves consistent agreement with the LES results, spanning over two orders of magnitude in $\beta$ (Fig. \ref{fig:wingstat}(a)). The comprehensive validation underscores the model's robustness in capturing wall-normal stress transport phenomena across diverse flow configurations. Only minor discrepancies occur near the trailing edge ($x/c=0.966$ and $0.983$ in Fig. \ref{fig:wing_tau_model_valid}(h)) for the NACA0012 airfoil at zero AoA. The underlying causes of these localized deviations remain unidentified (discrepancies occur near the wall, indicating inconsistency in $P_w^+$). 

Within the current framework, we introduce a unified velocity scale $u_{**}$ to map complete TSS profiles from non-equilibrium APG TBLs to ZPG TBLs: 
\begin{equation}
  \tau^{**}=\frac{\tau}{\rho u_{**}^2}=1-\left(\frac{y}{\delta}\right)^{1.5},
  \label{eq:tau_mapping}
\end{equation}
where $u_{**}$ is defined as
\begin{align}
  u_{**}=&u_\tau\sqrt{1+y^+\left[{1 + {\left(y^+/y_P^+\right)^4}}\right]^{-1/4}u_p^3/u_\tau^3 }\nonumber\\
 &\times\sqrt{1+c_m\left[{1 + {\left(y_m^+/y^+\right)^{2}}} \right]^{-1.5/2} }.
  \label{eq:new_velo_scale}
\end{align}
For the investigated wing-section TBLs, $y_P^+=0.491\delta^+$. Close to the wall, $u_{**}$ reduces to the inner velocity scale $u_*$ defined by Skote and Henningson \cite{skote2002direct} and the hybrid inner velocity scale $u_{hyb}$ proposed by Romero et al.\cite{Romero2022fluid} Near the boundary layer edge, $u_{**}$ becomes constant: $u_{**}\approx u_{**}(\delta)\approx u_\tau\sqrt{(1+c_m)\left[1+0.491\left(\delta/\delta_1\right)\beta\right]}$. Thus $u_{**}(\delta)$ represents a new outer velocity scale defined by the current theory.  

Fig. \ref{fig:Skote_scaling} compares the TSS profiles scaled by $u_{*}$ and $u_{**}$ for the wing-section TBLs. With the $u_{*}$-scaling, $\tau^{*}$ preserves unity in the inner region as $\tau^{+}$ does for ZPG TBLs. The distinct peak shear stresses shown in Fig. \ref{fig:wing_tau_model_valid} are greatly suppressed by employing the $u_{*}$-scaling. However, $\tau^{*}$ does not collapse in the outer region. In contrast, the $u_{**}$-scaling successfully aligns the non-equilibrium APG TBLs' TSS on the NACA4412 airfoil with the ZPG TBLs' 3/2 defect power law (Fig. \ref{fig:Skote_scaling}(b)), covering two orders of magnitude in $\beta$. Fig. \ref{fig:Skote_scaling}(c) illustrates the wall-normal profiles of $u_{*}$ and $u_{**}$ applied in panels (a) and (b). Close to the wall $u_{*}$ collapses with $u_{**}$. While $u_{*}$ continues to grow when $y$ approaches $\delta$, $u_{**}$ saturates to $u_{**}(\delta)$ near the boundary layer edge. This plateau of $u_{**}$ near $\delta$ resembles the behavior of a turbulent mixing layer, which is widely considered to share similar characteristics with the outer flow of APG TBLs.\cite{devenport2022equilibrium,bradshaw1967turbulence}

\begin{figure}
    \centering
    \includegraphics[width=0.75\linewidth]{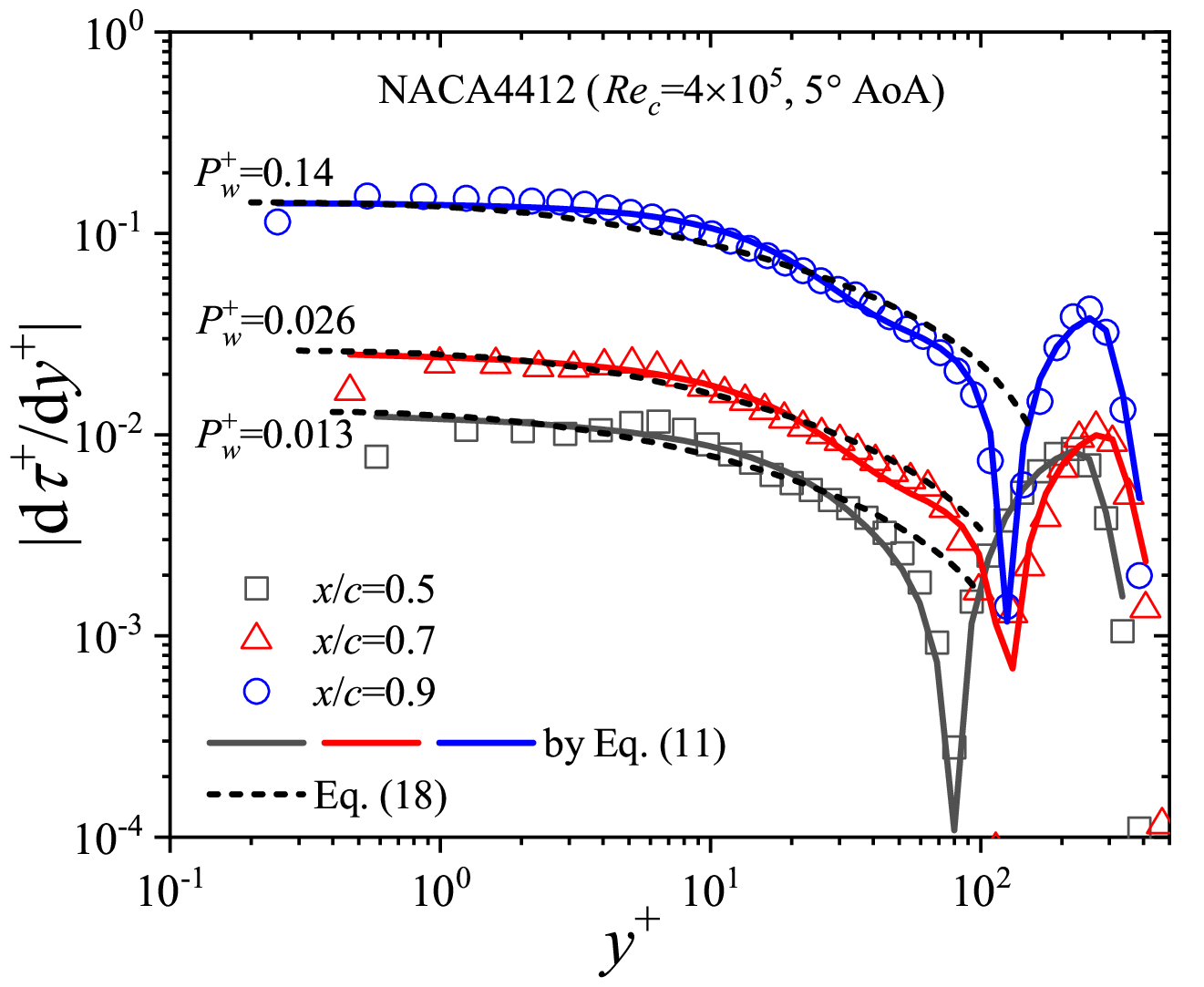}\\(a)\\
    \includegraphics[width=0.75\linewidth]{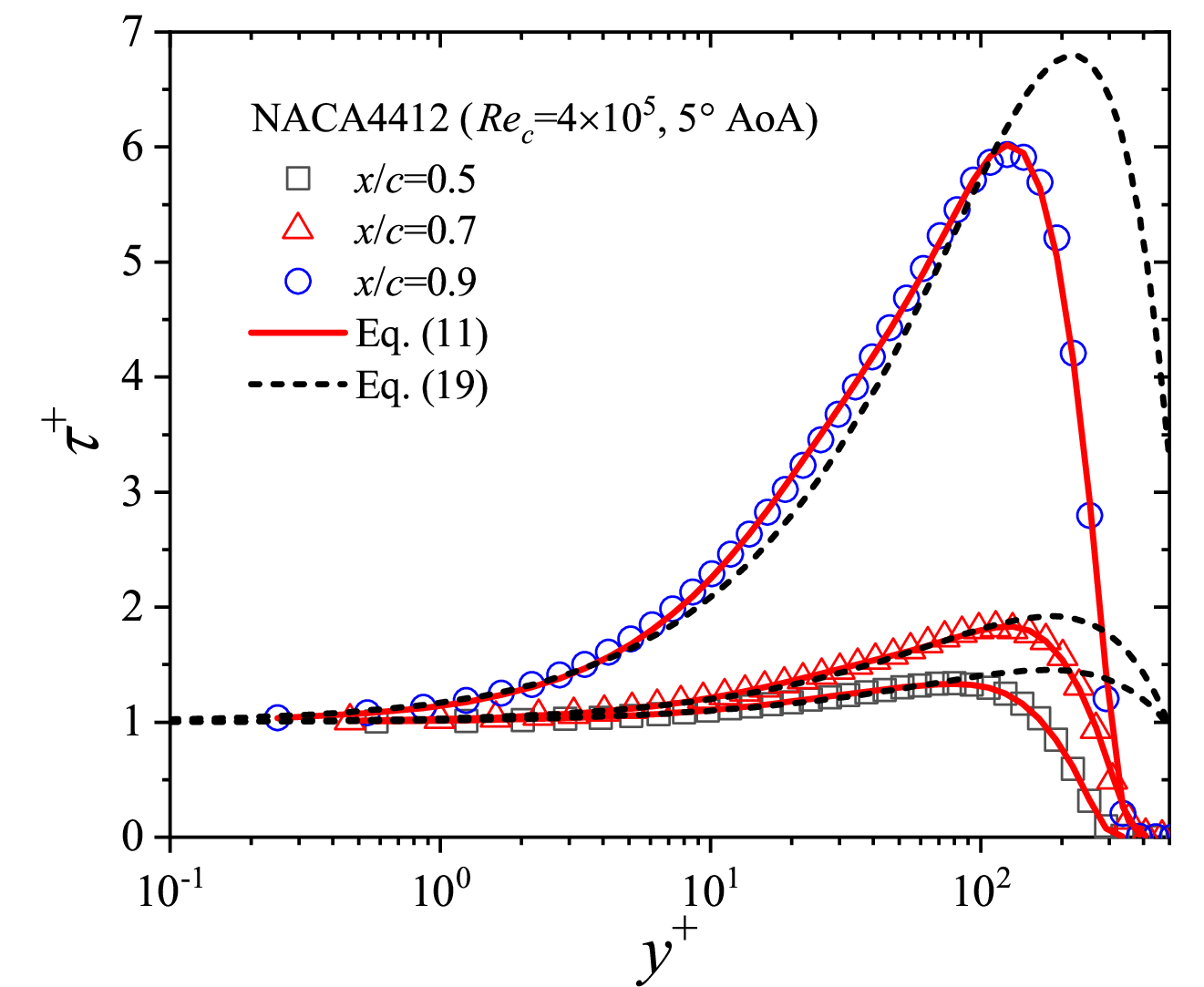}\\(b)
  \caption{(a) Profiles of $|d\tau^+/dy^+|$ predicted by Ma et al.'s model (Eq. (\ref{eq:dtdy_Ma}), short-dashed lines) and the current model (Eq. (\ref{eq:tau_NonEPG_threelayer}), solid lines), compared with the LES data (symbols) for the NACA4412 airfoil at $5^\circ$ AoA with $Re_c=4.0\times 10^5$. (b) Profiles of $\tau^+$ predicted by Ma et al.'s inner scaling (Eq. (\ref{eq:tau_inner_Ma}), short-dashed lines) and the current model (Eq. (\ref{eq:tau_NonEPG_threelayer}), solid lines), compared with the LES data (symbols).}
  \label{fig:Ma_inner_scaling}
\end{figure}

Via an asymptotic expansion method,\cite{Afzal2008} Ma et al. \cite{MaYan2024} recently derived an inner scaling for TSS of equilibrium APG TBLs as well as non-equilibrium APG TBLs before separation. Their formulation for ${\rm d} \tau^+/{\rm d} y^+$ reads
\begin{equation}
  \frac{{\rm d} \tau^+}{{\rm d} y^+}=P_w^+-\frac{\alpha P_w^+}{{\rm ln} (0.1\delta^+)}\frac{{\rm ln} (y^+ +1)}{1+0.005\delta^+/y^+},
  \label{eq:dtdy_Ma}
\end{equation}
where $\alpha=1-({\rm d}\tau^+/{\rm d}y^+)_{0.1\delta}/P_w^+$, being the modulation factor of the PG parameter. $\alpha$ represents the ratio of decayed energy to the wall shear stress gradient,\cite{MaYan2024} and should be measured with experiment or simulation. The inner profile of $\tau^+$ is integrated as
\begin{equation}
  \tau^+=1+P_w^+y^+-\frac{\alpha P_w^+}{{\rm ln} (0.1\delta^+)}(y^+{\rm ln}y^+-y^+).
  \label{eq:tau_inner_Ma}
\end{equation}
(\ref{eq:dtdy_Ma}) improves the linear approximation of TSS in the overlap region by quantifying the advection contribution in the mean momentum equation. 

Fig. \ref{fig:Ma_inner_scaling}(a) illustrates the profiles of $|{\rm d} \tau^+/{\rm d} y^+|$ predicted by (\ref{eq:tau_NonEPG_threelayer}) and (\ref{eq:dtdy_Ma}) for the TBLs at three chord positions on the suction surface of the NACA4412 airfoil (AoA$=5^\circ$ and $Re_c=4.0\times 10^5$). While (\ref{eq:dtdy_Ma}) captures well the variation of ${\rm d} \tau^+/{\rm d} y^+$ in the inner region, the current model accurately reproduces the complete profile. Notably, (\ref{eq:tau_NonEPG_threelayer}) precisely predicts the wall-normal location of the peak TSS (sharp features in Fig. \ref{fig:Ma_inner_scaling}(a)). Similar performance is observed in Fig. \ref{fig:Ma_inner_scaling}(b) for $\tau^+$ profiles predicted by (\ref{eq:tau_NonEPG_threelayer}) and (\ref{eq:tau_inner_Ma}). The agreement in the inner region allows using (\ref{eq:tau_inner_Ma}) as a constraint for estimating the empirical parameters in (\ref{eq:tau_NonEPG_threelayer}). It is critical to note, however, that (\ref{eq:tau_inner_Ma}) relies on an empirical parameter $\alpha$ requiring its independent calibration.

\begin{figure*}
    \centering
    \includegraphics[width=0.33\linewidth]{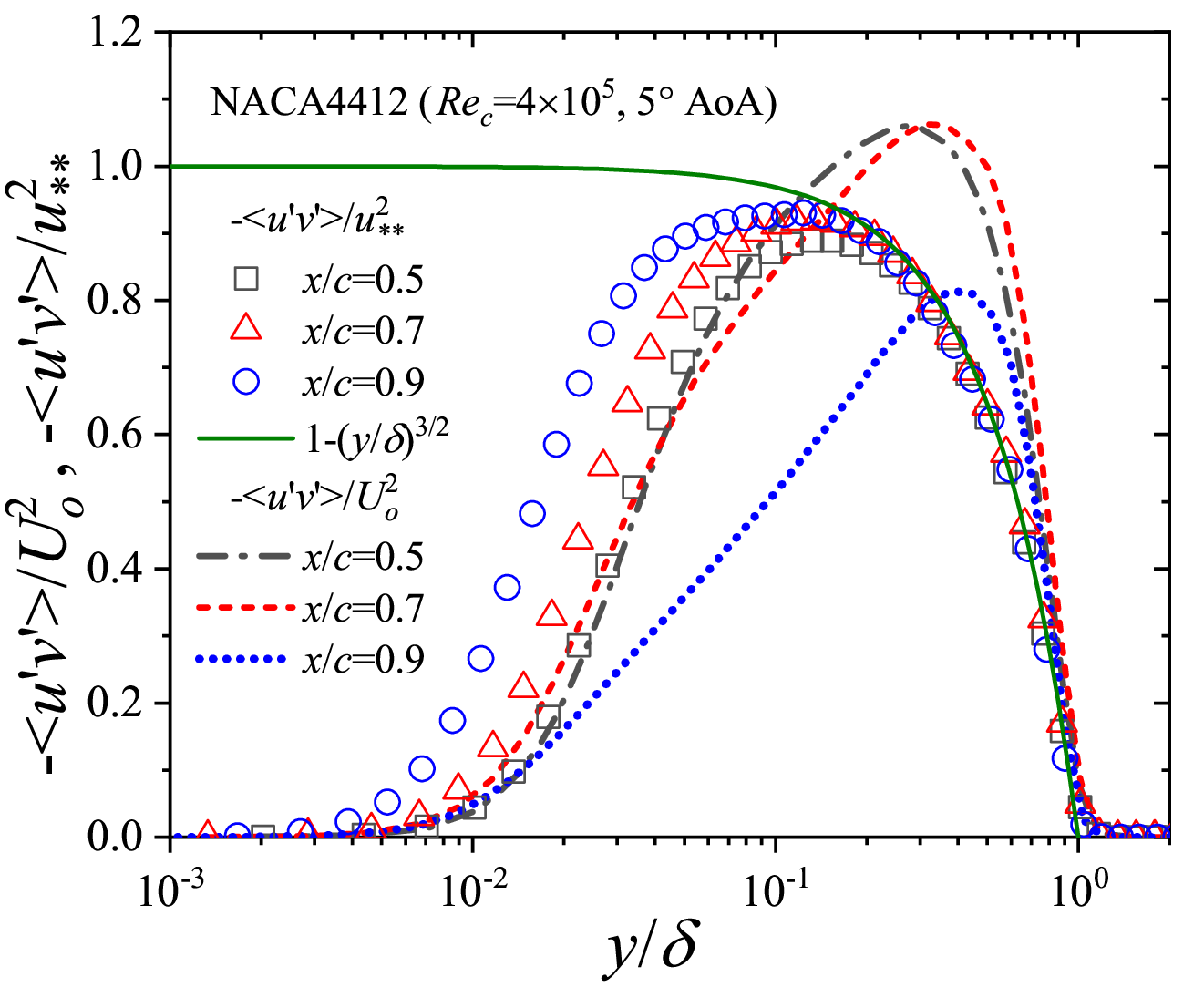}
    \includegraphics[width=0.33\linewidth]{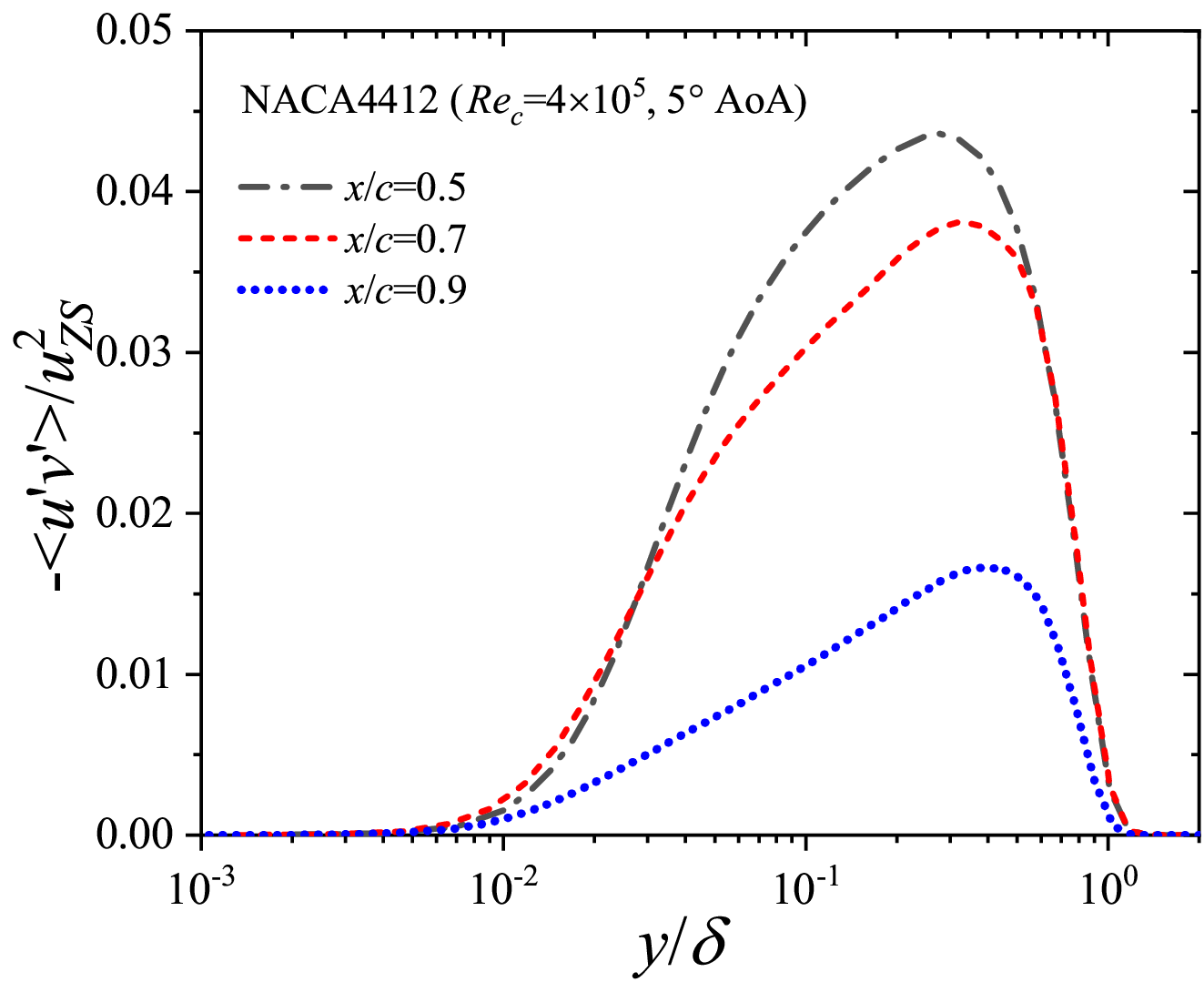}
    \includegraphics[width=0.33\linewidth]{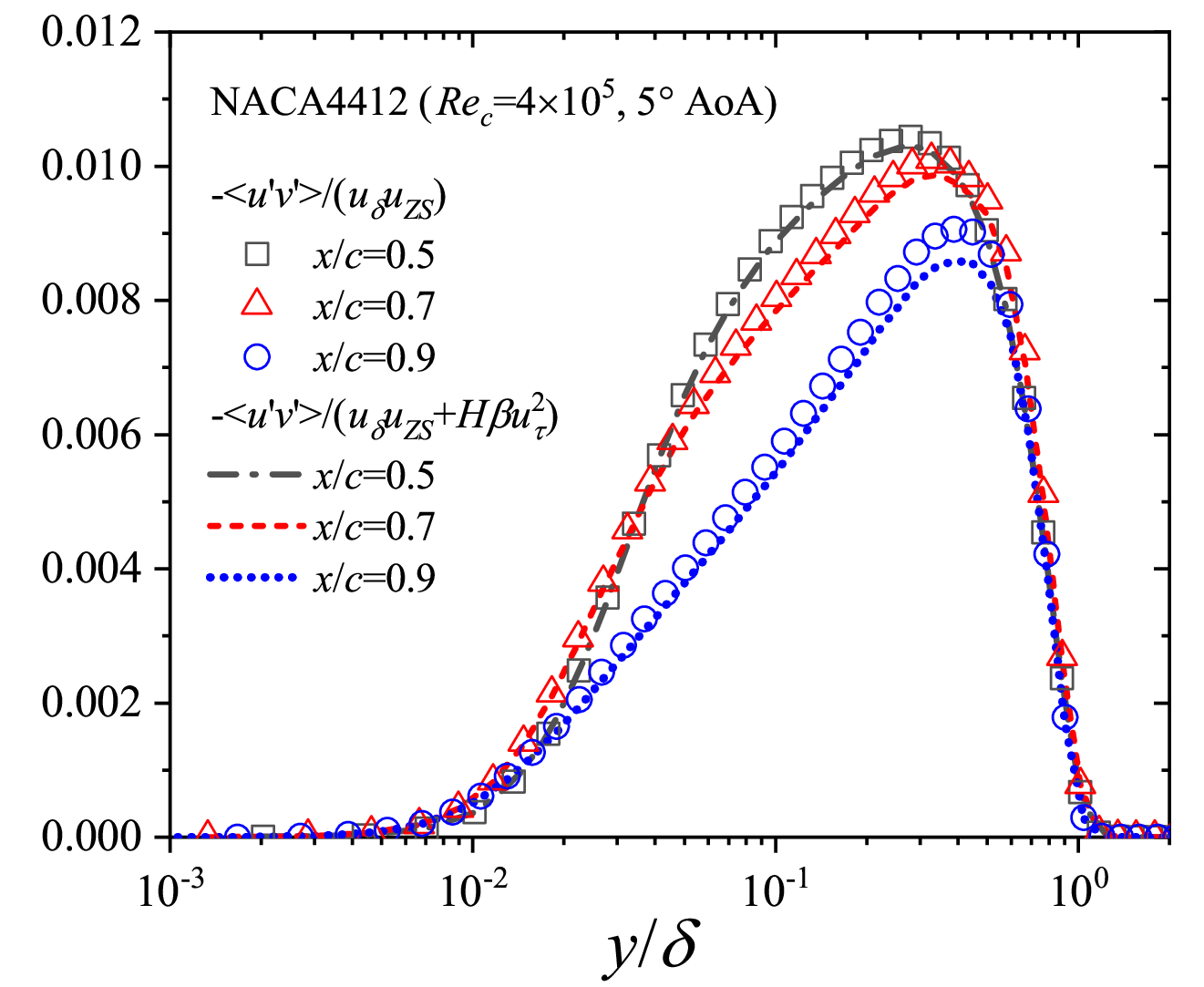}
    \\\quad(a)\quad\quad\quad\quad\quad\quad\quad\quad\quad\quad\quad\quad\quad\quad\quad\quad\quad\quad(b)\quad\quad\quad\quad\quad\quad\quad\quad\quad\quad\quad\quad\quad\quad\quad\quad\quad(c)
  \caption{(a) $U_o$- and $u_{**}$-scaling of Reynolds shear stress in NACA4412 TBLs, compared to the 3/2 defect law of ZPG TBLs. (b) $u_{ZS}$-scaling of Reynolds shear stress. (c) The consistent outer scaling of Reynolds shear stress, with and without the history correction. }
  \label{fig:Ma_outer_scaling}
\end{figure*}

Ma et al. \cite{MaYan2024} also proposed an outer velocity scale, $U_o=u_\tau\sqrt{1+(\delta_1/\delta)\beta}$, to characterize the mean velocity defect and Reynolds shear stress in the outer region. They demonstrated that $-\left<u'v'\right>/U_o^2=O(1)$ holds for both equilibrium APG TBLs and non-equilibrium APG TBLs before separation, outperforming multiple existing scalings such as the Zagarola-Smits scaling.\cite{zagarola1998mean} Fig. \ref{fig:Ma_outer_scaling}(a) compares Reynolds shear stress scalings in the outer region using $U_o$- and $u_{**}$-normalizations. While Ma et al.'s $U_o$-scaling preserves $O(1)$ magnitude, the proposed $u_{**}$-scaling maintains the $3/2$ defect power law of ZPG TBLs. Fig. \ref{fig:Ma_outer_scaling}(b) demonstrates the Zagarola-Smits scaling for Reynolds shear stress, where $u_{ZS}$-scaled stresses fail to collapse throughout the boundary layer under different APGs. In addition, the peak stresses are at least one order of magnitude smaller than unity.\cite{Maciel2018} 

Han et al. \cite{YanJFM2024} recently proposed consistent outer scalings for near-equilibrium and non-equilibrium APG TBLs. Based on the linearized boundary-layer equation,\cite{Lumley} they argued that Reynolds shear stress in the outer region is invariant when scaled with 
$u_\delta u_{ZS}$. By introducing a correction term, they established that 
\begin{equation}
  \frac{-\left<u'v'\right>}{u_\delta u_{ZS}+H\beta u_\tau^2}=f(y/\delta), \quad y/\delta>0.45,
  \label{eq:tau_outer_Han}
\end{equation}
where $f$ is independent of APG and flow history when $y/\delta>0.45$, and $H\beta u_\tau^2$ corrects $u_\delta u_{ZS}$ to account for history effects. Fig. \ref{fig:Ma_outer_scaling}(c) validates this consistent outer scaling (Eq. (\ref{eq:tau_outer_Han})) using non-equilibrium APG TBL data from the NACA4412 wing section. Although the rescaled Reynolds shear stress is about two orders of magnitude below unity, (\ref{eq:tau_outer_Han}) holds for $y/\delta>0.45$. We find the correction term $H\beta u_\tau^2$ is negligible for the current databases. 

\begin{figure}
    \centering
    \includegraphics[width=0.75\linewidth]{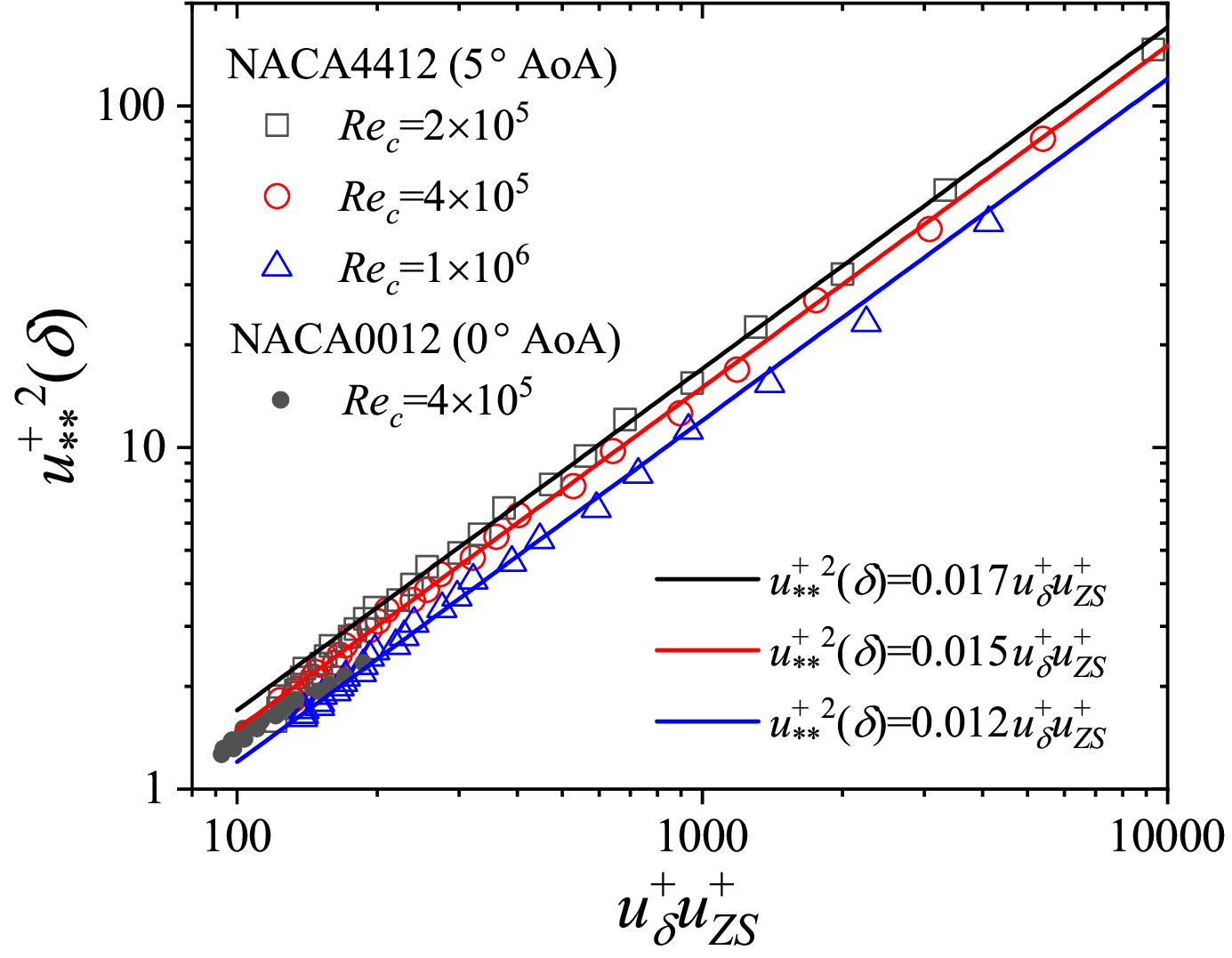}\\(a)\\
    \includegraphics[width=0.75\linewidth]{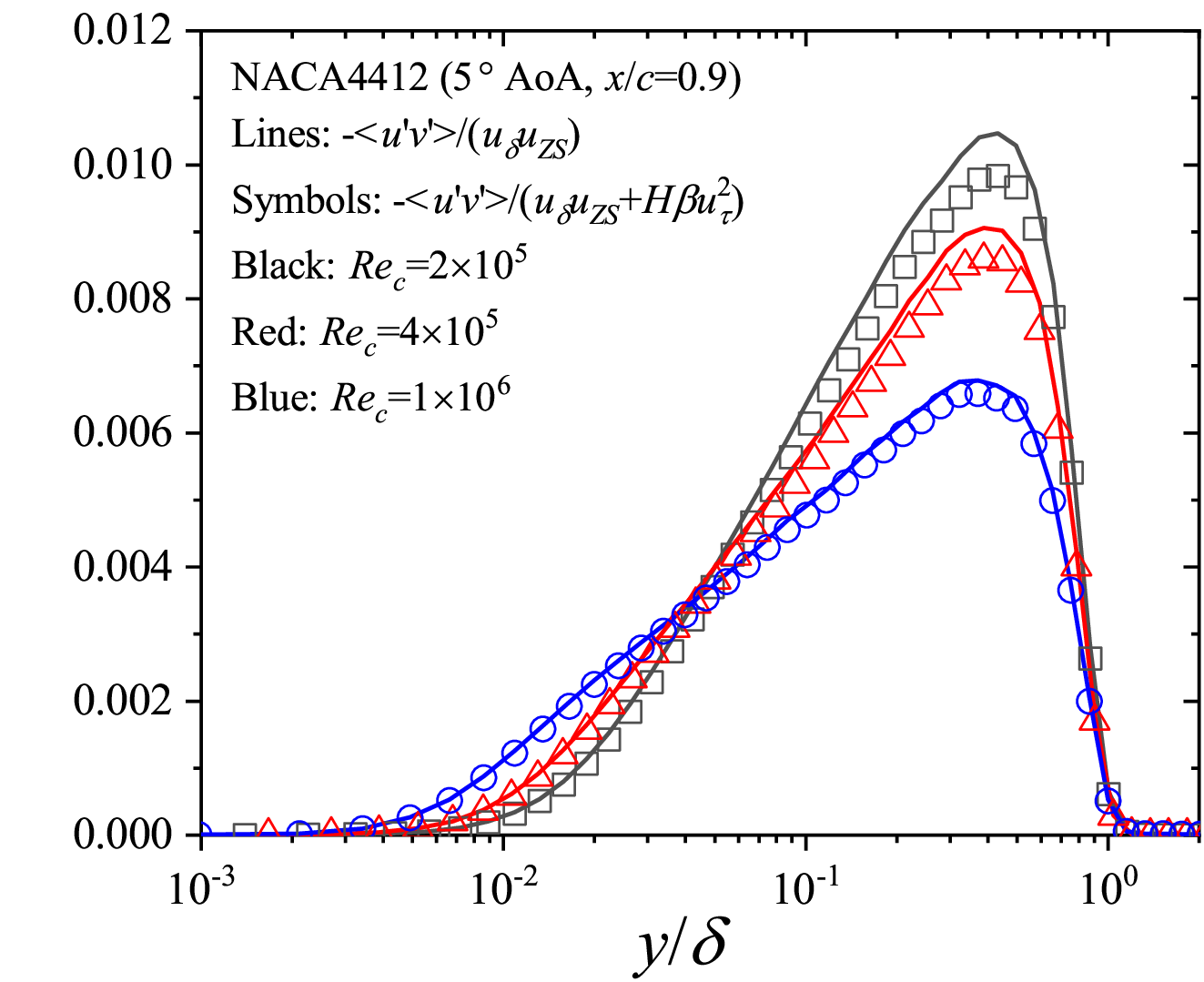}\\(b)
  \caption{(a) Validation of the proportional relationship $u_{**}^2(\delta)=\sigma u_\delta u_{ZS}$ using the NACA4412 and NACA0012 TBL data (symbols). Data distribution shows $\sigma$ varies between datasets. Each dataset (same chord Reynolds number and airfoil type) contains 28 chord positions evenly distributed in $0.5<x/c<1.0$. Lines show least-squares fits. (b) The consistent outer scaling of Reynolds shear stress (with/without history correction) for NACA4412 TBL at $x/c=0.9$ with three $Re_c$. The scaling fails to collapse for $y/\delta>0.45$, indicating $Re_c$-dependence.}
  \label{fig:cm_correlation}
\end{figure}

Combining the $u_{**}$- and $u_\delta u_{ZS}$-scaling for Reynolds shear stress in the outer region, we derive 
\begin{equation}
u_{**}^2(\delta)=\sigma u_\delta u_{ZS}, 
  \label{eq:ustarstar_uduZS}
\end{equation}
where $\sigma$ is a constant coefficient. This relationship is validated in Fig. \ref{fig:cm_correlation}(a) using the NACA4412 and NACA0012 TBL data. Interestingly, (\ref{eq:ustarstar_uduZS}) holds for each dataset (different chord positions but same $Re_c$ and airfoil type), but $\sigma$ varies between datasets. Least-squares fitting yields $\sigma=0.017$, 0.015, and 0.012 for the NACA4412 TBL at $Re_c=2\times10^5$, $4\times10^5$, and $1\times10^6$, respectively, and $\sigma=0.014$ for the NACA0012 TBL at $Re_c=4\times10^5$. This $Re_c$-dependence is attributed to the consistent outer scaling, as validated in Fig. \ref{fig:cm_correlation}(b) using the NACA4412 TBL data at $x/c=0.9$ with three $Re_c$. The consistent outer scaling of Reynolds shear stress fails to collapse different $Re_c$ data for $y/\delta>0.45$. The reason and quantitative behavior of this $Re_c$-dependence should be clarified with future studies.

\begin{figure}
    \centering
    \includegraphics[width=0.75\linewidth]{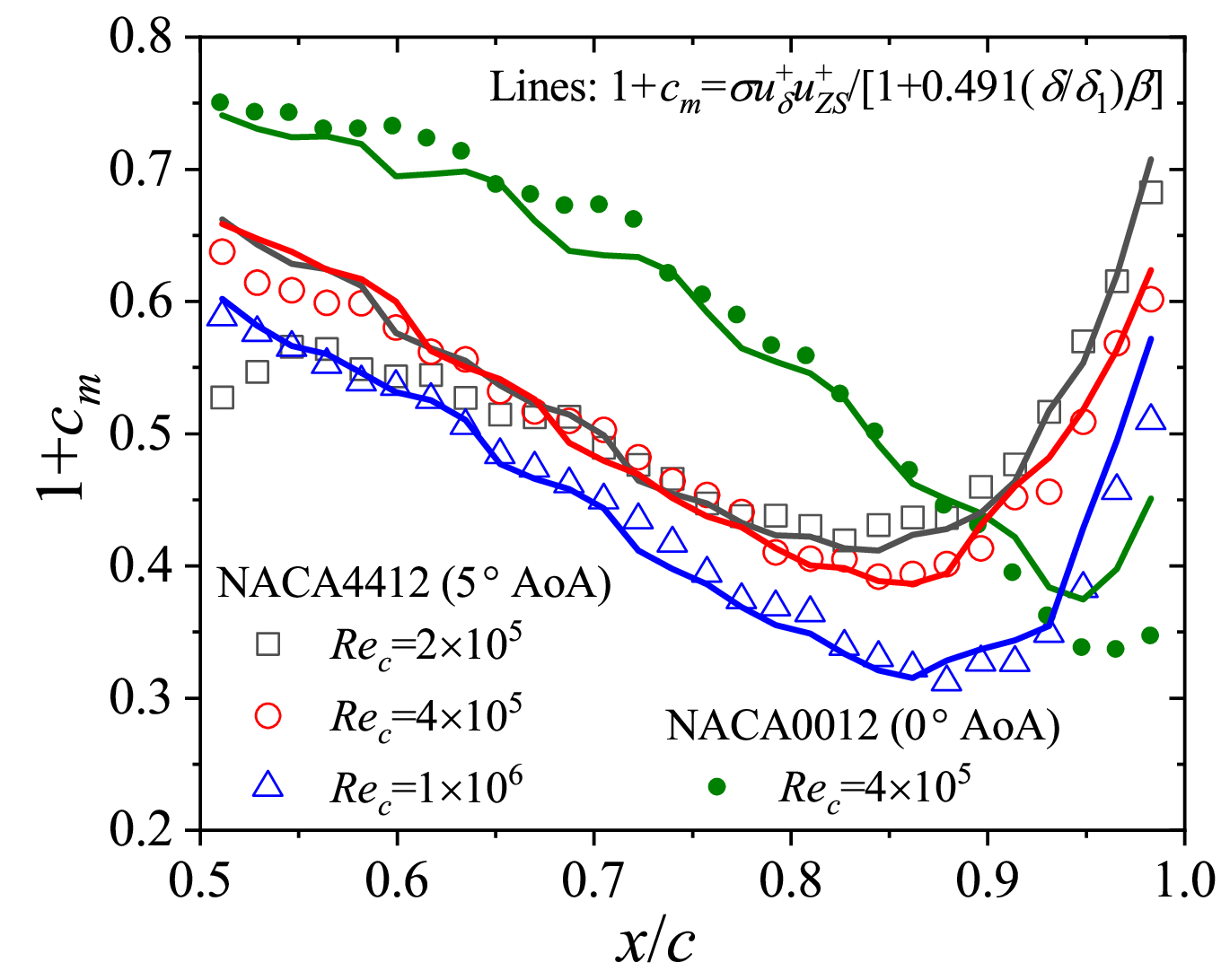}\\(a)\\
    \includegraphics[width=0.75\linewidth]{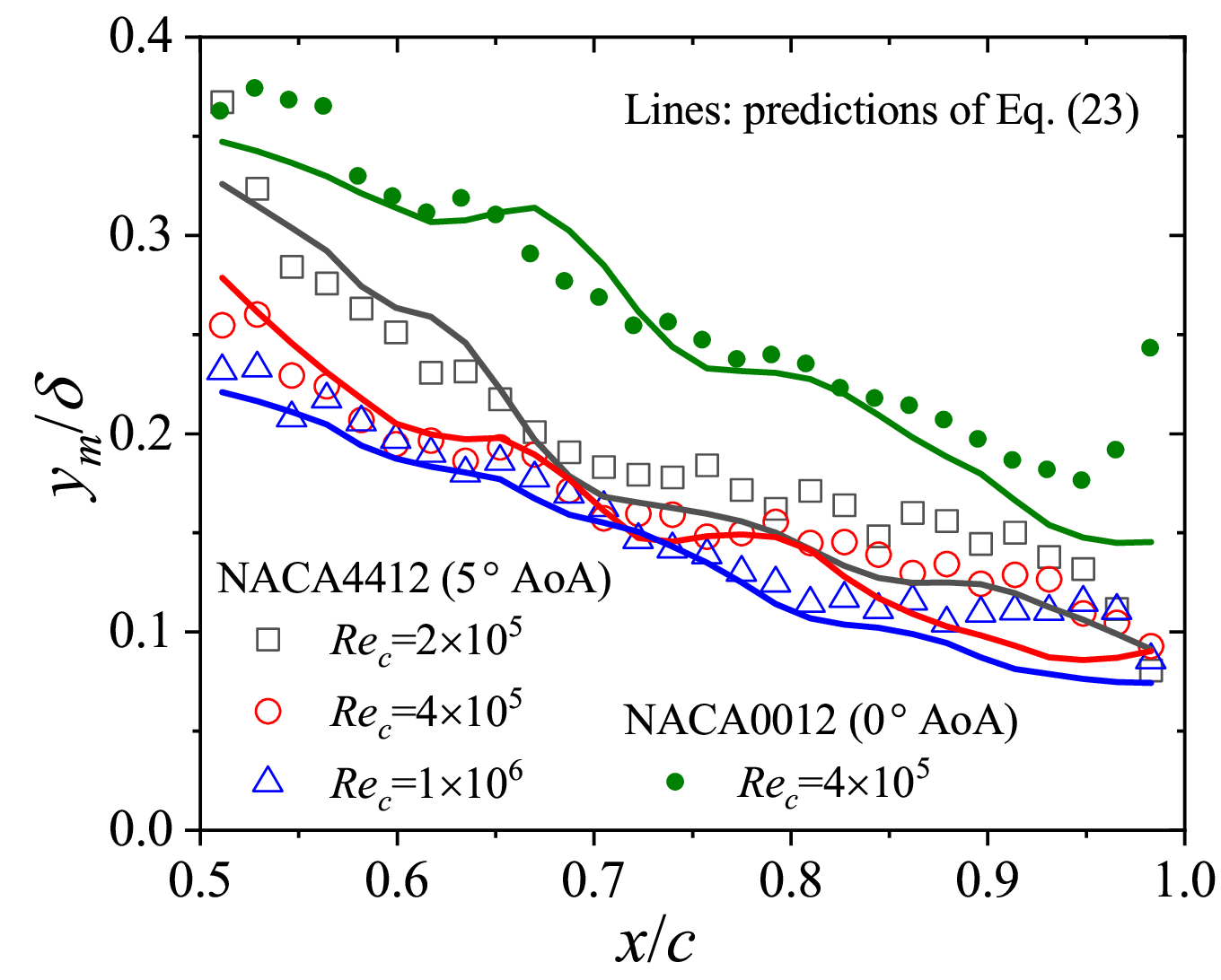}\\(b)
  \caption{Evolution of (a) $1+c_m$, and (b) $y_m/\delta$ in the chordwise direction on the suction surface of the NACA4412 and NACA0012 airfoils. Symbols denote results from least-squares optimization. Lines illustrate predictions of (\ref{eq:cm_model}) and (\ref{eq:ym_model}).}
  \label{fig:cm_ym_wing}
\end{figure}

To remedy the finite-$Re$ effect in the consistent outer scaling, we allow $\sigma$ to depend on $Re_c$. Based on this, a model for parameter $c_m$ in (\ref{eq:tau_NonEPG_threelayer}) is derived from (\ref{eq:ustarstar_uduZS}) as
\begin{equation}
  1+c_m=\frac{\sigma u_\delta^+ u_{ZS}^+}{1+0.491\left(\delta/\delta_1\right)\beta}.
  \label{eq:cm_model}
\end{equation}
The streamwise development of $1+c_m$ is presented in Fig. \ref{fig:cm_ym_wing}(a) for all datasets. Here, $1+c_m$ quantifies the ratio of the non-equilibrium TSS to the reference equilibrium TSS near the boundary layer edge, thereby serving as a metric for outer-region delay strength. As illustrated in Fig. \ref{fig:cm_ym_wing}(a), $1+c_m$ associated with the NACA4412 airfoil exhibits a gradual decline with increasing $x/c$, accompanied by rising $\beta$, $\beta_{ZS}$, and $P_w^+$ (Fig. \ref{fig:wingstat}). This pattern suggests a progressive accumulation of equilibrium deviation in the outer flow region along the chordwise direction. This trend undergoes an abrupt reversal at approximately 85\% chord length near the trailing edge, which is consistent with $\beta_{ZS}$ but not $\beta$ and $P_w^+$. It is documented in Ref.~\onlinecite{Vinuesa2018wing} that the TBLs are in incipient separation near the trailing edge. Therefore, the current parameterization captures the reduced non-equilibrium intensity as TBLs approach separation conditions, agreeing with previous observations.\cite{Alving1996,CastilloASME} In contrast, the NACA0012 airfoil at zero AoA maintains attached flow throughout its chord length, resulting in a monotonic decrease of $1+c_m$ along the entire wing chord. The apparent anomalies at $x/c=0.966$ and 0.983 positions (detailed in Fig. \ref{fig:wing_tau_model_valid}(h)) likely represent localized flow phenomena rather than characteristic trends. We observe that (\ref{eq:cm_model}) accurately captures the evolution of $1+c_m$ in the chordwise direction for all datasets.

\begin{figure}
    \centering
    \includegraphics[width=0.75\linewidth]{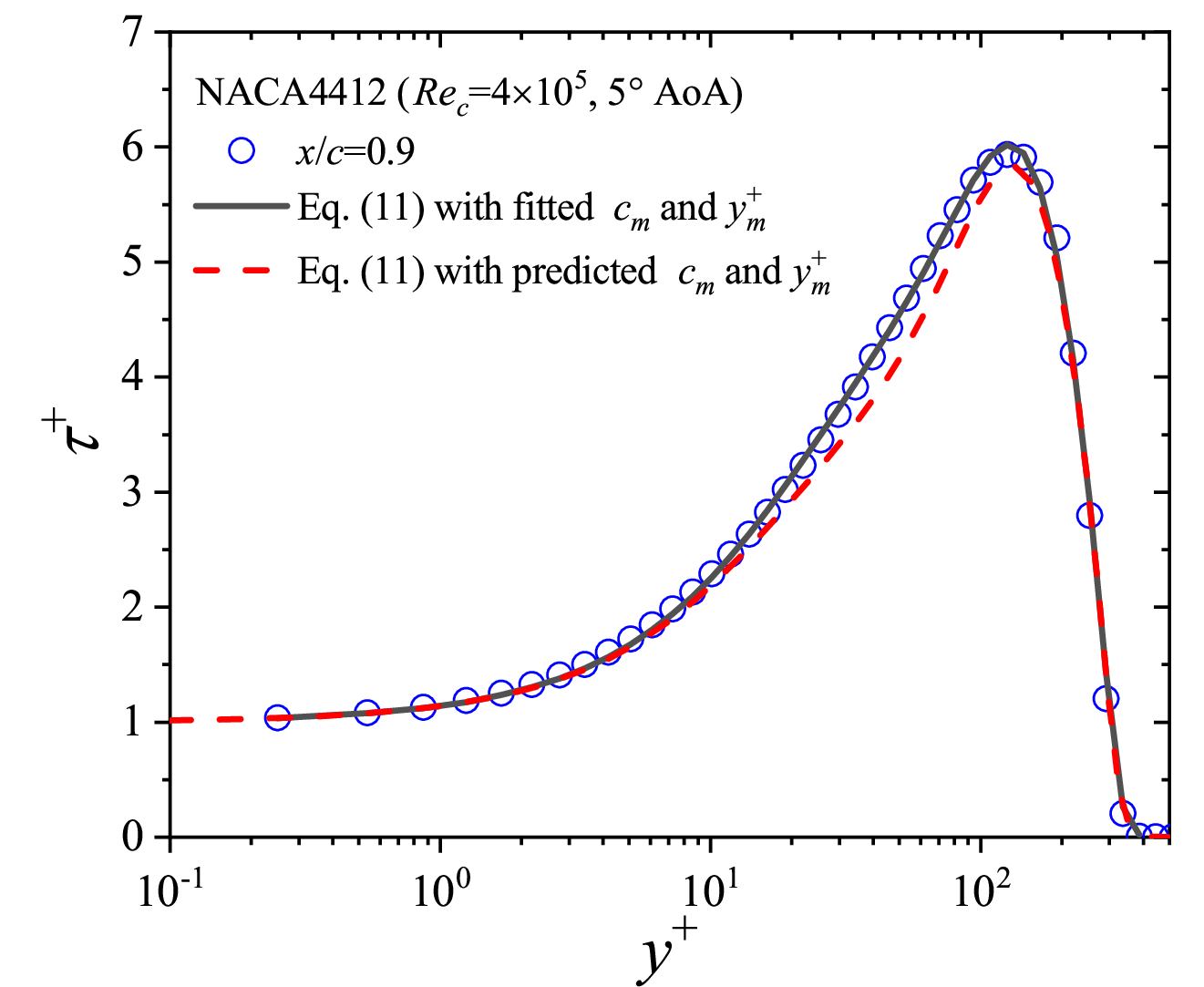}
  \caption{Comparison between the predictions of (\ref{eq:tau_NonEPG_threelayer}) with $c_m$ and $y_m^+$ obtained via least-squares optimization (solid line), and with $c_m$ and $y_m^+$ predicted by (\ref{eq:cm_model}) and (\ref{eq:ym_model}) (dashed line). Open circles denote the LES data of TSS at $x/c=0.9$ for NACA4412 airfoil at AoA=$5^\circ$ and $Re_c=4.0\times 10^5$.}
  \label{fig:wing_tau_model}
\end{figure}

The other parameter, $y_m^+$, in (\ref{eq:tau_NonEPG_threelayer}) can be derived by equating the current prediction for TSS to Ma et al.'s inner scaling, as mentioned earlier. As shown in Fig. \ref{fig:Ma_inner_scaling}(b), (\ref{eq:tau_NonEPG_threelayer}) and (\ref{eq:tau_inner_Ma}) nearly collapse below the location of peak shear stress. Since $y_m^+$ is smaller than this peak location when the non-equilibrium effect is beyond mild levels, equating (\ref{eq:tau_NonEPG_threelayer}) and (\ref{eq:tau_inner_Ma}) at $y_m^+$ yields
\begin{align}
&1+P_w^+y_m^+-\frac{\alpha P_w^+}{{\rm ln} (0.1\delta^+)}(y_m^+{\rm ln}y_m^+-y_m^+)\nonumber\\
&=(1+2^{-0.75}c_m)\left(1+P_w^+y_m^+\right)\left[1-\left({y_m^+}/{\delta^+}\right)^{1.5}\right],
  \label{eq:ym_model}
\end{align}
where $c_m$ is described by (\ref{eq:cm_model}) and $\alpha$ should be measured from empirical data. The solution of (\ref{eq:ym_model}) is implicit and thus should be solved numerically.

Fig. \ref{fig:cm_ym_wing}(b) illustrates the chordwise evolution of $y_m/\delta$ for all datasets. This parameter defines the dimensionless wall-normal position separating the near-wall instant response region from the outer delayed zone. Thus, the smaller $y_m/\delta$ is, the greater the cumulative effect of the non-equilibrium PG. For all datasets, $y_m/\delta$ decreases progressively in the chordwise direction (except the anomalies at $x/c=0.966$ and 0.983 of the NACA0012 airfoil), unlike $1-c_m$, which displays significant reversal near the trailing edge. (\ref{eq:ym_model}) captures the decreasing trend of $y_m/\delta$, but appears quantitatively insufficient. A possible explanation is that $y_m$ is a less sensitive parameter for predicting TSS with (\ref{eq:tau_NonEPG_threelayer}). This explanation is validated in Fig. \ref{fig:wing_tau_model}, in which predictions of (\ref{eq:tau_NonEPG_threelayer}) with $c_m$ and $y_m^+$ obtained via least-squares optimization, and with $c_m$ and $y_m^+$ predicted by (\ref{eq:cm_model}) and (\ref{eq:ym_model}), are compared with LES data for a typical chord position of the NACA4412 TBL. The discrepancy between fitted $y_m/\delta$ and predicted $y_m/\delta$ shown in Fig. \ref{fig:cm_ym_wing}(b) is proved to result in negligible differences for predicting TSS by (\ref{eq:tau_NonEPG_threelayer}). Consequently, we obtain a closed model (Eq. (\ref{eq:tau_NonEPG_threelayer})) for predicting complete profiles of TSS in non-equilibrium TBLs under gradually varying PG conditions. 

\subsection{Relaxing TBLs subjected to rapidly decreasing APG and/or APG-FPG transition}\label{subsubsec:decreaseAPG}
Relaxing TBLs refer to the evolution process of APG TBLs undergoing sudden APG removal or abrupt transition to FPG conditions.\cite{White2005} This phenomenon is observed on the windward surface of a Gaussian bump flow configuration. To evaluate the predictive capability of current TSS models in characterizing relaxing TBL dynamics, we employ high-fidelity hybrid DNS-LES datasets on Gaussian bump flows from Uzun and Malik.\cite{Uzun2022AIAA}

Uzun and Malik's numerical simulation employs a new benchmark case termed the ``speed bump'' flow, which examines the interaction of an incoming TBL with the strong FPG and APG generated by a Gaussian bump. The bump, centered at $x/L=0$, has a height of $0.085L$ and a streamwise width of $0.195L$, where $L$ is the reference length. The Reynolds number based on the bump height is 170,000, and the momentum-thickness Reynolds number of the incoming TBL at $x/L=-0.8$ is $Re_\theta=1035$.\cite{Uzun2022AIAA} 

\begin{figure*}
    \centering
    \includegraphics[width=0.33\linewidth]{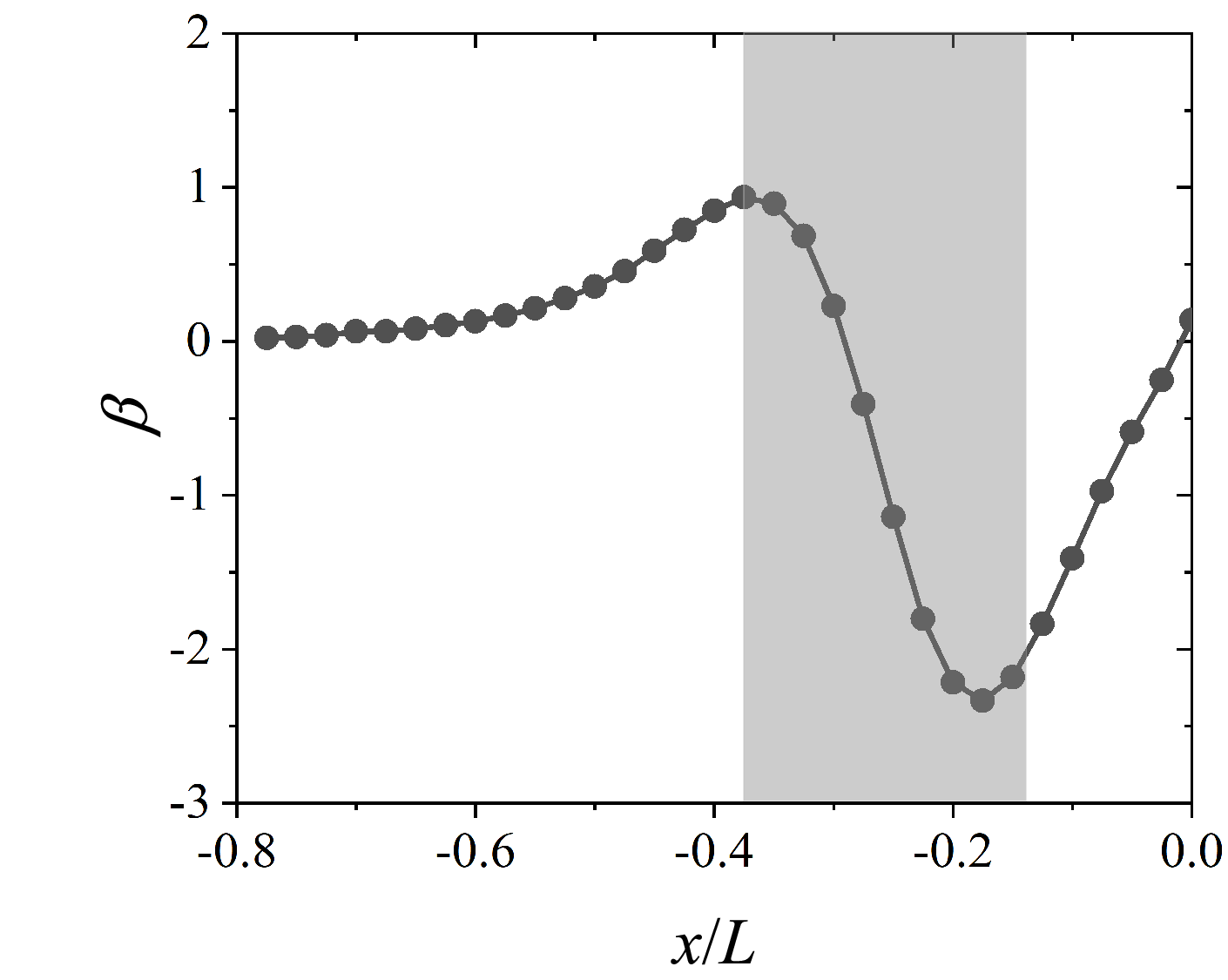}
    \includegraphics[width=0.33\linewidth]{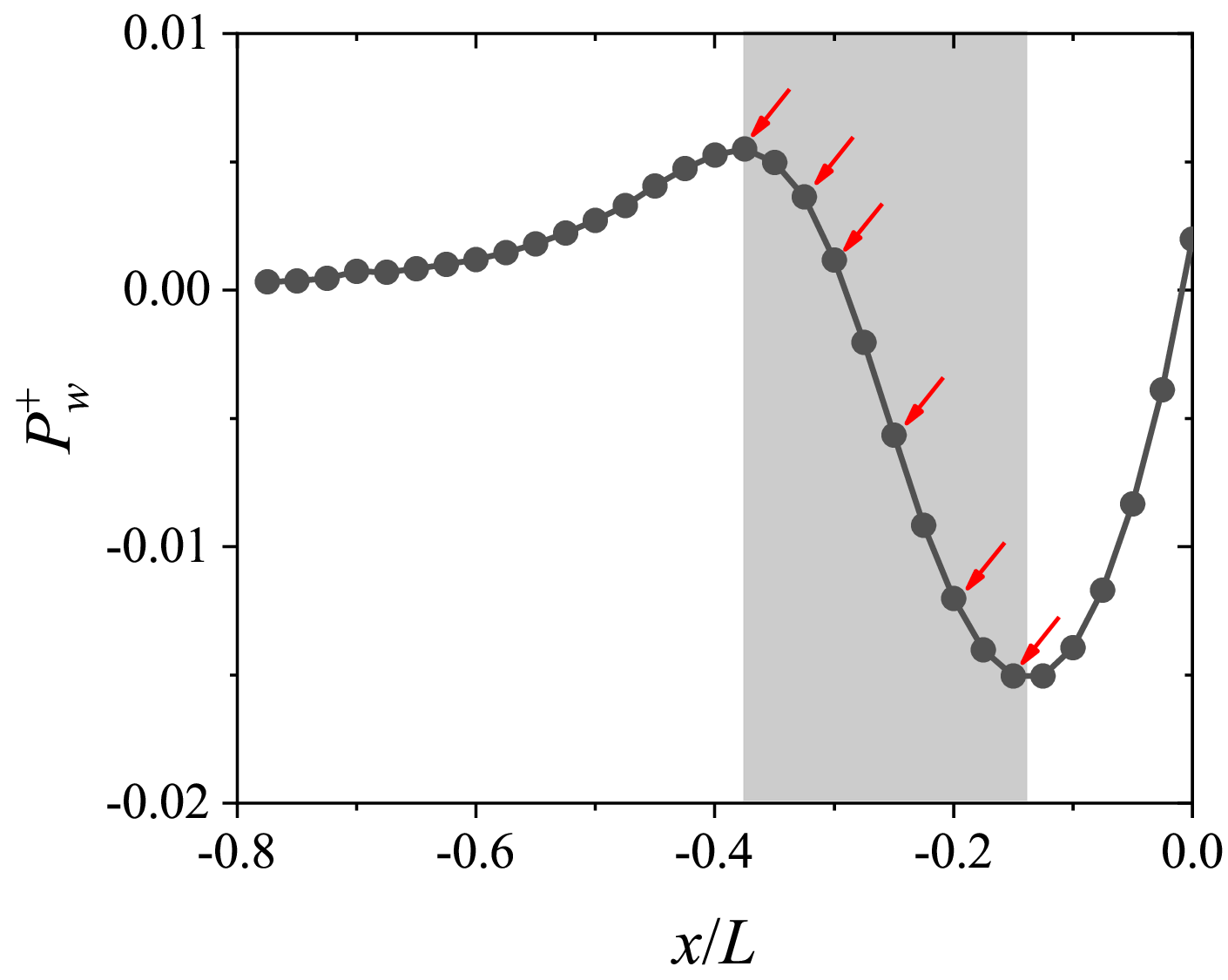}
    \includegraphics[width=0.33\linewidth]{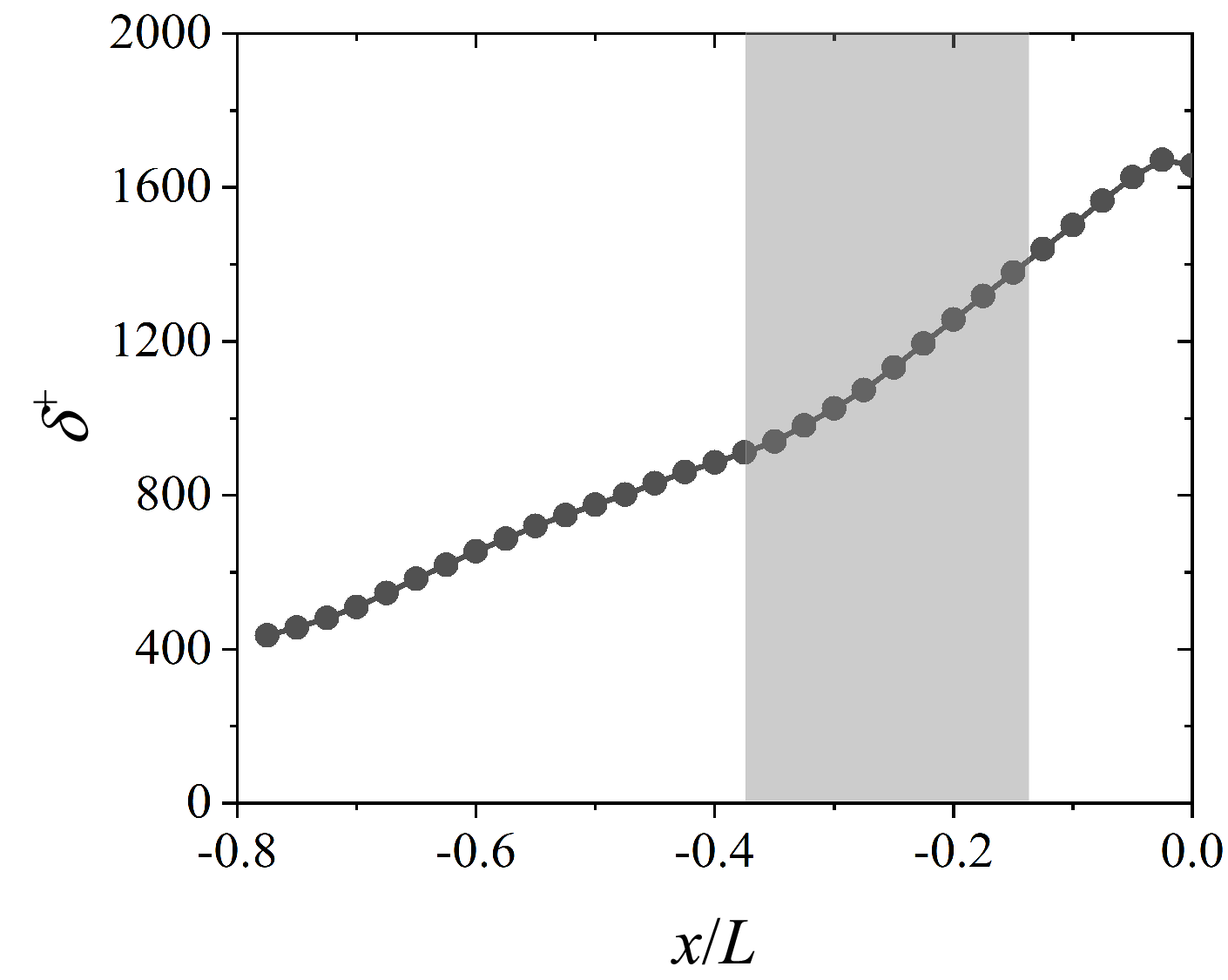}
    \\\quad(a)\quad\quad\quad\quad\quad\quad\quad\quad\quad\quad\quad\quad\quad\quad\quad\quad\quad\quad(b)\quad\quad\quad\quad\quad\quad\quad\quad\quad\quad\quad\quad\quad\quad\quad\quad\quad(c)
  \caption{Evolution of (a) the Clauser PG parameter $\beta$, (b) the PG parameter $P_w^+$, and (c) the friction Reynolds number $Re_\tau$ (i.e., $\delta^+$) on the front surface of the Gaussian bump. Data are acquired from the hybrid DNS-LES datasets from Uzun and Malik.\cite{Uzun2022AIAA} The region in gray marks the relaxing-TBL regime defined by negative ${\rm d}P_w^+/{\rm d}x$. The streamwise positions analyzed in this study are labelled with arrows in (b).}
  \label{fig:bumpstat}
\end{figure*}
\begin{table*}
\caption{\label{tab:parameters_bump}Boundary-layer conditions and model parameters at the investigated streamwise positions of the relaxing TBL.}
\begin{ruledtabular}
\begin{tabular}{lccccccccccr}
      $x/L$ & $P_w^+$ & $\tau_{\rm {max\mbox{-}o}}^+$ (\footnotemark[1])  & $y_P^+/\delta^+$ (\footnotemark[2]) & $c_m$  & $y_m^+/\delta^+$  &  $y_P^+$ (\footnotemark[3])  & $\delta_i^+/\delta^+$ & $\delta_w^+/\delta^+$ & $W_{\rm {max}}^+$ & $\delta^+$  & $\beta$ \\[3pt]
     -0.375 & 0.0055  & 1.3 & 0.32 & -0.45   & 0.19  & - & - & - & -  & 912 & 0.937 \\
     -0.325 & 0.0036  & 1.44 & 0.91 & -1.68   & 1.36  & - & - & - & -  & 981 & 0.685 \\
     -0.3 & 0.0012  & 1.48 & 0 & 0.93   & 0.18  & - & - & - & -  & 1,025 & 0.23 \\
     -0.25 & -0.0056  & 1.4 & 0.021 & 1.28   & 0.23  & - & - & - & -  & 1,132 & -1.14 \\
     -0.2 & -0.012  & 1.08 & 0.019 & 1.37   & 0.33  & 9.92 & 0.62 & 0.36 & 1.36  & 1,256 & -2.22 \\
     -0.15 & -0.015  & 0.68 & - & -   & -  & 14.1 & 0.29 & 0.32 & 1.0  & 1,379 & -2.18 \\
\end{tabular}
\end{ruledtabular}
\footnotetext[1]{$\tau_{\rm {max\mbox{-}o}}^+$ represents the outer peak of TSS.}
\footnotetext[2]{$y_P^+$ in the three-layer TSS model (Eq. (\ref{eq:tau_NonEPG_threelayer})).}
\footnotetext[3]{$y_P^+$ in the dual-boundary-layer TSS model (Eq. (\ref{eq:tau_noneq})).}
\end{table*}

Fig. \ref{fig:bumpstat} illustrates the distributions of $\beta$, $P_w^+$, and $\delta^+$ along the bump's windward surface, which demonstrates greater complexity in PG variation than those on the suction surface of wing sections. Gungor et al. \cite{Gungor2024} argued that characteristic parameters based only on local variables cannot fully capture the physics of non-equilibrium boundary layers, and finding appropriate parameters and response length scales
for non-equilibrium TBLs remains an open question. Here, we temporarily use $P_w^+$ as an indicator for flow classification because $P_w^+$ is the driving factor under the wall-unit normalization (see Eq. (\ref{eq:tau_nearwall})) and the boundary layer responds to its variation. As shown in Fig. \ref{fig:bumpstat}(b), the $P_w^+$ evolution reveals three distinct regimes: (1) increasing APG for $x/L<-0.375$, (2) decreasing APG transitioning to increasing FPG in $-0.375<x/L<-0.14$, and (3) decaying FPG transitioning to increasing APG for $x/L>-0.14$. The first regime belongs to the non-equilibrium process studied in section \ref{subsubsec:TBL_wing}, while the third regime belongs to the non-equilibrium process studied in section \ref{subsubsec:increaseAPG}. The second regime (shown in gray in Fig. \ref{fig:bumpstat}) features a relaxing TBL, which is the primary focus of this discussion. In the following analysis, the boundary layer thickness $\delta^+$ follows the 99.5\% velocity criterion (Method B in Ref.~\onlinecite{Uzun2022AIAA}) established by Uzun and Malik,\cite{Uzun2022AIAA} ensuring consistent scaling throughout the analysis.

Given that the PG and its streamwise variation are moderate, we utilize the three-layer model (Eq. (\ref{eq:tau_NonEPG_threelayer})) to investigate the TSS evolution in the relaxing TBL over the Gaussian bump. Since $\beta<1$, the parameter $y_P^+$ in the reference equilibrium TSS profile ($\tau_{\rm {E\mbox{-}APG}}^+$ in (\ref{eq:tau_NonEPG_threelayer})) must be estimated empirically from simulation data, leading to quantification of three empirical parameters: $y_P^+$, $c_m^+$, and $y_m^+$. These parameters are optimized simultaneously using a least-squares fitting procedure applied to the complete TSS profile, with results listed in Table \ref{tab:parameters_bump} for all streamwise positions analyzed.

\begin{figure*}
    \centering
    \includegraphics[width=0.33\linewidth]{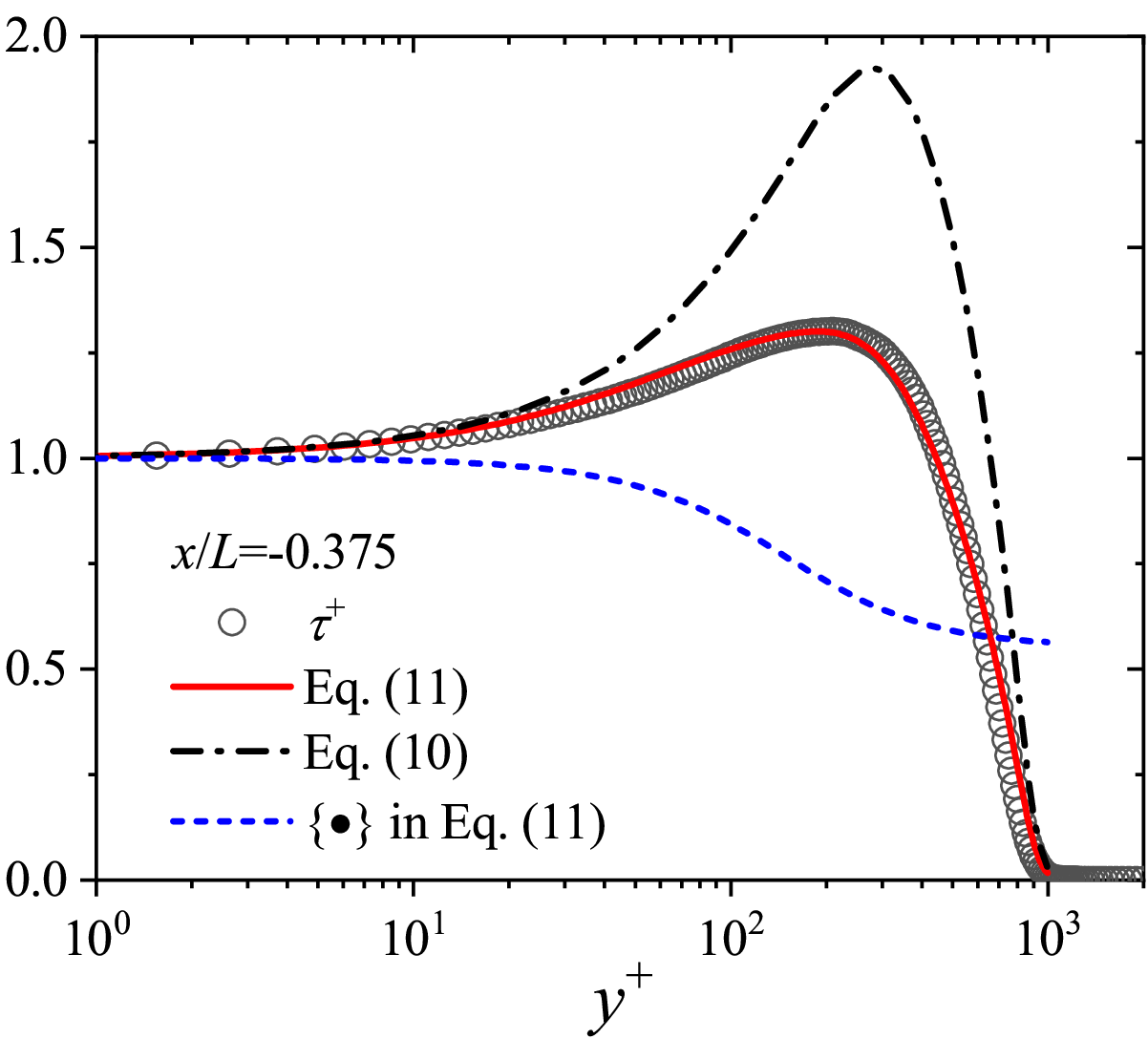}
    \includegraphics[width=0.33\linewidth]{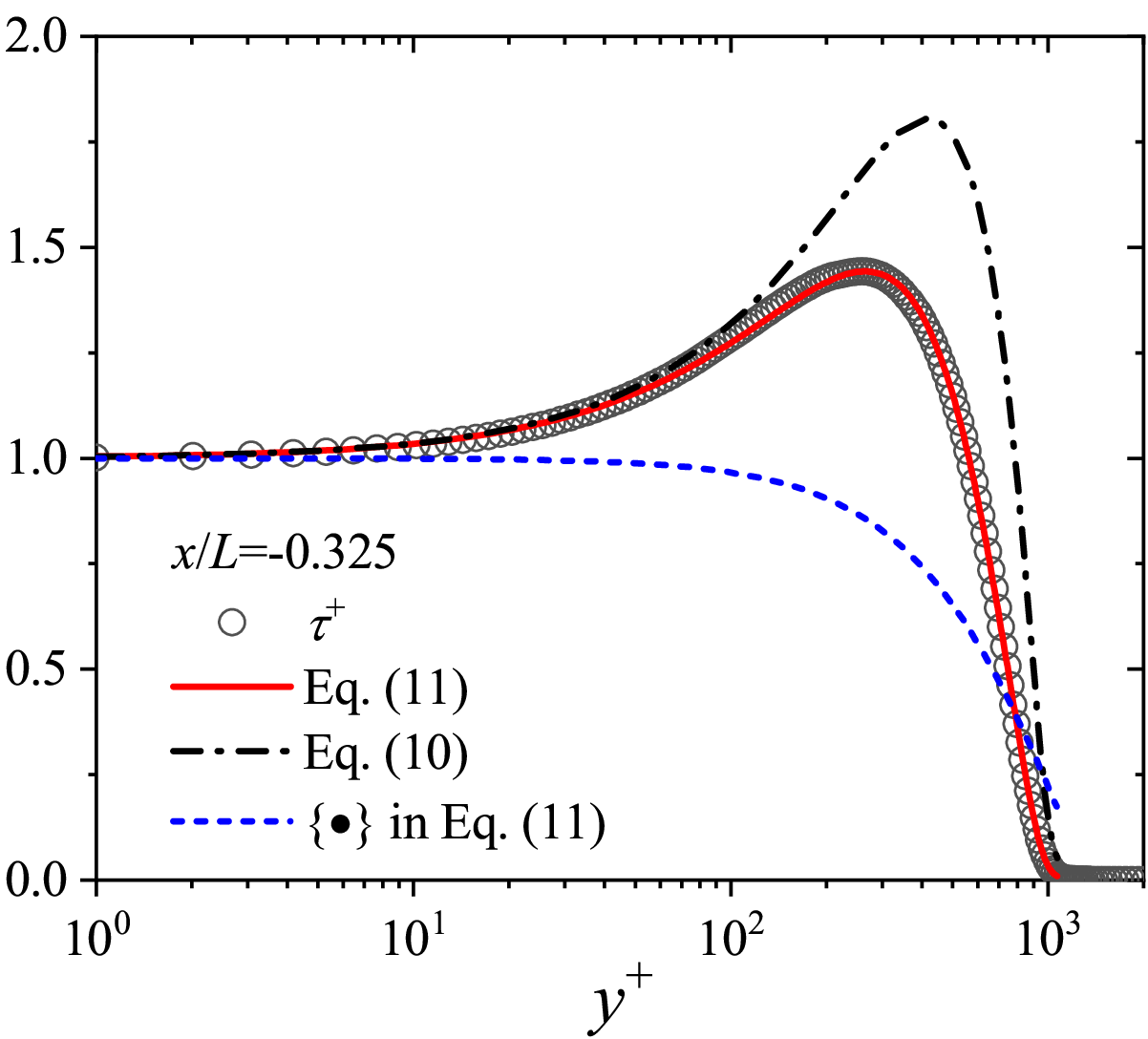}
    \includegraphics[width=0.33\linewidth]{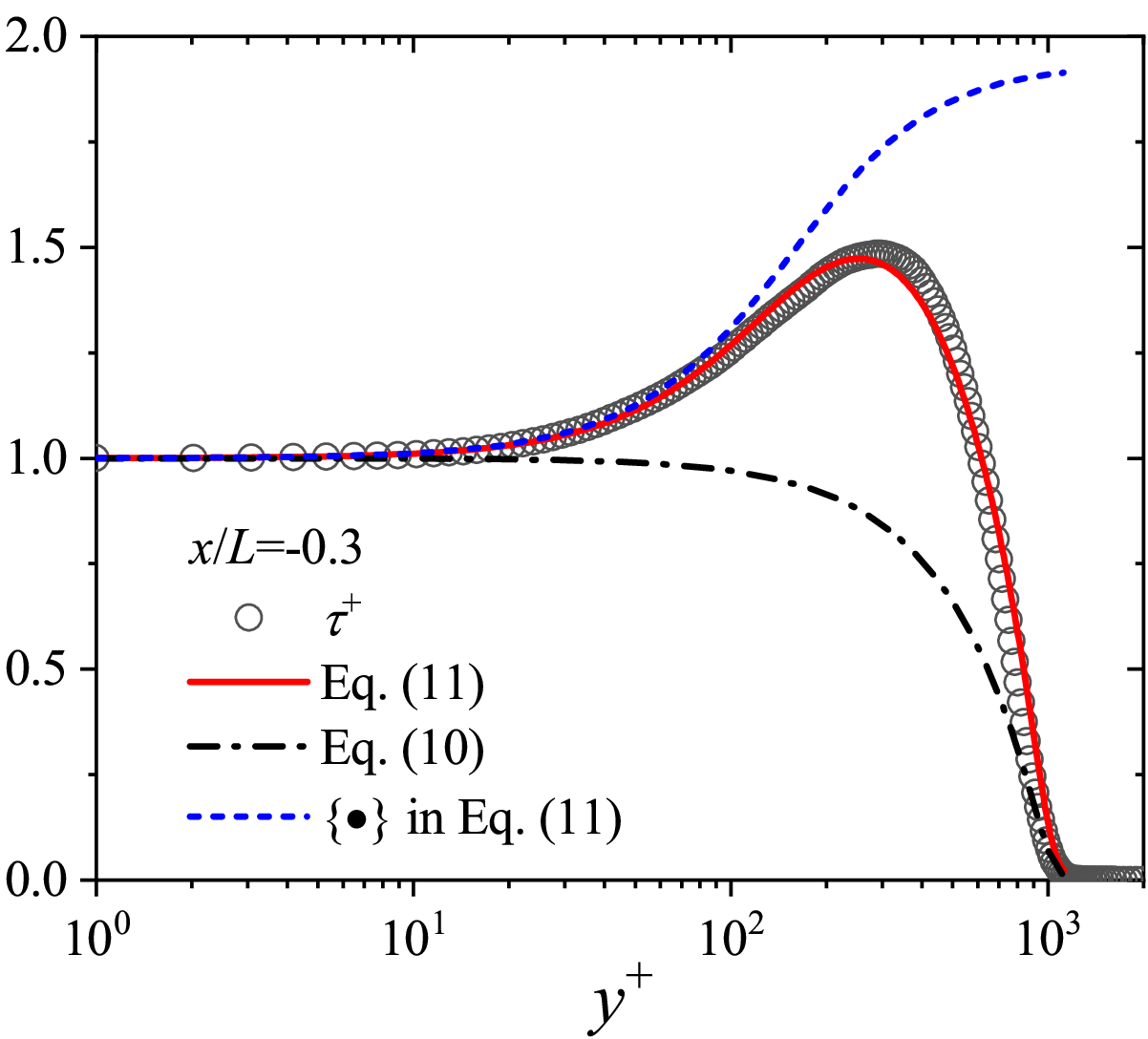}
    \\\quad(a)\quad\quad\quad\quad\quad\quad\quad\quad\quad\quad\quad\quad\quad\quad\quad\quad\quad\quad(b)\quad\quad\quad\quad\quad\quad\quad\quad\quad\quad\quad\quad\quad\quad\quad\quad\quad(c)
    \includegraphics[width=0.33\linewidth]{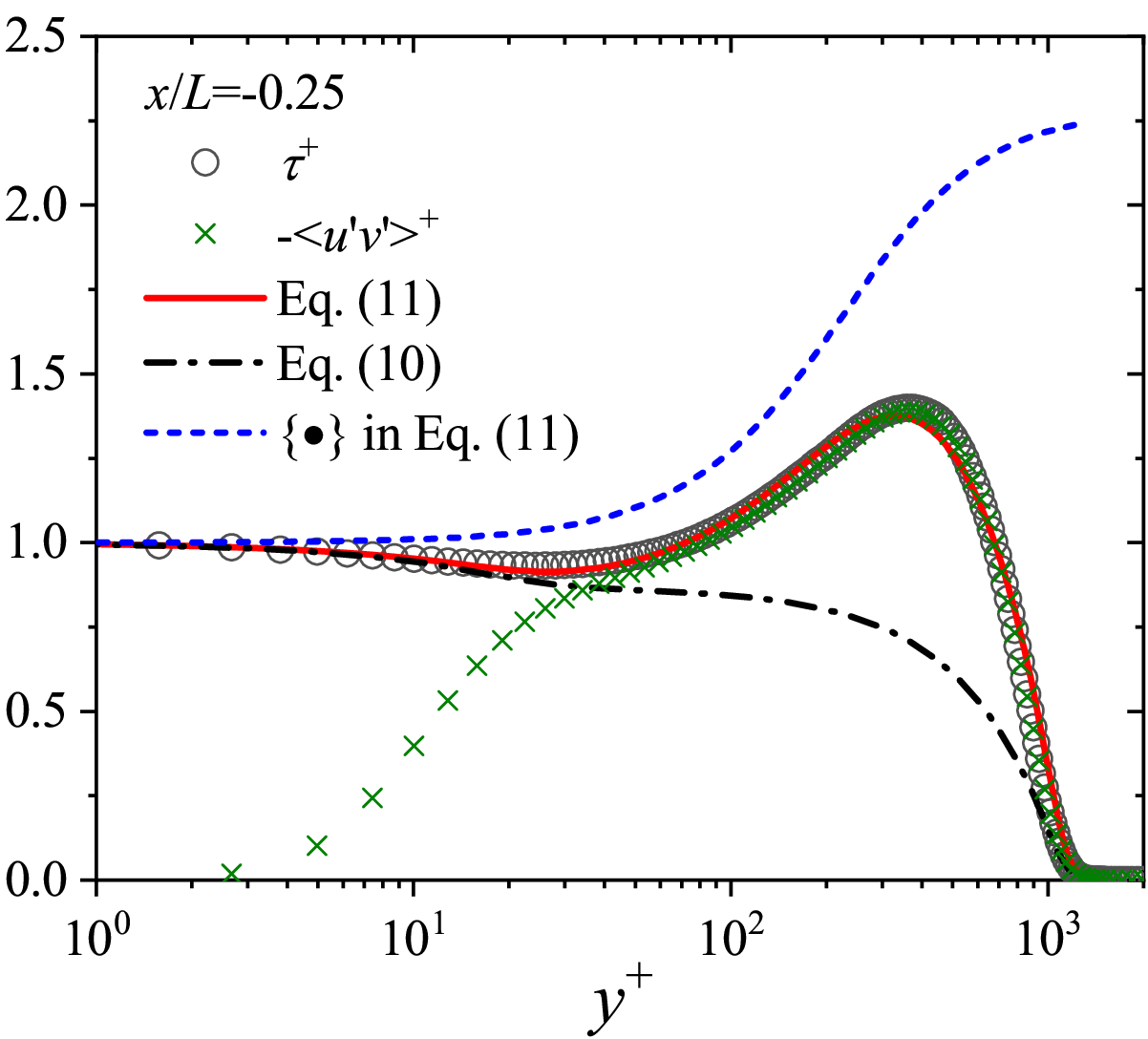}
    \includegraphics[width=0.33\linewidth]{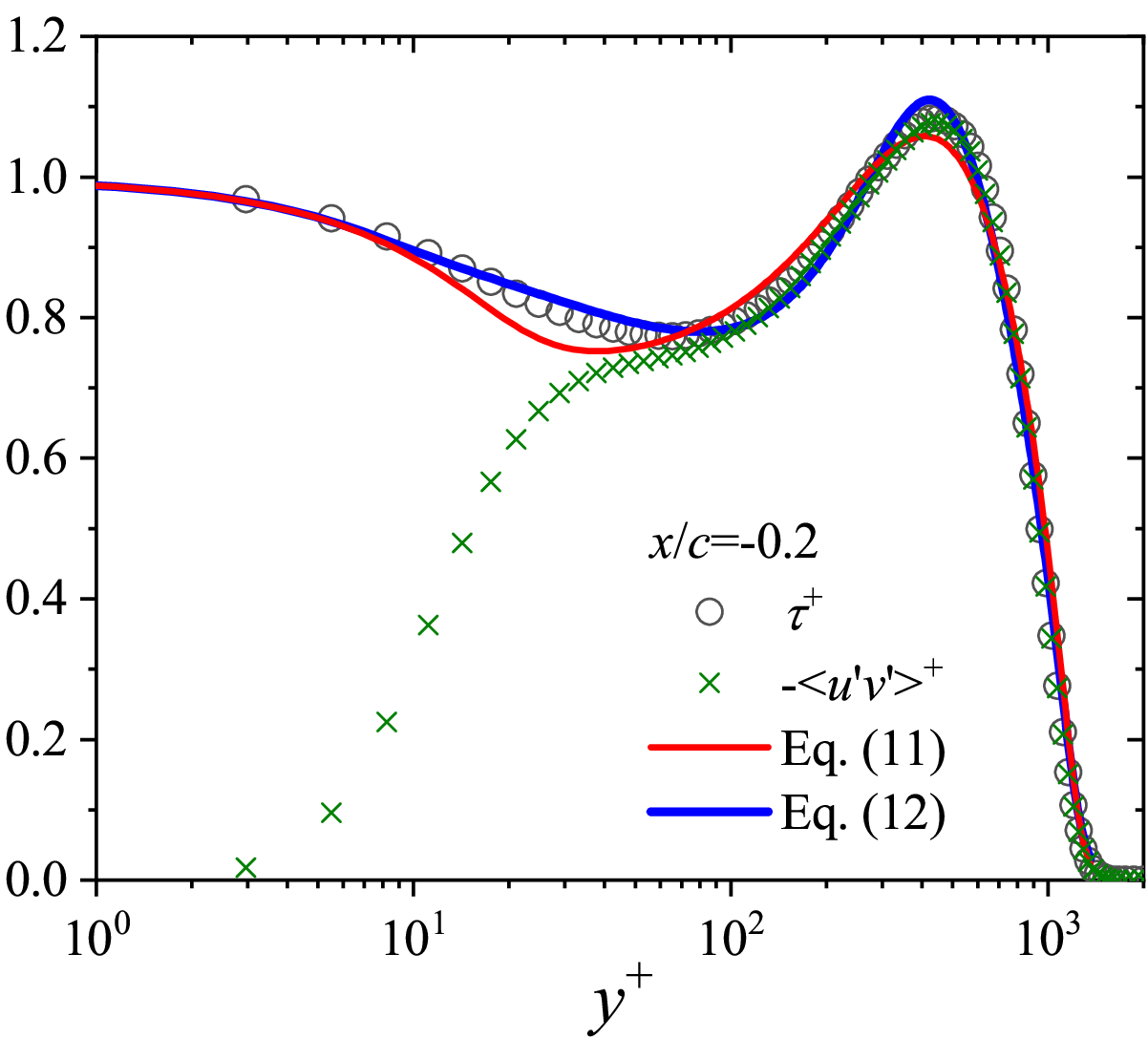}
    \includegraphics[width=0.33\linewidth]{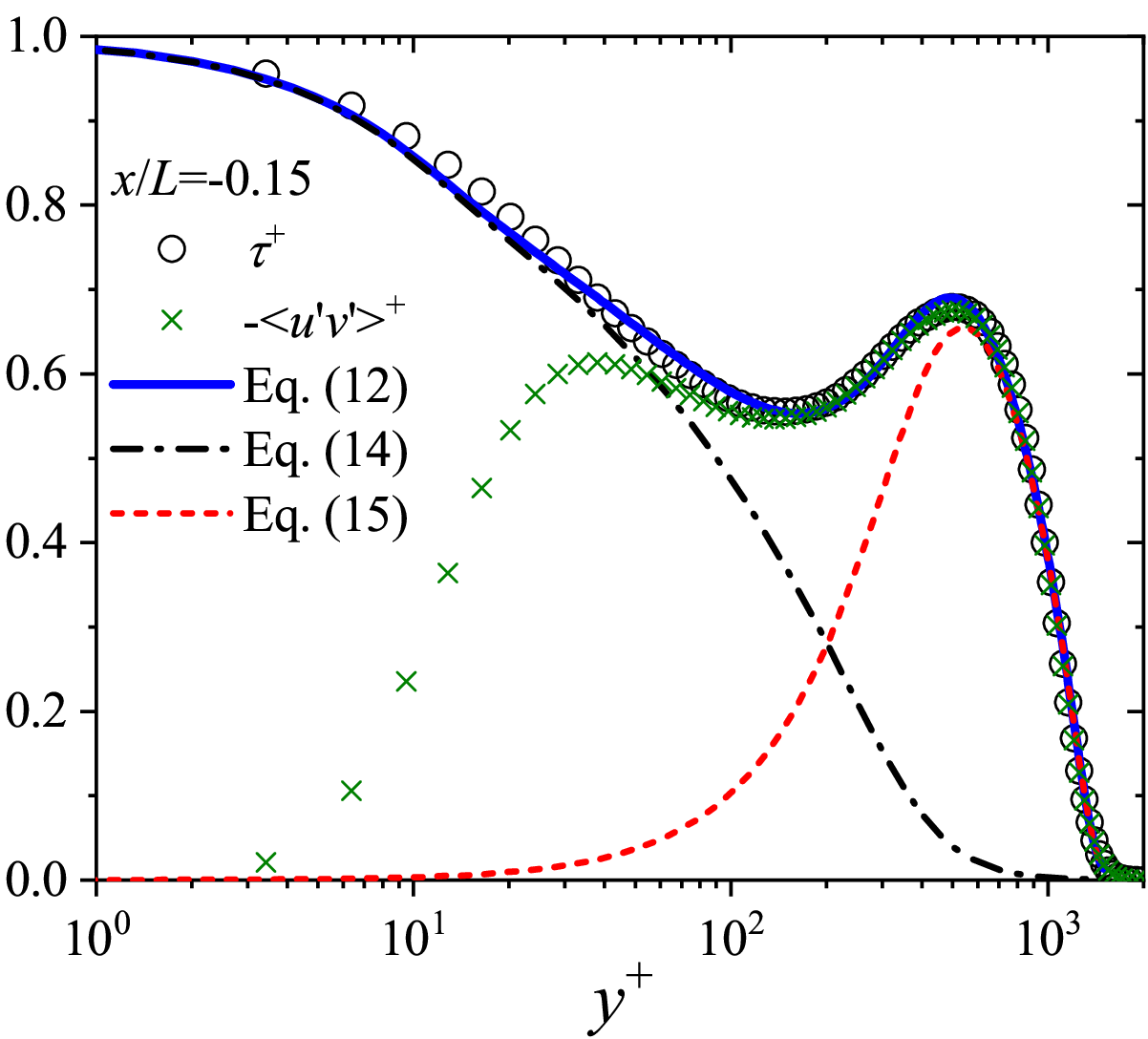}
    \\\quad(d)\quad\quad\quad\quad\quad\quad\quad\quad\quad\quad\quad\quad\quad\quad\quad\quad\quad\quad(e)\quad\quad\quad\quad\quad\quad\quad\quad\quad\quad\quad\quad\quad\quad\quad\quad\quad(f)
\caption{Comparisons between the TSS models and numerical simulation data (circles) at different streamwise positions in the relaxing TBL. In panels (a)-(d), the three-layer TSS model predictions (Eq. (\ref{eq:tau_NonEPG_threelayer}), solid lines) are compared with reference equilibrium TSS solutions (Eq. (\ref{eq:tau_equili_APG}), dashed-dotted lines). The delay function (the brace term in Eq. (\ref{eq:tau_NonEPG_threelayer})) is illustrated as the short-dashed line. Panel (e) presents a comparative analysis of both the three-layer TSS model and the dual-boundary-layer TSS model (Eq. (\ref{eq:tau_noneq})) against numerical simulation data at $x/L=-0.2$. In panel (f), the dual-boundary-layer TSS formulation (solid line) is validated at $x/L=-0.15$, with comparative references to the equilibrium FPG TSS solution of the IBL (Eq. (\ref{eq:tau_noneq_in_FPG}), dashed-dotted line) and the outer-flow Reynolds shear stress profile (Eq. (\ref{eq:tau_residue_stress}), short-dashed line). Cross symbols denote the Reynolds shear stress from numerical simulation. The model parameters, optimized through least-squares regression, are systematically summarized in Table \ref{tab:parameters_bump}.}     
  \label{fig:bump_tau_model_valid}
\end{figure*}

\begin{figure*}
    \centering
    \includegraphics[width=0.33\linewidth]{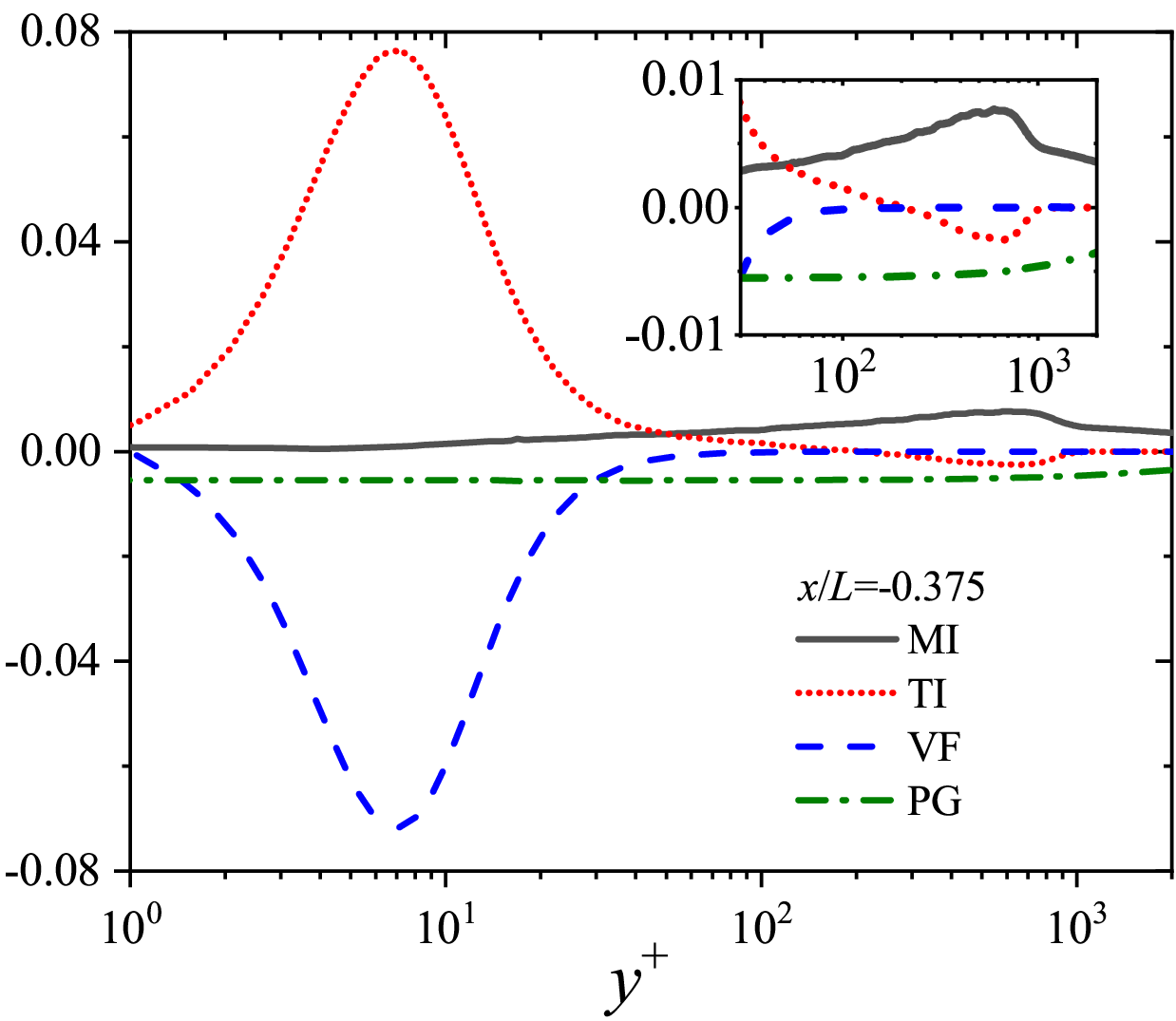}
    \includegraphics[width=0.33\linewidth]{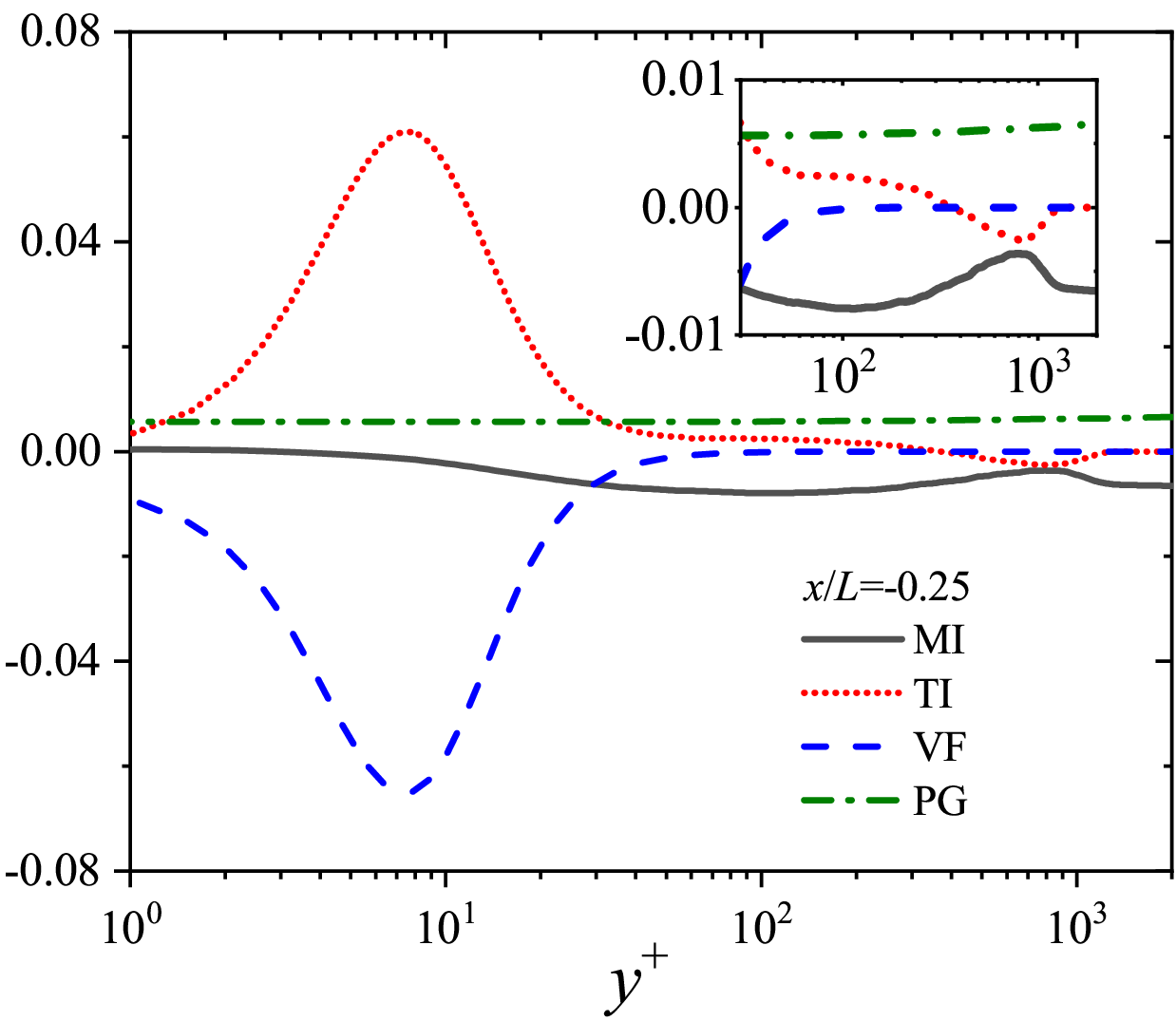}
    \includegraphics[width=0.33\linewidth]{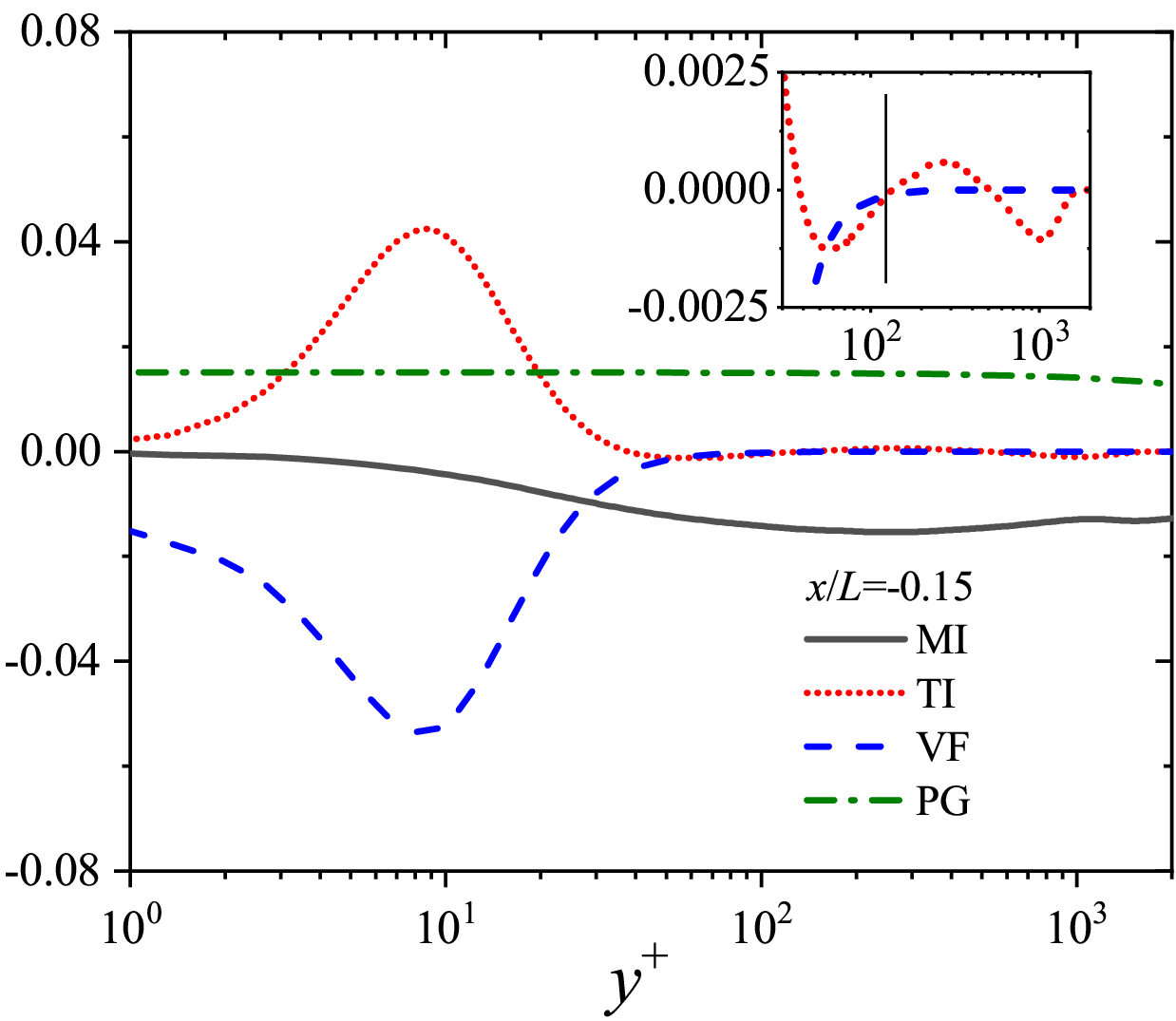}
    \\\quad(a)\quad\quad\quad\quad\quad\quad\quad\quad\quad\quad\quad\quad\quad\quad\quad\quad\quad\quad(b)\quad\quad\quad\quad\quad\quad\quad\quad\quad\quad\quad\quad\quad\quad\quad\quad\quad(c)
  \caption{Balance of the streamwise mean momentum equation (Eq. (\ref{eq:x_momentum})) at (a) $x/L=-0.375$, (b) $x/L=-0.25$, and (c) $x/L=-0.15$ in the relaxing TBL regime. $y$ is the wall-normal coordinate. MI is the mean inertia term, PG is the pressure gradient term, TI is the turbulent inertia term, and VF is the viscous stress term. Insets are zoomed in on the outer region. In the panel (c) inset, the vertical solid line marks the IBL boundary.}
  \label{fig:MMB_bump}
\end{figure*}

We aim to quantify the evolution dynamics of the relaxing TBL with the symmetry-based TSS models. Six streamwise positions in the relaxing TBL regime are studied, as labeled with arrows in Fig. \ref{fig:bumpstat}(b). As illustrated in Fig. \ref{fig:bump_tau_model_valid}, the near-wall shear stress gradient ($P_w^+$, listed in Table \ref{tab:parameters_bump}) exhibits progressive attenuation under decaying APG and subsequent FPG development, inducing concave profile formation in the intermediate region between inner and outer flow regimes. This curvature emerges from differential response timescales: while the near-wall region rapidly adjusts to $P_w^+$ changes, outer flow turbulence exhibits a delayed reaction to diminishing $P_w^+$. Notably, Table \ref{tab:parameters_bump} and panels (a)-(c) in Fig. \ref{fig:bump_tau_model_valid} reveal sustained growth of outer peak Reynolds shear stress ($\tau_{\rm {max\mbox{-}o}}^+$) persisting even as $P_w^+$ approaches vanishingly small values.\cite{Gungor2024} This apparent paradox implies self-sustaining turbulence mechanisms in the outer flow during relaxation. Such evolving interaction culminates in pronounced flow restructuring, characterized by structural decoupling between inner and outer regions as seen in Fig. \ref{fig:bump_tau_model_valid}(e) and (f).

This profound evolution can be quantified through the current TSS models. During the initial relaxation phase, the TBL exhibits non-equilibrium APG characteristics as in the wing section TBL. Fig. \ref{fig:bump_tau_model_valid}(a) demonstrates the TSS profile at $x/L=-0.375$ modeled through (\ref{eq:tau_NonEPG_threelayer}) using empirical parameters $y_P^+$, $c_m$, and $y_m^+$ from Table \ref{tab:parameters_bump}, compared against reference equilibrium profile and delay function. Significant TSS depletion in the outer region relative to the reference equilibrium stress reveals outer flow hysteresis, quantified through the delay function's damping effect. At $x/L=-0.325$ (Fig. \ref{fig:bump_tau_model_valid}(b)), this pattern persists but with reduced TSS-equilibrium discrepancies due to diminishing $P_w^+$, indicating attainment of transitional equilibrium soon after $x/L=-0.325$.

At $x/L=-0.3$ (Fig. \ref{fig:bump_tau_model_valid}(c)), the TBL has gone through the transitional equilibrium state. The positive $c_m$ values in the delay function lead the TSS profile to exceed equilibrium predictions, demonstrating outer flow hysteresis in response to diminishing $P_w^+$. This hysteresis persists at $x/L=-0.25$ where $P_w^+$ transitions to negative values (Fig. \ref{fig:bump_tau_model_valid}(d)). Notably, the reference equilibrium TSS profile in (\ref{eq:tau_NonEPG_threelayer}) remains governed by (\ref{eq:tau_equili_APG}) despite FPG dominance, which confirms that the formulation with negative $P_w^+$ in (\ref{eq:tau_equili_APG}) captures a pseudo-equilibrium state in FPG TBLs, as proposed by Zheng et al. \cite{ZhengBi2025} 

At $x/L=-0.2$ (Fig. \ref{fig:bump_tau_model_valid}(e)), the three-layer TSS model exhibits marked deviations from numerical simulation data in the intermediate regime between inner and outer flows, revealing limitations of the single dilation-breaking mechanism in capturing cumulative radical non-equilibrium effects. This insufficiency arises from progressive flow separation between inner and outer regions, prompting adoption of the dual-boundary-layer framework (Eq. (\ref{eq:tau_noneq})). The dual-boundary-layer model successfully reproduces TSS profiles at both $x/L=-0.2$ and $x/L=-0.15$ (Fig. \ref{fig:bump_tau_model_valid}(f)), using DNS-calibrated parameters $y_P^+$, $\delta_i^+/\delta^+$, $\delta_w^+/\delta^+$, and $W_{\rm {max}}^+$ detailed in Table \ref{tab:parameters_bump}. Notably, the thickness ratio reversal ($\delta_i^+>\delta_w^+$ at $x/L=-0.2$ versus $\delta_i^+<\delta_w^+$ at $x/L=-0.15$, see Table \ref{tab:parameters_bump}) signals a structural transition from partial overlap to distinct separation between IBL and outer flow. 

It is critical to identify the onset condition where the relaxing TBL splits into an IBL and an outer flow, triggering the TSS model transition from (\ref{eq:tau_NonEPG_threelayer}) to (\ref{eq:tau_noneq}). As shown in Fig. \ref{fig:bump_tau_model_valid}(e), this split initiates near $x/L=-0.2$ where the Reynolds shear stress exhibits a nascent inner peak. Thus, the onset of boundary layer split is defined by the critical condition where the Reynolds shear stress forms a knee profile. Such knee profiles are widely observed in radical non-equilibrium PG TBLs and accepted as evidence of IBL existence.\cite{Tsuji1976,Baskaran1987,Cavar2011} Here, this knee structure becomes fully evident at $x/L=-0.15$ (cross symbols in Fig. \ref{fig:bump_tau_model_valid}(f)) for the relaxing TBL.

She et al.\cite{she2010new,she2017quantifying,chen2018quantifying} established that multilayer structures arise from competing balance mechanisms in transport equations. The multilayer defect scaling of TSS stems from wall-normal changes of the mean momentum balance. Within Eq. (\ref{eq:x_momentum}), the dominant terms contributing to the momentum balance are:\cite{YanJFM2024} mean inertia (MI): $-u^+{\partial u^+}/{\partial x^+}-v^+{\partial u^+}/{\partial y^+}$; pressure gradient (PG): $-P^+$; viscous force (VF): ${\partial^2 u^+}/{\partial {y^+}^2}$; and turbulent inertia (TI): $-{\partial \langle u'v' \rangle^+}/{\partial y^+}$. In equilibrium PG TBLs, VF, PG, and TI dominate near the wall, with VF-PG balance closest to the wall and TI contributing above. Such a balance yields the near-wall linear TSS law (Eq. (\ref{eq:tau_nearwall})). Near the boundary layer edge, MI, PG, and TI dominate as TI decays to zero, resulting in the 3/2 defect scaling of TSS. In the intermediate regime, MI progressively replaces VF as a key player. Critically, TI crosses zero here, producing the distinct TSS peak.\cite{YanJFM2024} For the relaxing TBL, Fig. \ref{fig:MMB_bump}(a) and (b) show that this balance mechanism persists at $x/L=-0.375$ and $x/L=-0.25$. However, term magnitudes and key locations (e.g., zero-crossing or peak points) shift from equilibrium values due to non-equilibrium effects, which are captured by the delay function in (\ref{eq:tau_NonEPG_threelayer}). At $x/L=-0.15$, however, the balance mechanism completes its wall-normal development around $y^+\approx110$ (marked by the vertical solid line in Fig. \ref{fig:MMB_bump}(c) inset), signifying a fully developed IBL. The breakdown of the conventional balance mechanism necessitates quantification via the dual-boundary-layer framework.


\subsection{TBLs subjected to FPG-APG transition and/or rapidly increasing APG} \label{subsubsec:increaseAPG}
TBLs experiencing rapid FPG-APG transitions also exhibit dual-boundary-layer behavior, which is widely observed in flows around wings, smooth obstacles, and bumps, and through converging-diverging channels. To validate our dual-boundary-layer TSS model in such flows, we employ DNS data from Laval et al.'s converging-diverging channel study.\cite{Laval2012JoT}

\begin{figure*}
    \centering
    \includegraphics[width=1\linewidth]{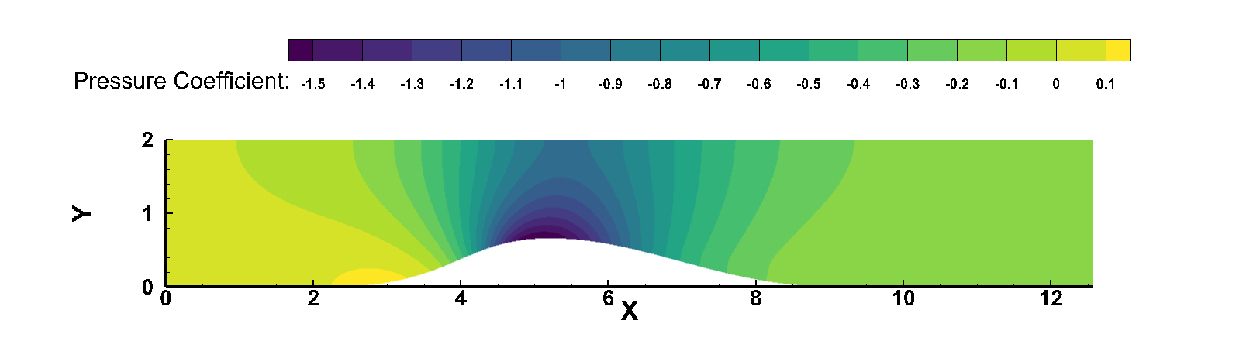}
  \caption{Configuration of the converging-diverging channel flow. The contour represents the pressure coefficient distribution.}
  \label{fig:configuration}
\end{figure*}

Laval et al.'s simulation focuses on an incompressible converging-diverging channel flow configuration, consisting of a lower curved wall and an upper flat wall (Fig. \ref{fig:configuration}). The lower wall features a bump with a convex central region and concave sections at the leading and trailing edges, ensuring geometric continuity with the flat wall sections. The simulation domain spans $4\pi$ in the streamwise $x$-direction, $2$ in the normal $y$-direction, and $\pi$ in the spanwise $z$-direction. The bump is located within the range $2<x<9$ and possesses a height of $2/3$ at $x\approx5.22$. The inlet flow is a fully-developed channel flow whose Reynolds number, defined based on the inlet friction velocity and half channel height, is $Re_\tau = 617$.

In this study, we have conducted RANS simulations on Laval et al.'s converging-diverging channel flow to investigate whether turbulence models can also capture the non-equilibrium behavior of TSS. Two classical turbulence models were investigated: the Spalart-Allmaras (SA) turbulence model and the Shear Stress Transport (SST) $k\mbox{-}\omega$ model. The SA turbulence model \cite{SA} formulates a single transport equation for modified eddy viscosity. This simplified framework balances computational efficiency with robust predictive capability, enabling its wide applications in RANS-based aerodynamic simulations. The SST $k\mbox{-}\omega$ model \cite{MenterSST} is a dominant two-equation turbulence closure for separated flows. It blends the $k\mbox{-}\omega$ (known for its near-wall accuracy \cite{Wilcox2006}) and $k\mbox{-}\epsilon$ (known for its competence in free-shear flows \cite{Kepsilon}) formulations through zonal blending functions. This hybrid architecture leverages the respective strengths of the parent models while maintaining computational tractability.

Numerical solutions were obtained using NASA's structured-grid RANS solver CFL3D, replicating Laval et al.'s DNS configuration.\cite{Laval2012JoT} As illustrated in Fig. \ref{fig:configuration}, the domain ($L_x\times L_y=4\pi\times2$) employed a 2D DNS-derived grid ($3361\times345$ nodes) with multiblock partitioning across 16 zones. Boundary conditions matched the reference DNS: no-slip adiabatic walls (upper/lower surfaces), extrapolation outflow, and DNS-sourced inflow profiles. Temporal integration used implicit approximate factorization, while spatial discretization combined second-order central differencing for viscous fluxes with upwind flux-difference splitting for inviscid terms. Grid convergence studies (omitted) confirmed solution robustness.

\begin{figure}
    \centering
    \includegraphics[width=0.75\linewidth]{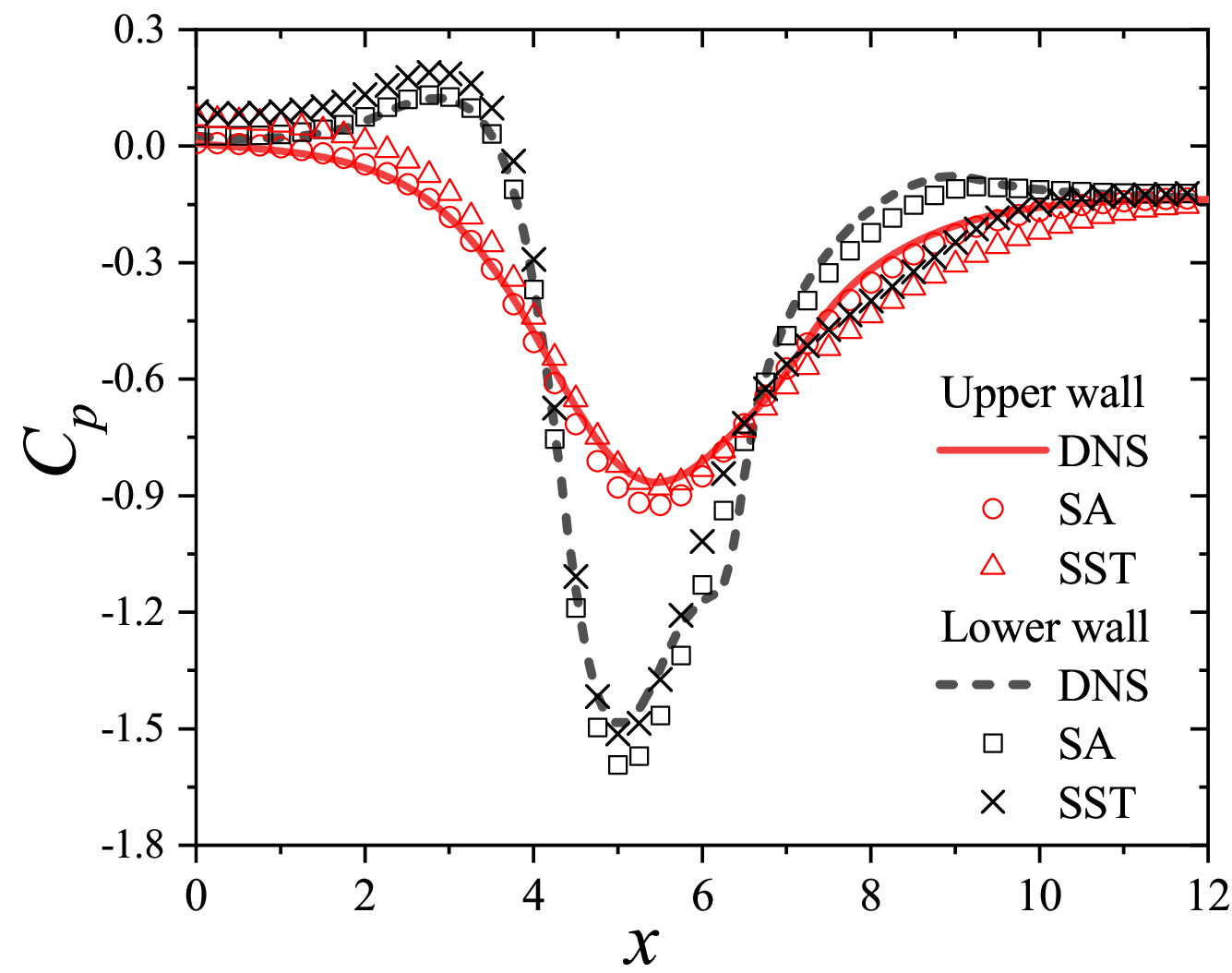}\\(a)\\
    \includegraphics[width=0.75\linewidth]{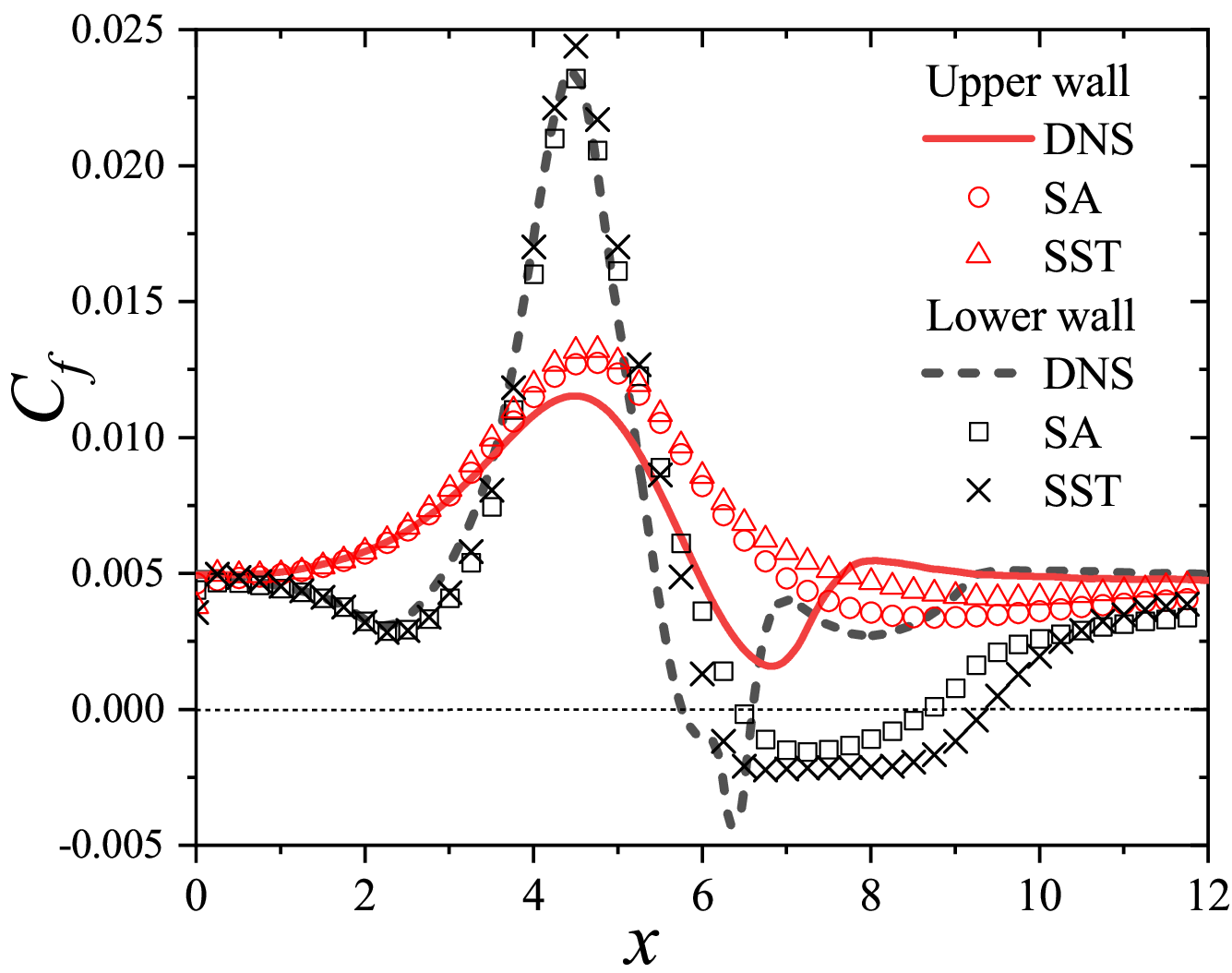}\\(b)
  \caption{Distributions of (a) surface pressure coefficient ($C_p$) and (b) skin friction coefficient ($C_f$) along the upper and lower walls of the converging-diverging channel, comparing DNS results \cite{Laval2012JoT} with RANS predictions using the SA and SST turbulence models.}
  \label{fig:Cp_Cf_channel_RANS}
\end{figure}

Fig. \ref{fig:Cp_Cf_channel_RANS} illustrates the distributions of the surface pressure gradient coefficient $C_p$ and skin friction coefficient $C_f$ computed by DNS and RANS. The RANS models approximate the $C_p$ distributions along both the upper and lower channel walls reasonably well. However, in the channel's diverging portion, the $C_f$ distributions predicted by both the SA and SST $k\mbox{-}\omega$ models show apparent deviations from the DNS. This discrepancy arises because both models incorrectly predict a delayed and significantly more extensive boundary layer separation on the lower wall, which subsequently influences the $C_f$ distribution on the upper wall.

\begin{figure*}
    \centering
    \includegraphics[width=0.33\linewidth]{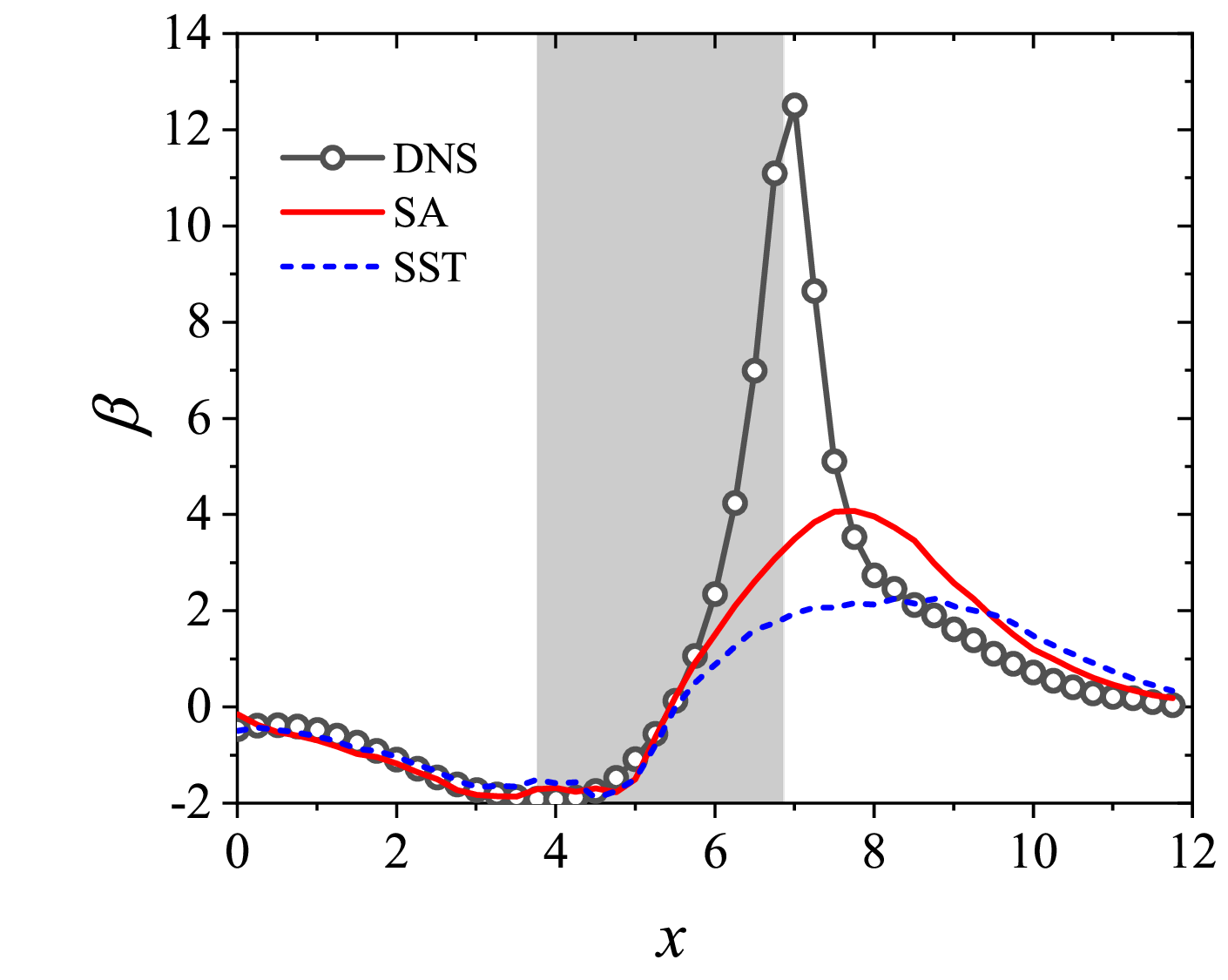}
    \includegraphics[width=0.33\linewidth]{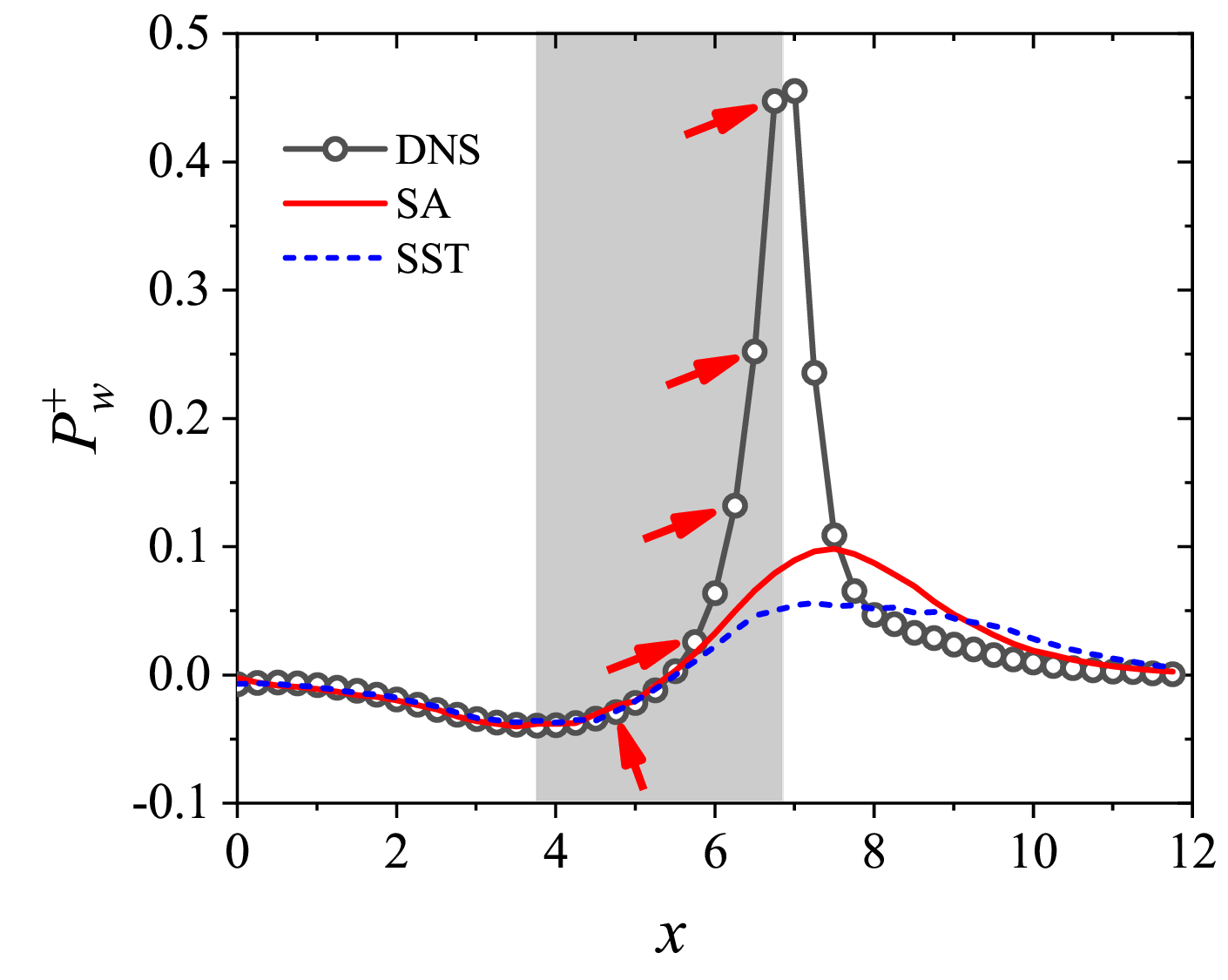}
    \includegraphics[width=0.33\linewidth]{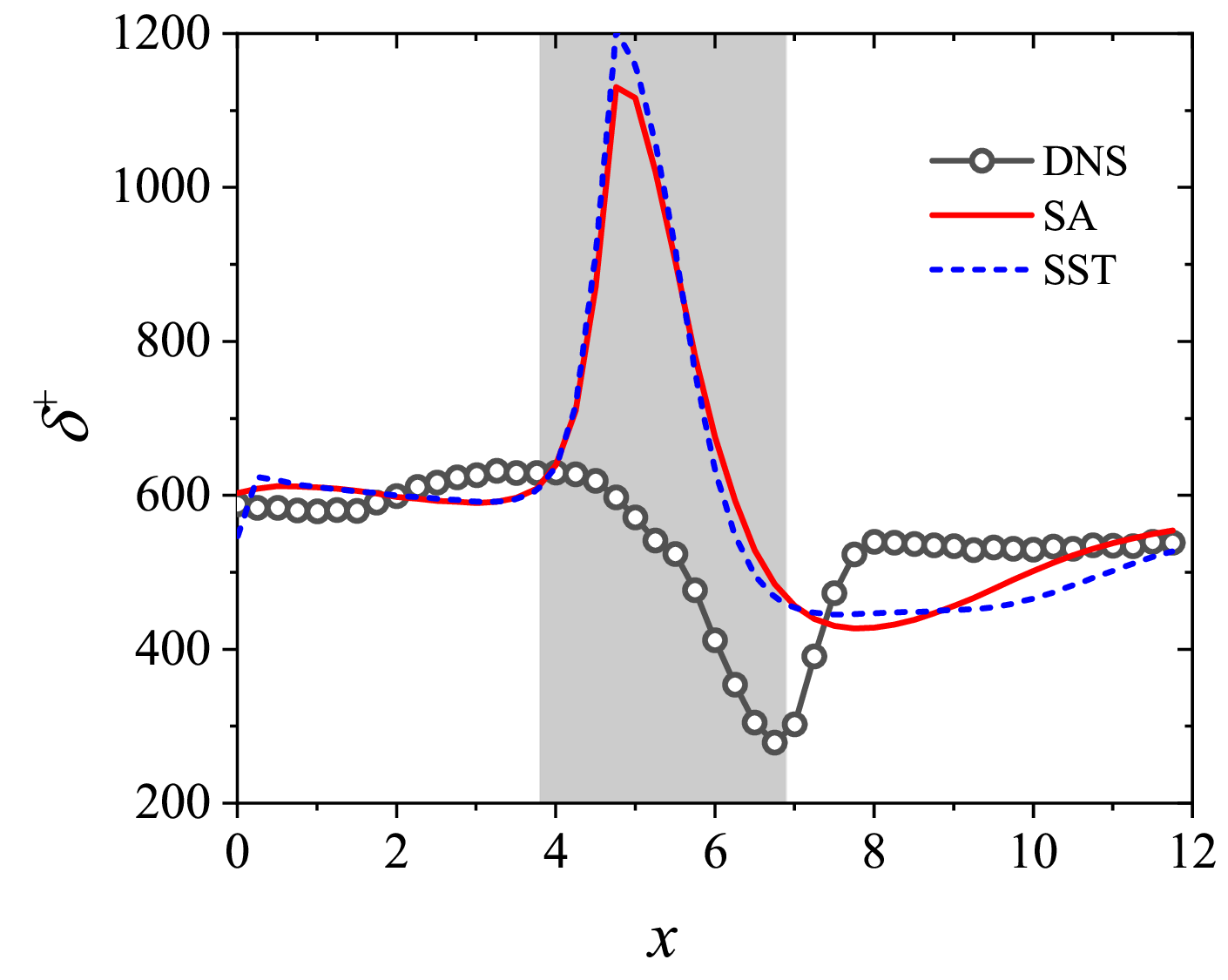}
    \\\quad(a)\quad\quad\quad\quad\quad\quad\quad\quad\quad\quad\quad\quad\quad\quad\quad\quad\quad\quad(b)\quad\quad\quad\quad\quad\quad\quad\quad\quad\quad\quad\quad\quad\quad\quad\quad\quad(c)
  \caption{Streamwise evolution of (a) the Clauser PG parameter $\beta$, (b) the PG parameter $P_w^+$, and (c) the friction Reynolds number $Re_\tau$ (i.e., $\delta^+$) along the upper wall of the converging-diverging channel. DNS data \cite{Laval2012JoT} are compared against RANS predictions using the SA and SST $k\mbox{-}\omega$ turbulence models. The gray-shaded region demarcates the investigated flow regime. Arrows in panel (b) indicate streamwise positions corresponding to the TSS profiles analyzed in Fig. \ref{fig:model_para_channel}.}
  \label{fig:channelstat}
\end{figure*}

Fig. \ref{fig:channelstat} shows the distributions of $\beta$, $P_w^+$, and $\delta^+$ (defined as the thickness with zero Reynolds shear stress because of channel flow) along the upper wall of the channel. As indicated by the $P_w^+$ distribution, the upper-wall TBL can be discernibly partitioned into three distinct regions: (1) an initial region ($0 < x < 3.8$) characterized by increasing FPG due to the converging channel, (2) an intermediate region ($3.8 < x < 6.9$, marked in gray) exhibiting a transition from decreasing FPG to rapidly increasing APG, which is the primary focus here, and (3) a region of relaxing TBL ($x > 6.9$) where the APG diminishes rapidly. Significant discrepancies between the DNS and RANS predictions occur in the intermediate and early relaxing regions (Fig. \ref{fig:channelstat}), highlighting the limitations of classical turbulence models for highly non-equilibrium PG TBLs and separated flows.

In the first region, the FPG generated by the converging channel significantly suppresses turbulence intensity within the boundary layer. Consequently, as the upper TBL undergoes FPG-APG transition, an IBL develops instantly in the near-wall region, with its peak shear stress rapidly exceeding the substantially attenuated Reynolds shear stress of the outer region. This TSS profile corresponds to the dual-boundary-layer TSS model given by Eq. (\ref{eq:tau_noneq}). Since the IBL grows from the wall, we define its thickness ($\delta_i$) as equivalent to the lower boundary of the outer flow, such that $\delta_i = \delta_w$. Consequently, Eq. (\ref{eq:tau_noneq}) contains three empirical parameters: $y_P^+$, $\delta_i^+$, and $W_{\rm{max}}^+$, which are determined through least-squares fitting of the $\tau^+$ profile at each streamwise location.

\begin{figure}
    \centering
    \includegraphics[width=0.75\linewidth]{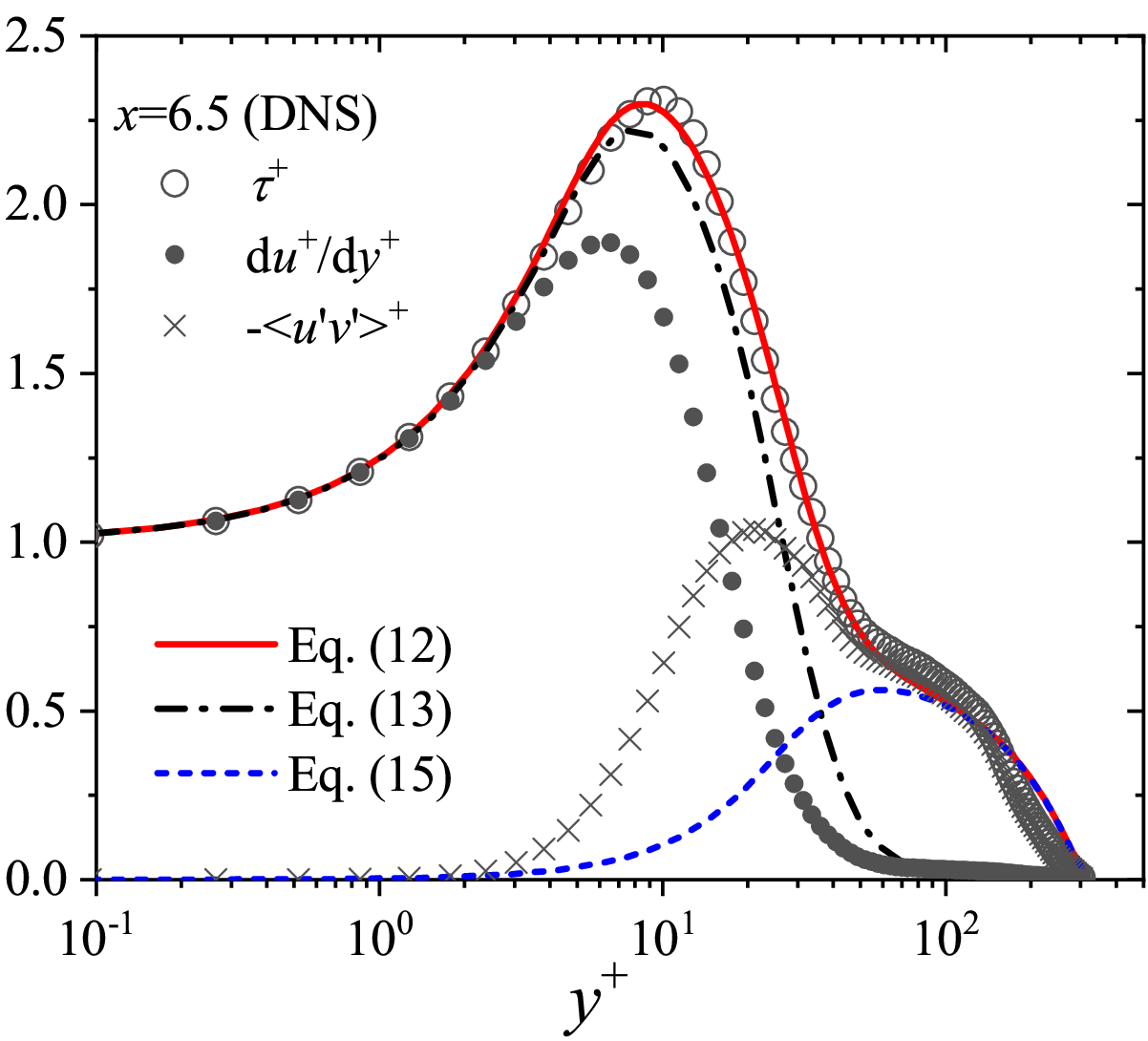}
  \caption{Profiles of the mean viscous shear stress (solid circles), the Reynolds shear stress (cross symbols), and the TSS (open circles) at $x=6.5$ for the upper wall TBL in the converging-diverging channel. The dual-boundary-layer TSS model (Eq. (\ref{eq:tau_noneq}), solid line) is validated, with comparative references to the equilibrium APG TSS solution of the IBL (Eq. (\ref{eq:tau_noneq_in_APG}), dashed-dotted line) and the outer-flow Reynolds shear stress profile (Eq. (\ref{eq:tau_residue_stress}), short-dashed line).  The model parameters, optimized through least-squares regression, are summarized in Fig. \ref{fig:model_para_channel}.}
  \label{fig:x27_tau_valid}
\end{figure}

Fig. \ref{fig:x27_tau_valid} illustrates the dual-boundary-layer structure of $\tau^+$ in the upper TBL at $x = 6.5$. The $\tau^+$ profile comprises two distinct components: an IBL exhibiting characteristic APG features with a pronounced peak shear stress, and an outer flow region retaining residual Reynolds shear stress inherited from upstream turbulence. Notably, the peak shear stress location at $y^+ \approx 10$ resides significantly closer to the wall than typically observed in equilibrium APG TBLs. The IBL demonstrates remarkable thinness ($\delta_i^+\approx32$), resulting in a significant viscous contribution to the TSS. 

The dual-boundary-layer TSS model (Eq. (\ref{eq:tau_noneq})), calibrated using DNS data, demonstrates excellent agreement with the complete $\tau^+$ profile measured in the DNS (Fig. \ref{fig:x27_tau_valid}). Decomposition analysis reveals distinct mechanistic contributions: the equilibrium-modeled internal APG component ($\tau_{\rm{in}}^+$ via Eq. (\ref{eq:tau_noneq_in_APG})) accurately reproduces near-wall stress distribution, confirming the equilibrium state of the IBL. Simultaneously, the outer flow's Reynolds shear stress profile is reasonably captured by the residual stress formulation $W_{\rm{out}}^+$ (Eq. (\ref{eq:tau_residue_stress})), with minor discrepancies near the boundary layer edge stemming from deviations from the $3/2$ defect power law in channel flow and uncertainties in boundary layer thickness definition. The observed equivalence $\delta_i^+=\delta_w$ suggests limited interaction between the IBL and outer flow during this non-equilibrium process. This spatial segregation allows independent modeling of the two flow regions while maintaining composite solution accuracy.

\begin{figure}
    \centering
    \includegraphics[width=0.75\linewidth]{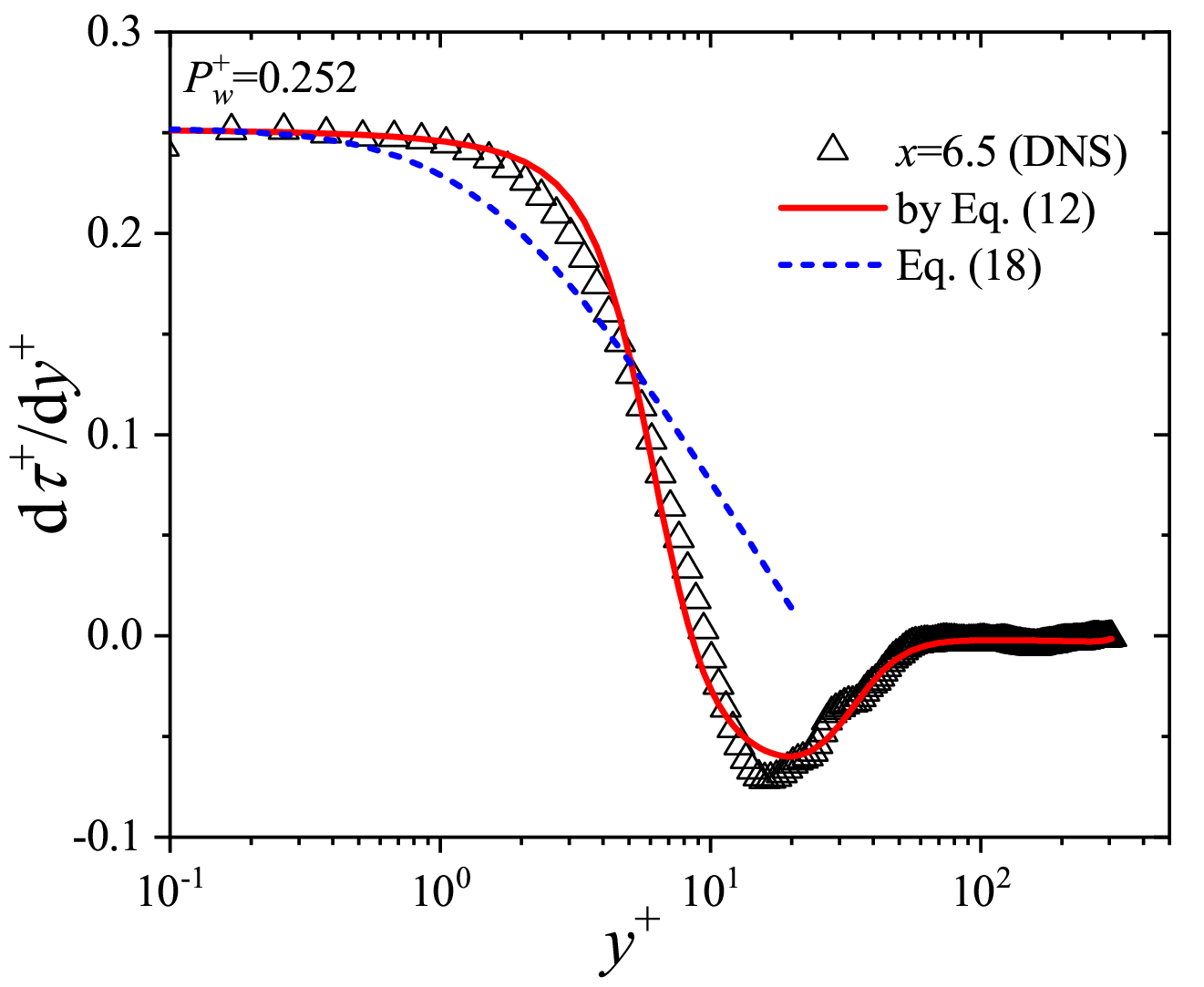}
  \caption{Comparison analysis of the ${\rm d}\tau^+/{\rm d}y^+$ profiles predicted by the dual-boundary-layer TSS model (Eq. (\ref{eq:tau_noneq}), solid line) and Ma et al.'s inner scaling (Eq. (\ref{eq:dtdy_Ma}), short-dashed line). Open triangles represent the DNS data from Laval et al. \cite{Laval2012JoT} }
  \label{fig:x27_dtaudy_valid}
\end{figure}

\begin{figure*}
    \centering
    \includegraphics[width=0.33\linewidth]{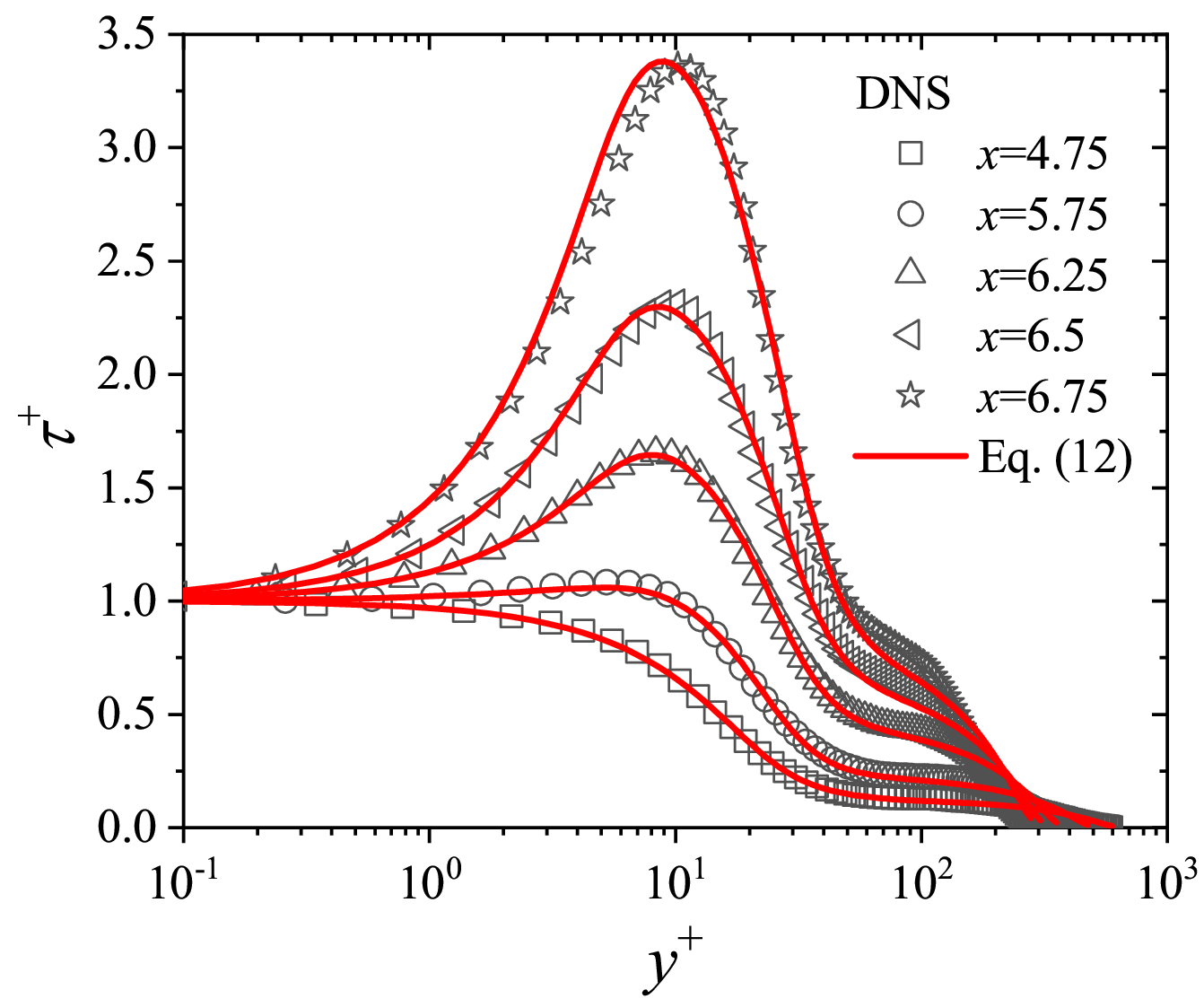}
    \includegraphics[width=0.33\linewidth]{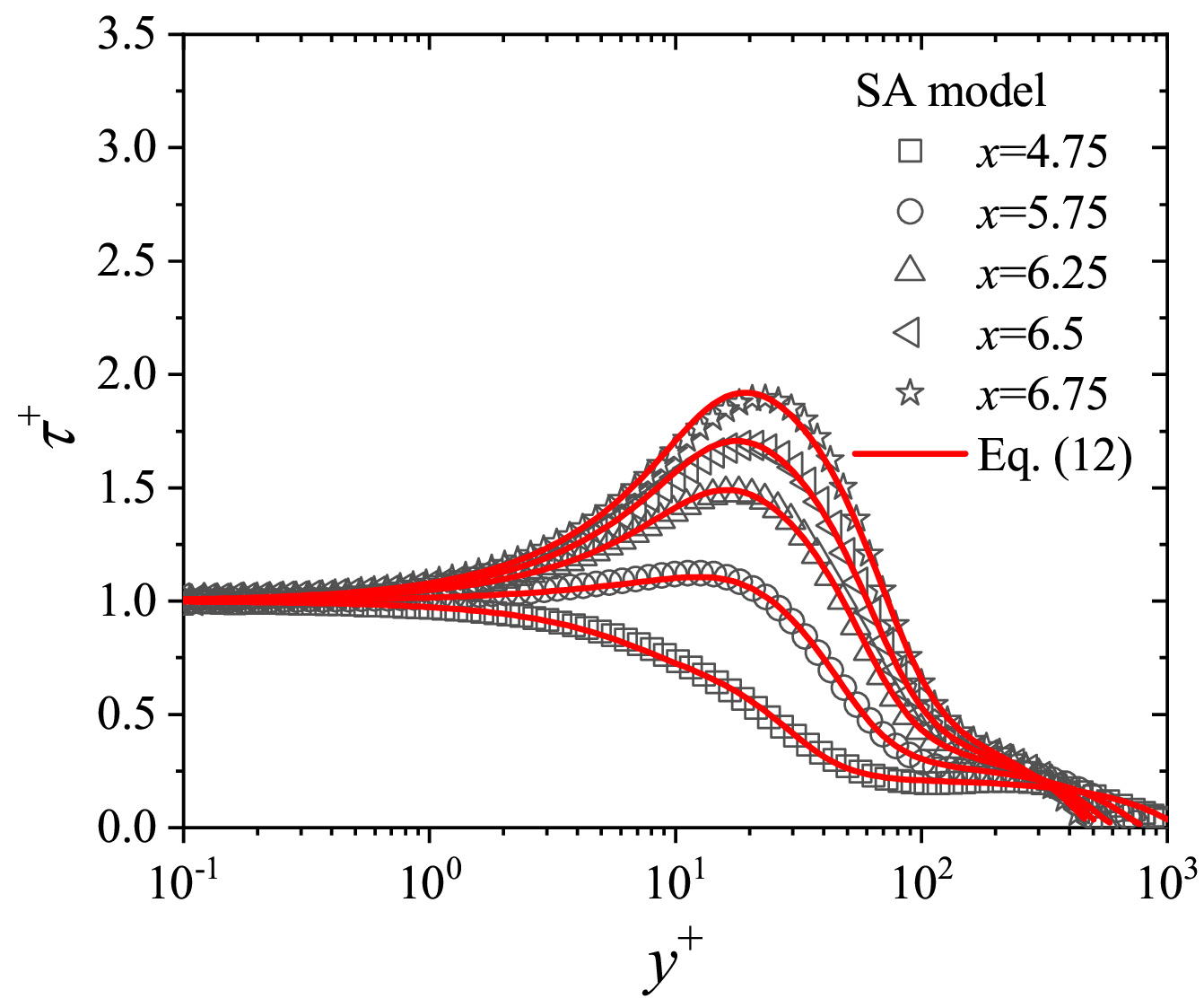}
    \includegraphics[width=0.33\linewidth]{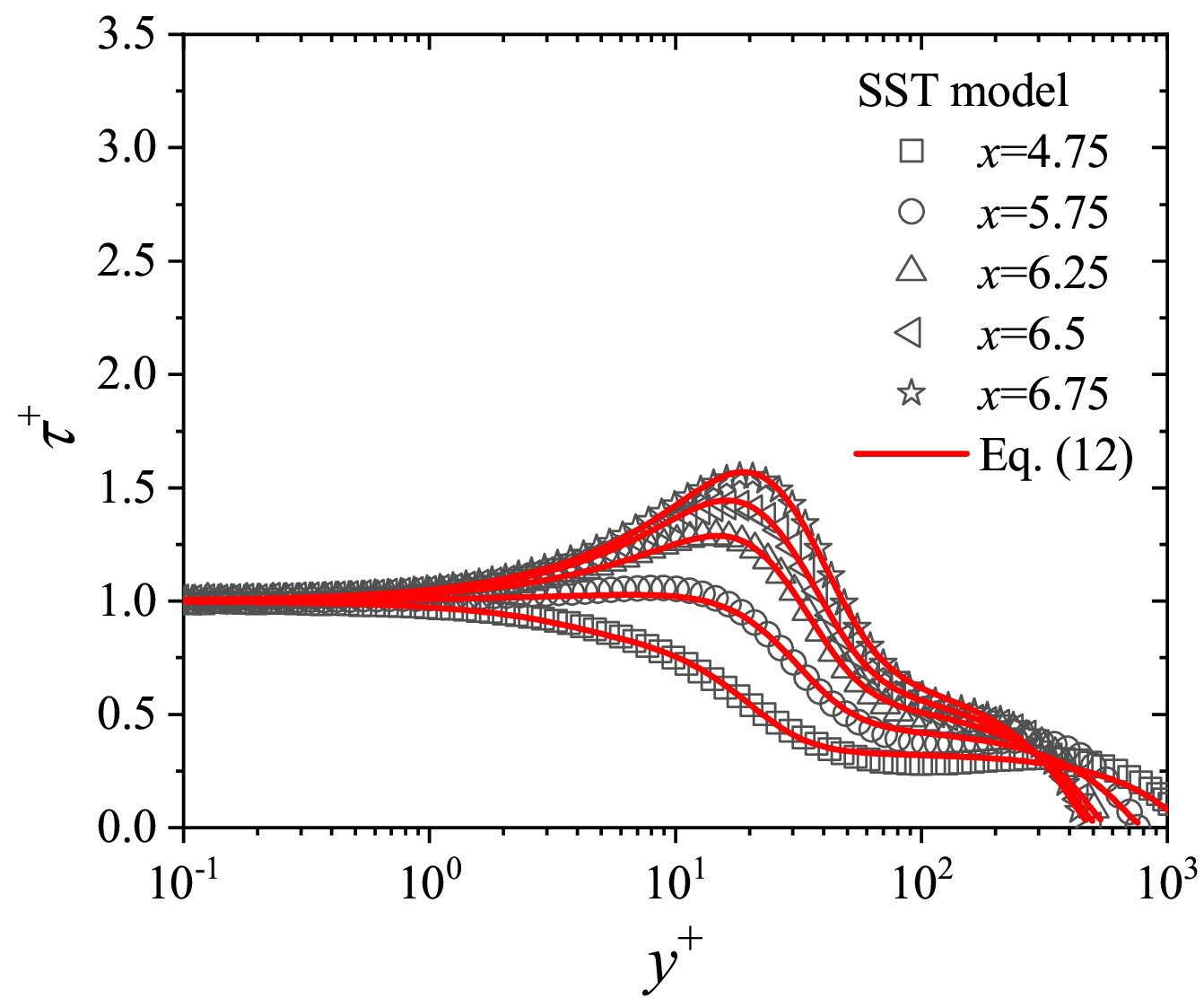}
    \\\quad(a)\quad\quad\quad\quad\quad\quad\quad\quad\quad\quad\quad\quad\quad\quad\quad\quad\quad\quad(b)\quad\quad\quad\quad\quad\quad\quad\quad\quad\quad\quad\quad\quad\quad\quad\quad\quad(c)
    \includegraphics[width=0.33\linewidth]{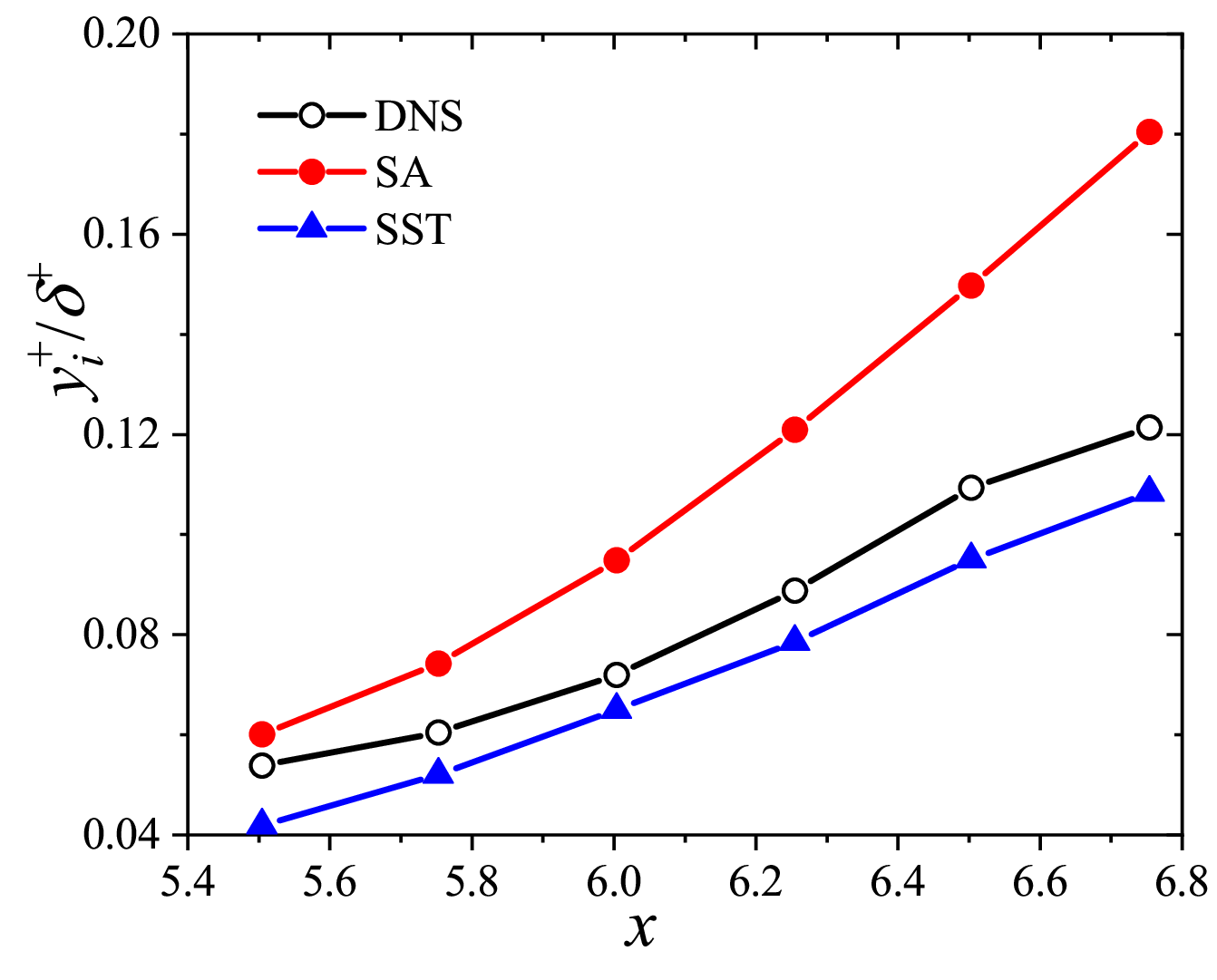}
    \includegraphics[width=0.33\linewidth]{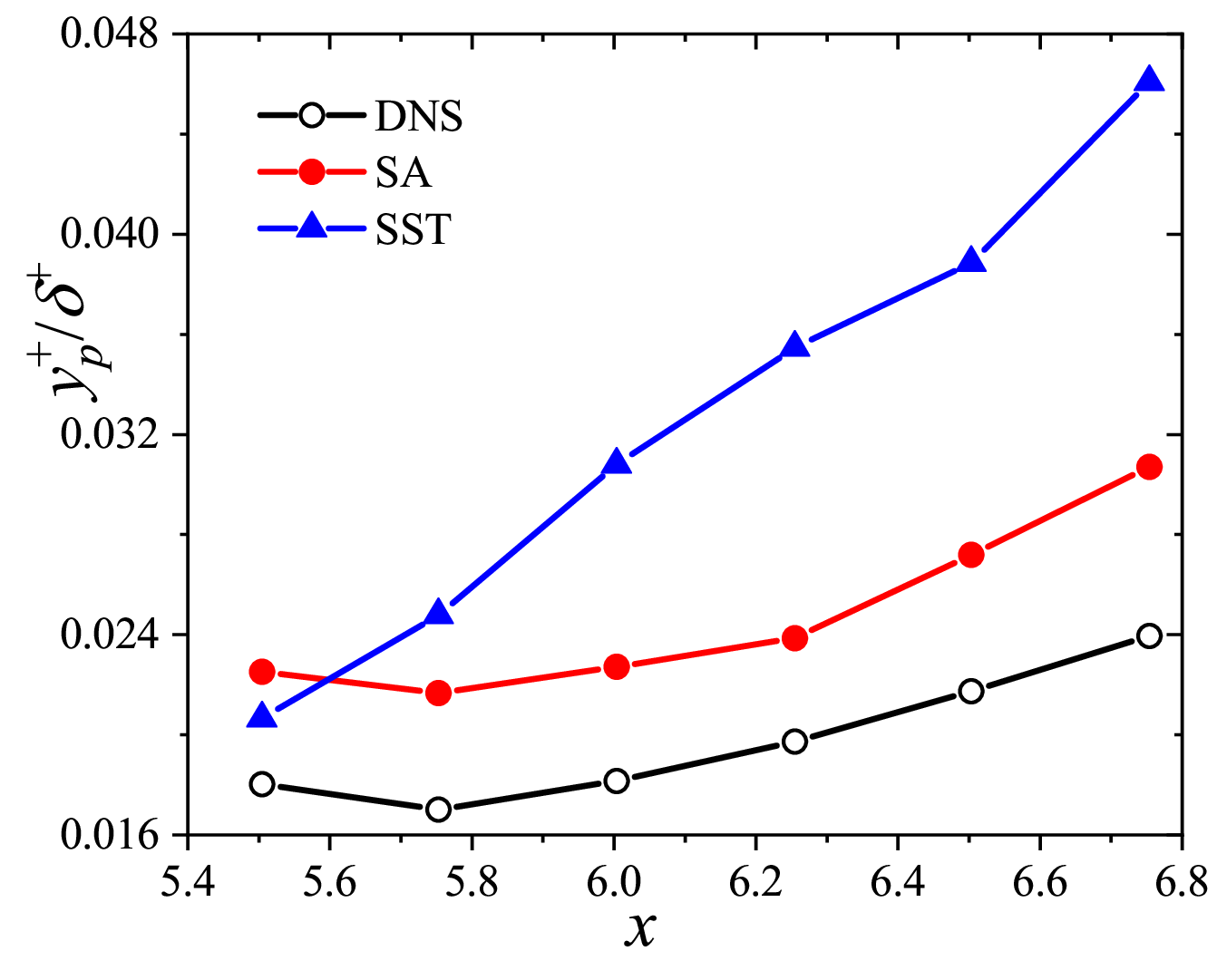}
    \includegraphics[width=0.33\linewidth]{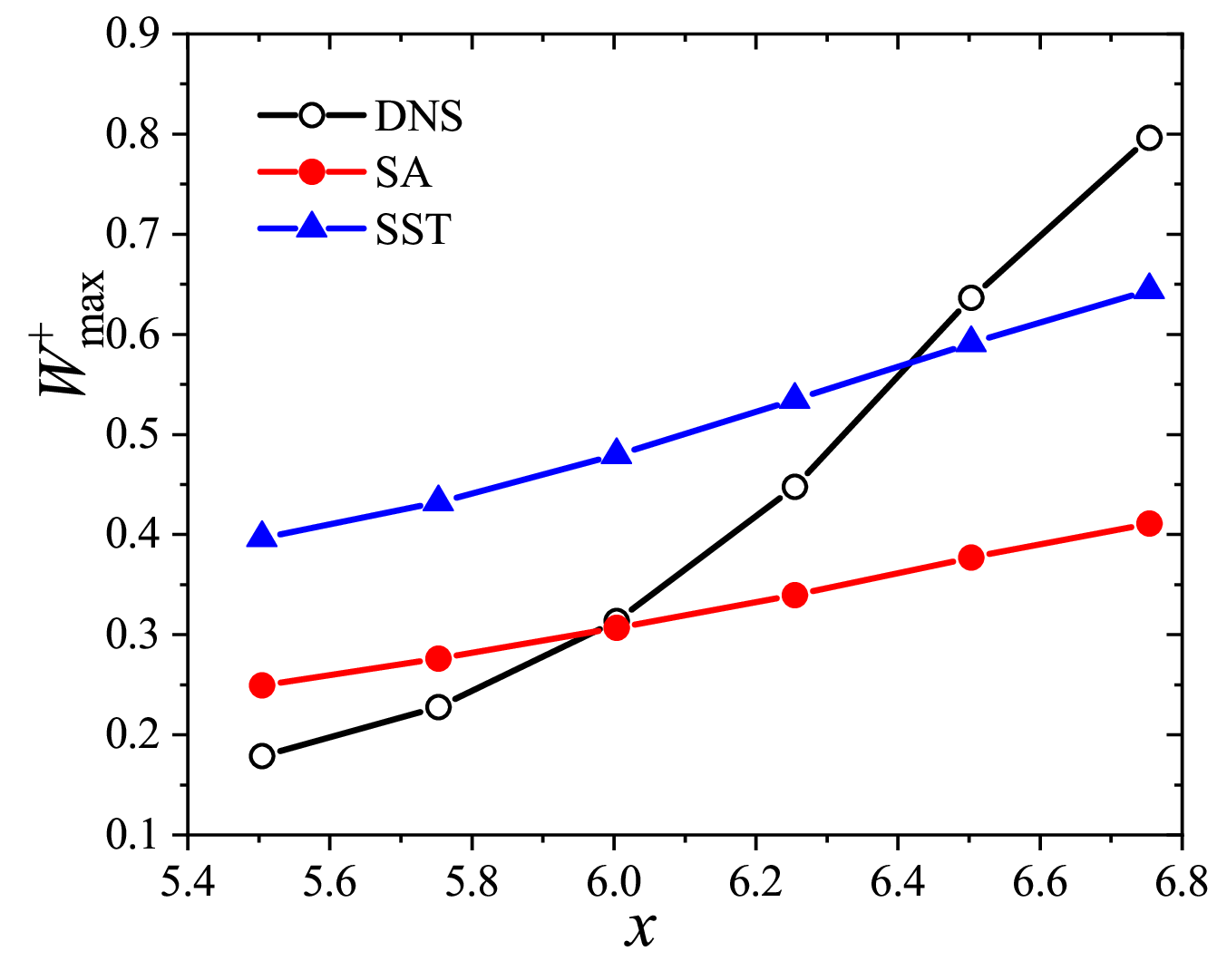}
    \\\quad(d)\quad\quad\quad\quad\quad\quad\quad\quad\quad\quad\quad\quad\quad\quad\quad\quad\quad\quad(e)\quad\quad\quad\quad\quad\quad\quad\quad\quad\quad\quad\quad\quad\quad\quad\quad\quad(f)
  \caption{Validation of the dual-boundary-layer TSS model (Eq. (\ref{eq:tau_noneq}), solid line) against TSS profiles at selected streamwise positions along the upper wall of the converging-diverging channel: (a) DNS, (b) the SA model, and (c) the SST $k\mbox{-}\omega$ model. Panels (d)-(f) present the streamwise evolution of the model parameters: $y_P^+$, $\delta_i^+$ (with $\delta_w^+=\delta_i^+$ enforced), and $W_{\rm{max}}^+$.}
  \label{fig:model_para_channel}
\end{figure*}

The emergence of IBL disrupts the inner scaling of $\tau^+$ established by Ma et al. for both equilibrium TBLs and pre-separation non-equilibrium APG TBLs. As Fig. \ref{fig:x27_dtaudy_valid} shows, while Ma et al.'s scaling formulation (Eq. (\ref{eq:dtdy_Ma})) fails to capture the ${\rm d}\tau^+/{\rm d}y^+$ profile at $x=6.5$, our dual-boundary-layer TSS model (\ref{eq:tau_noneq}) accurately reproduces the complete derivative profile. This discrepancy highlights a fundamental limitation of equilibrium-based scaling approaches: their inability to capture abrupt flow restructuring caused by IBL formation. In contrast, it demonstrates the predictive capability of our dual-mechanism model under strongly non-equilibrium conditions.

Panels (a)-(c) of Fig. \ref{fig:model_para_channel} present the streamwise evolution of $\tau^+$ computed by DNS and the two turbulence models for the upper-wall TBL undergoing FPG-APG transition and subsequent rapidly-intensifying APG conditions. In Fig. \ref{fig:model_para_channel}(a), as the APG intensifies downstream, a characteristic near-wall peak shear stress emerges and grows rapidly, signaling APG-induced IBL formation. The overlying outer flow maintains relatively low Reynolds shear stress magnitudes that gradually increase under APG influence. This differential response stems from the near-wall region's faster equilibrium adjustment to PG changes compared to the outer flow's delayed response.\cite{Smits1985} Progressive boundary layer development reveals an upward migration of the wall-normal position of the peak-stress concurrent with the thinning outer flow, suggesting that sustained APG would eventually lead to IBL dominance over the outer flow. Crucially, the dual-boundary-layer TSS model shows precise agreement with the DNS data across all streamwise positions and throughout the boundary layer depths, validating its capability to capture non-equilibrium transitional processes as the TBL evolves from initially FPG to rapidly intensifying APG conditions.

Interestingly, while the TSS magnitudes predicted by both SA and SST $k\mbox{-}\omega$ models are substantially lower than DNS results, their profile shapes resemble DNS shapes, and are captured by the current dual-boundary-layer model (Fig. \ref{fig:model_para_channel}(b) and (c)). This similarity raises a critical question: Does the multilayer defect scaling in TSS fundamentally require realistic turbulence physics? Scaling laws are statistical emergences from nonlinear interactions among numerous multi-scale turbulent eddies, a mechanism inherently absent in RANS simulations. However, as established in Wilcox's famous book \cite{Wilcox2006} and our recent analysis of $k\mbox{-}\omega$ model equations for pipe flow,\cite{AMM2023} multilayer scaling structures can be embedded within turbulence models, which is achieved through judicious calibration of the balance mechanisms in the model transport equations against canonical flows. In non-equilibrium flows, these inherent multilayer scaling characteristics appear qualitatively preserved by RANS models (explaining the shape similarity). However, quantitative accuracy deteriorates significantly due to the departure from calibrated equilibrium conditions.

Fig. \ref{fig:model_para_channel}(d)-(e) presents the streamwise evolution of the three model parameters. Both $y_P^+/\delta^+$ and $\delta_i^+/\delta^+$ exhibit progressive growth with downstream development, indicating continuous expansion of the IBL. This expansion persists despite the decrease in $\delta^+$ caused by viscous length scale enlargement under intensifying APG conditions. Concurrently, $W_{\rm{max}}^+$ exhibits a gradual increase due to APG effects, although its growth rate remains substantially lower than the intensification of the IBL's peak shear stress. The DNS shows the latter enhancing dramatically from 1 to 3.4 over the observed domain (Fig. \ref{fig:model_para_channel}(a)). While RANS simulations exhibit similar parameter trends in the streamwise evolution, their parameters differ quantitatively from the DNS results. 

\begin{figure}
    \centering
    \includegraphics[width=0.75\linewidth]{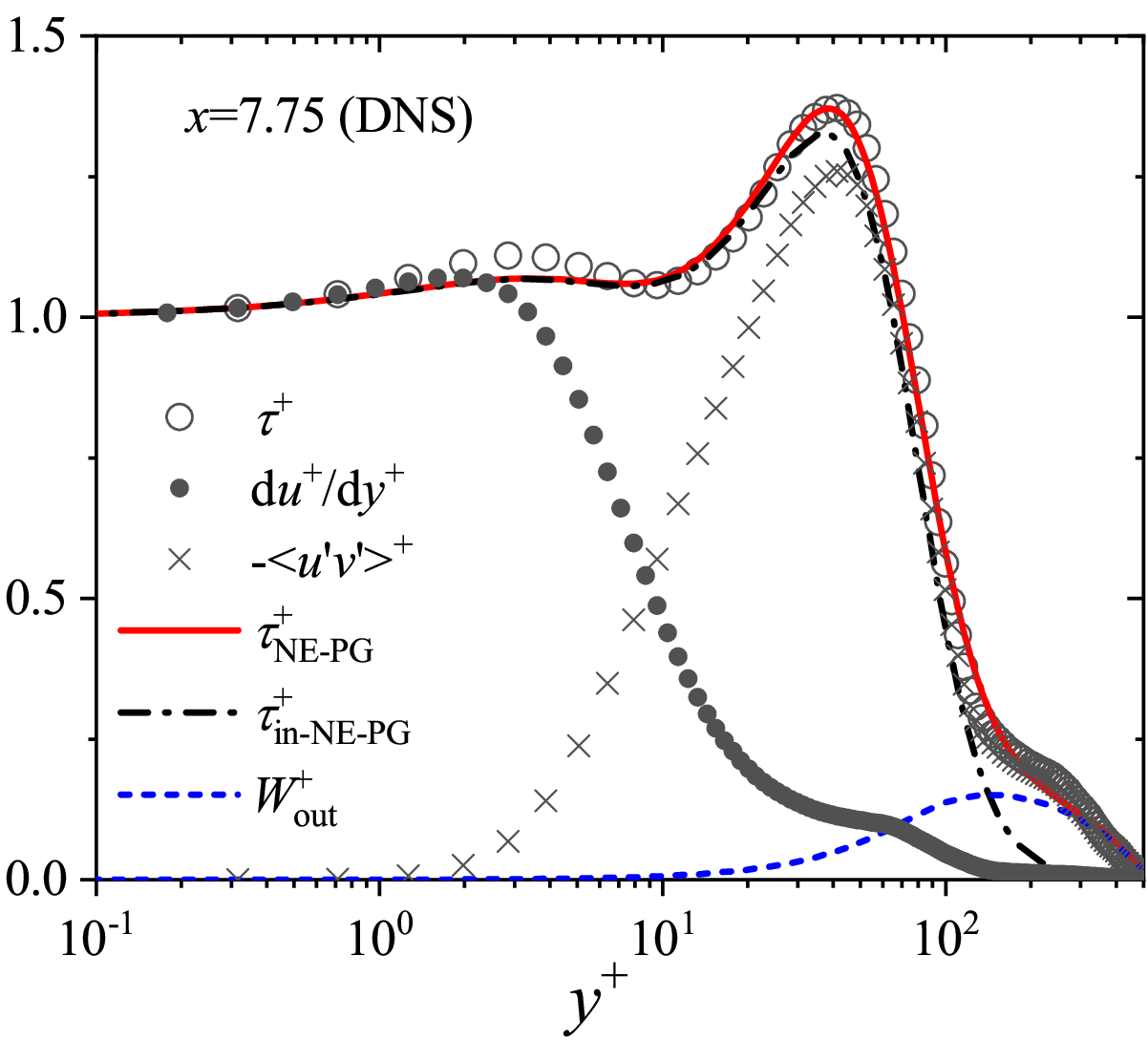}
  \caption{Profiles of the mean viscous shear stress (solid circles), the Reynolds shear stress (cross symbols), and the TSS (open circles) at $x=7.75$ on the upper surface of the converging-diverging channel. The TBL develops a three-boundary-layer structure when the IBL undergoes relaxation. The modified TSS model (Eq. (\ref{eq:three_BL_TSS_model}), solid line) is validated, with comparative references to $\tau_{\rm{in\mbox{-}NE\mbox{-}PG}}^+$ of the non-equilibrium IBL and $W_{\rm {out}}^+$ of the outer flow. The model parameters, $c_m=-0.48$, $y_m^+=7.8$, $\delta_i^+=97$, and $W_{\rm{max}}^+=0.19$ ($y_P^+=0.491\delta_i^+$ and $\delta_w^+=\delta_i^+$ are enforced to reduce the parameter number), are optimized through least-squares regression of the complete TSS profile.}
  \label{fig:x32_channel_valid}
\end{figure}

Finally, we consider the TBL's subsequent development in the relaxing regime ($x>6.9$) featured by rapid APG attenuation - a process addressed in Section \ref{subsubsec:decreaseAPG}. As Fig. \ref{fig:x32_channel_valid} illustrates, the IBL undergoes relaxation during APG decay while maintaining partial segregation from the outer flow. This transient process exposes limitations in the current dual-boundary-layer framework, which inherently presumes IBL equilibrium. A straightforward modification resolves this limitation: substituting the equilibrium component $\tau_{\rm{in}}^+$ in (\ref{eq:tau_noneq}) with the non-equilibrium formulation (\ref{eq:tau_NonEPG_threelayer}): 
\begin{equation}
  \tau_{\rm {NE\mbox{-}PG}}^+=\tau_{\rm{in\mbox{-}NE\mbox{-}PG}}^++W_{\rm {out}}^+,
  \label{eq:three_BL_TSS_model}
\end{equation}
where $\tau_{\rm{in\mbox{-}NE\mbox{-}PG}}^+$ preserves the functional structure of (\ref{eq:tau_NonEPG_threelayer}) but adopts $\delta_i^+$ as the characteristic IBL thickness. This extension enables modeling of transient relaxing dynamics while preserving the original framework's residual stress treatment. Validation in Fig. \ref{fig:x32_channel_valid} confirms that the enhanced model accurately captures the complex TSS profile using only four empirically calibrated parameters, demonstrating an optimal balance between physical fidelity and parametric efficiency.

\section{Conclusion and discussion}\label{sec:conclusion}
In this paper, we introduce a novel symmetry-based modeling approach for TSS profiles in general non-equilibrium PG TBLs. The objective of this research is twofold: first, to generate new predictions of the TSS distributions, and second, to elucidate the underlying mean-flow structures that characterize these complex flows. Given that mean-flow properties of TBLs are difficult to deduce directly from first principles, scaling laws (or dilation symmetries), which arise from the self-organization of turbulent flows, can be identified from empirical observations. These empirically derived scaling laws are then employed to enhance TBL modeling. Unlike prior approaches, our symmetry-based framework formulates the scaling properties of TSS, which enable new models to quantify general non-equilibrium PG TBLs. This approach not only advances the theoretical understanding of complex PG TBLs but also paves the way for more accurate predictions.

The framework postulates a multilayer defect scaling property for TSS profiles. While our recent research validated that a two-layer formulation successfully captures equilibrium flows,\cite{ZhengBi2025} this study considers that non-equilibrium effects activate dilation symmetry breaking in the two-layer structure of equilibrium TSS. With a single symmetry breaking, a three-layer defect scaling accurately describes the TSS profiles in gradually varying APG TBLs on the suction surfaces of NACA4412 and NACA0012 airfoils under different aerodynamic conditions. The new layer acts as a delay function. Its parameter $y_m$ characterizes the thickness separating the instant-response inner layer from the delayed-response outer region, while $c_m$ quantifies the cumulative hysteresis effect of the outer flow. Both parameters are modeled by incorporating existing scalings. A new velocity scale $u_{**}$ is introduced based on the three-layer defect scaling model, extending the hybrid inner velocity scale $u_*$ and $u_{hyb}$ to apply to the entire boundary layer. Normalized with $u_{**}$, entire TSS profiles in airfoil TBLs transform to those of canonical ZPG TBLs, exceeding previous inner and outer scalings.

The framework suggests that radical non-equilibrium PG TBLs (characterized by IBL emergence) involve more symmetry breaking. For these flows, the current study introduces a simple dual-boundary-layer formulation. This model proposes that the TSS is decomposed into an equilibrium IBL part and an outer history-dependent flow part, arising from the differential response time between the IBL and outer inertia-dominated turbulence. The efficacy of the dual-boundary-layer model is demonstrated through its application in capturing the TSS profiles and their streamwise evolution in relaxing TBLs over a Gaussian bump and TBLs experiencing abrupt FPG-APG transition in a converging-diverging channel flow. Both flows are radical non-equilibrium PG TBLs with IBL presence, beyond the scope of previous descriptions. Furthermore, the SA and SST $k\mbox{-}\omega$ turbulence models are demonstrated to reproduce the dual-boundary-layer structures. However, substantial discrepancies occur, revealing the limitations of these models owing to departures from calibrated equilibrium conditions in radical non-equilibrium PG TBLs.

Critically, the success of these models underscores the significance of dilation symmetry as a fundamental property of TBLs. This property can be employed for constructing a mean-field theory for complex TBLs. In particular, the universal dilation-symmetry-breaking function is identified as a critical element for capturing the crossover scaling phenomena prevalent in TBLs. 

Currently, the theory does not extend to separated TBLs and three-dimensional flows. Additionally, the dual-boundary-layer TSS model for radical non-equilibrium PG TBLs is not exhaustive, given the extensive diversity of these flows. For 2D attached TBLs, however, the theory provides a robust parameterization. This parameterization has a solid physical basis and high descriptive accuracy, establishing a novel foundation for understanding the streamwise development and mean-flow structures of these flows. Importantly, it offers a new physics-based parameter system for machine learning applications aiming at capturing complex wall turbulence.

\begin{acknowledgments}
The authors are grateful to Prof. J. P. Laval, Prof. R. Vinuesa, and Dr. A. Uzun for sharing publicly their numerical simulation data, without which this research could not have been conducted. The study was supported by the CSSC Open Fund under Grant No. NKLH2024KF02, and the NSFC under Grant No. 91952201.
\end{acknowledgments}

\section*{Data Availability Statement}
Data available on request from the authors.

\bibliography{NEqPGTBL}

\end{document}